\documentclass[review,hidelinks]{elsarticle}

\usepackage[displaymath,mathlines,pagewise]{lineno}
\usepackage{hyperref}
\usepackage{amsmath}
\usepackage{amssymb}
\usepackage{booktabs}
\usepackage{multirow}
\usepackage{color}
\usepackage{subfig}
\modulolinenumbers[1]

\DeclareSymbolFont{TXlettersA}{U}{txmia}{m}{it}
\SetSymbolFont{TXlettersA}{bold}{U}{txmia}{bx}{it}
\DeclareFontSubstitution{U}{txmia}{m}{it}
\DeclareMathSymbol{\deltaup}{\mathord}{TXlettersA}{14}

\newcommand{\pd}[2]{\frac{\partial #1}{\partial #2}}
\newcommand{\pdF}[2]{\frac{\deltaup #1}{\deltaup #2}}
\newcommand{\pdFl}[2]{\deltaup #1/\deltaup #2}

\newcommand{\pdl}[2]{\partial #1 / \partial #2}

\newcommand{\Sec}[1]{Section\ \ref{#1}}
\newcommand{\Secs}[2]{Sections\ \ref{#1} -- \ref{#2}}

\newcommand{\Eq}[1]{Eq.\ (\ref{#1})}
\newcommand{\Eqs}[2]{Eqs.\ (\ref{#1}) and (\ref{#2})}
\newcommand{\Fig}[1]{Fig.~\ref{#1}}
\newcommand{\Figs}[2]{Figs.\ \ref{#1} -- \ref{#2}}

\newcommand{\lr}{\left(}
\newcommand{\rr}{\right)}

\newcommand{\ls}{\left[}
\newcommand{\rs}{\right]}

\newcommand{\av}[1]{\langle #1 \rangle}
\newcommand{\avG}[1]{\langle #1 \rangle_\Gamma}

\newcommand{\be}{\begin{equation}}
\newcommand{\blnm}{\begin{linenomath*}}
\newcommand{\ee}{\end{equation}}
\newcommand{\elnm}{\end{linenomath*}}

\newcommand{\bspl}{\begin{split}}
\newcommand{\espl}{\end{split}}
\newcommand{\bea}{\begin{eqnarray}}
\newcommand{\eea}{\end{eqnarray}}
\newcommand{\bagn}{\begin{align}}
\newcommand{\eagn}{\end{align}}
\newcommand{\bs}{\begin{split}}
\newcommand{\es}{\end{split}}
\newcommand{\bc}{\begin{center}}
\newcommand{\ec}{\end{center}}
\newcommand{\bp}{\begin{picture}(0,0)}
\newcommand{\ep}{\end{picture}}
\newcommand{\bfl}{\begin{flushleft}}
\newcommand{\efl}{\end{flushleft}}

\newcommand{\bx}{\mathbf{x}}

\newcommand{\bW}{\mathbf{W}}

\newcommand{\bu}{\mathbf{u}}

\newcommand{\bw}{\mathbf{w}}

\newcommand{\br}{\mathbf{r}}

\newcommand{\HGPsixt}{H_\Gamma \! \lr \Psi \lr \bx,t \rr \rr}

\newcommand{\Psixt}{\Psi  \left( \bx,t \right)}
\newcommand{\psixt}{\psi  \left( \bx,t \right)}

\newcommand{\Ip}{\mathcal{I \lr \psi \rr}}

\newcommand{\bnG}{\mathbf{n}_\Gamma}
\newcommand{\bng}{\mathbf{n}_\gamma}

\newcommand{\eph}{\epsilon_h}
\newcommand{\ephb}{\epsilon_{h,b}}
\newcommand{\ephxt}{\epsilon_h \lr \bx,t \rr}

\newcommand{\ephppxt}{\epsilon_h^{\psi'} \! \lr \bx,t \rr}
\newcommand{\ephtpxt}{\epsilon_h^{t' \psi} \! \lr \bx,t \rr}
\newcommand{\ephxtB}{\epsilon_{h,B} \lr \bx,t \rr}
\newcommand{\ephB}{\epsilon_{h,B}}
\newcommand{\ephxtS}{\epsilon_{h,S} \lr \bx,t \rr}
\newcommand{\ephxtSo}{\epsilon_{h,S1} \lr \bx,t \rr}
\newcommand{\ephxtSt}{\epsilon_{h,S2} \lr \bx,t \rr}
\newcommand{\ephS}{\epsilon_{h,S}}

\newcommand{\alpp}{\alpha \lr  \psi \rr}

\newcommand{\deltaa}{\tilde{\delta} \lr  \alpha \rr}

\newcommand{\deltaGp}{\delta_\Gamma \! \lr  \Psi \rr}

\newcommand{\psia}{\psi  \left( \alpha \right)}
\newcommand{\psix}{\psi  \left( \bx,t \right)}

\journal{International Journal of Multiphase Flow}

\biboptions{authoryear}

\begin{document}

\begin{frontmatter}


\title{Modeling of non-equilibrium effects in intermittency region between two  phases}


  \author{Tomasz Wac{\l}awczyk\fnref{}\corref{mycorrespondingauthor}}
\address{Warsaw University of Technology, Institute of Aeronautics and Applied Mechanics, Division of Aerodynamics\\  ul. Nowowiejska 24, 00653 Warszawa, Poland}


\cortext[mycorrespondingauthor]{Tomasz Wac{\l}awczyk}
\ead{tomasz.waclawczyk@pw.edu.pl}


\begin{abstract}
 This paper
 concerns
 modeling 
 of the evolution
 of the intermittency region 
 between two
 weakly miscible 
 phases
 due to 
 temporal
 and spatial
 variations 
 of its  
 characteristic
 length scale.
 First,
 the need for
 a more general 
 description
 allowing for 
 the evolution 
 of the intermittency
 region is
 rationalized.
 Afterwards,
 results of the previous
 work (Wac{\l}awczyk T., 2017, 
 On a relation between the volume of fluid, level-set and phase field interface models, Int. J. Multiphas. Flow, Vol. 97)
 are discussed
 in the context of 
 sharp interface
 models known in
 the literature
 and  new insight into droplet
 coalescence mechanism
 recently recognized
 in molecular dynamics
 studies  (Perumanath S., Borg M.K., Chubynsky M.V., Sprittles
 J.E., Reese J.M., 2019, Droplet coalescence is initiated by thermal motion, Phys. Rev. Lett., Vol. 122). 
 Finally, 
 physical
 and numerical models
 extending the applicability
 of the equilibrium
 solution
 to the case
 when 
 the intermittency
 region could also
 be in
 the non-equilibrium state
 are introduced
 and verified.
\end{abstract}


\end{frontmatter}

%
%
%
\section{Introduction} 
\label{sec1}
%
%

The gas-liquid 
interface  
is a domain
where
material
properties
of two adjacent
phases
are changing.
However,
``the exact
definition
of the gas-liquid
interface is
nebulous'' \citep{faust2018}.
In
fluid
dynamics,
there are
two
accepted
physical
models of
the gas-liquid
interface,
namely:
the dividing
surface model
\citep{gibbs1874}
and 
the
diffusive
interface model
\citep{waals1893}.
In a
recent
review
paper \citep{elgho19}
concerning
direct
numerical
simulation (DNS)
of
the turbulent,
dispersed
two-phase
flows,
numerical
methods inspired
by these two 
physical models
of the gas-liquid
interface
are listed
as
``the
tracking 
scalar approach''.
The
dividing
surface
model
of Gibbs 
is the
foundation
of 
the volume of fluid (VOF)
\citep{trygg11,lu17}
and standard
level-set (SLS) \citep{osher1988,osher03,sussman03,deike16}
sharp interface 
methods.
The 
diffusive
interface
model of 
van der Waals
stimulated
development
of the phase-field
methods based
on
the Cahn-Hilliard \citep{cahn1958,anderson1998,komrakova15,fedeli17, soligo19} 
and modified Allen-Cahn 
\citep{allen1979,olsson05,chiu11, mccaslin2014, twacl15, grusz20,  kajzer20, mirja20}
equations.

A common
feature
of the aforementioned
physical and
numerical
models 
is the assumption
that the
gas-liquid
interface
is a
boundary
(geometric object)
evolving
in the given
(turbulent) 
velocity field.
Consequently,
in accordance
with the
postulates 
of the DNS,
the velocity
field
has to be
resolved
to the Kolmogrov
length scale $\sim\!Re^{-3/4}$
to reconstruct
all time and
length scales
governing its
evolution. 
However,
millions
of droplets
or bubbles
created
in the effect of
violent topological
changes can
easily have 
the diameter
below 
the Kolmogrov
length scale
\citep{elgho19}.
To model 
their sub-grid
dynamics and
its impact
on the
resolved
flow
field,
phenomenological
models are used. 

To increase 
the range
of \emph{Reynolds}
numbers
where 
numerical
simulations
can offer useful
predictions,
some reduced models
are obtained 
in the course
of filtering
or ensemble
averaging 
of the
two-phase
flow governing
equations.
These operations 
result, respectively,
in the large-eddy (LES)
\citep{labourasse07, toutant07, aniszewski12, herrmann13, saeedi20}
and Reynolds averaged (RANS) 
\citep{hong00,  guoshen10}
formulations
of the one-fluid
model. 
After
filtering
or 
ensemble averaging
of
one-fluid
model equations,
phenomenological
models (often based on the DNS)
are used 
to close 
correlations
between 
the instantaneous
(sub-grid) 
macroscopic interface
and turbulent-velocity field
(sub-grid) fluctuations.
Regardless
of the
reduction 
in the number
of degrees 
of freedom,
the 
gas-liquid interface
in the filtered/averaged
one-fluid model
is approximated
in the same 
way as in
the DNS.
The characteristic 
scalar function defining
the gas-liquid interface
is transported
using filtered
or ensemble
averaged
fluid velocity.
Hence,
the gas-liquid
interface
is once again viewed 
as the passive
boundary between
gas-liquid 
phases.
As a consequence,
the gas-liquid
interface
model
in the DNS, 
LES or RANS
formulations
of 
the one-fluid
model  
does not
play an 
active role
in the modeling
process.
In addition
the aforementioned 
phenomenological
models
are 
often based
on a
different
modeling
strategy 
than the one(s) 
used in
the one-fluid
model,
e.g. two-fluid
or
Euler-Lagrange
frameworks 
\citep{prosperetti07,  elgho19}.
This 
introduces
coupling and
feedback 
problems
that must
be addressed
during
time-consuming
simulations.

Recently,
the present
author
\citep{twacl17}
has shown
the mathematical
models describing
the gas-liquid
interface listed 
by \citep{elgho19}
as
``the tracking
scalar approach'' 
are  
complementary
components
of the gas-liquid
interface
statistical
description.
Herein,
this result
is extended
and used
to 
propose
modeling 
framework
that is
natural for
the one-fluid
model of (turbulent)
two-phase
flow.

In 
the 
present
work it
is assumed that
the macroscopic
intermittency region
is a domain where 
the gas-liquid   
interface $\Gamma$
can be found with non-zero
probability.
This 
description 
was first
introduced 
for the modeling
of turbulence/gas-liquid interface
interactions
\citep{brocchini01a,brocchini01b}.  
Therein, 
the 
sharp 
interface
$\Gamma$ 
is
the 
gas-liquid
interface,
its
ensemble 
averaged
oscillations
create
the
macroscopic
intermittency
region
evolving
due to
the stochastic,
unsteady
nature 
of
the turbulent
flow.
In this
interpretation,
deformations of the sharp interface $\Gamma$
are caused 
by stochastic
forcing of
the turbulent
eddies 
described 
typically 
in terms
of the characteristic
time
and length
scales.
These characteristic
scales are
altered
by gravity 
and surface tension
forces.
The phenomenological 
model of Brocchini and Peregrine
was used 
by several
authors \citep{hong00, smolentsev05, hoehne09, mwaclawczyk11, skartlien14}
to propose
quantitative models
of turbulence/gas-liquid
interface interactions.
In particular,
Wac{\l}awczyk and Oberlack
have proposed 
the correlation
between
the local interface $\Gamma$
position
and velocity
fluctuation
in the normal 
direction $\bnG$
must be modeled
to account for
the evolution of
the intermittency
region.
This idea
was used 
to analyze
the intermittency
region evolution
based on 
\emph{a priori}
study 
of turbulent 
velocity field 
in the vicinity
of the sharp
interface
\citep{twaclawczyketal14, waclawczyk2015}.
Therein
it was found that
the intermittency
region characteristic
time and length scales
are not constant
but can vary 
in time and space.

Next,
the present
author
\citep{twacl17}
has shown that
the intermittency
region paradigm
can be used 
to derive 
the equilibrium
condition  
for
the
non-flat,
gas-liquid 
interface
$\gamma$
(mesoscopic
intermittency
region).
The main
argument 
therein
is based 
on the
analogy   
between
processes 
of
turbulence/gas-liquid interface
and 
thermal-fluctuations/mesoscopic
interface
interactions. 
As 
argued by 
\citep{brocchini01a,brocchini01b}
and recently 
confirmed
in molecular
dynamics studies 
by \citep{perumanath19}
both processes are 
stochastic
in their nature.
This means 
the derivation of
the macroscopic (averaged)
equations
governing their
evolution
requires
conditional
averaging taking
into account
the instantaneous 
position
of the sharp 
interface $\Gamma$.
Moreover,
one expects
that
in the limit
of vanishing
energy of
turbulent or
thermal fluctuations,
a more general,
statistical
model of 
the intermittency
region 
should be
reduced 
to
sharp
interface models
known in 
the literature.
\blnm
\begin{figure}[!ht] \nonumber
	\centering\includegraphics[width=0.6\textwidth,height=0.4\textwidth,angle=0]{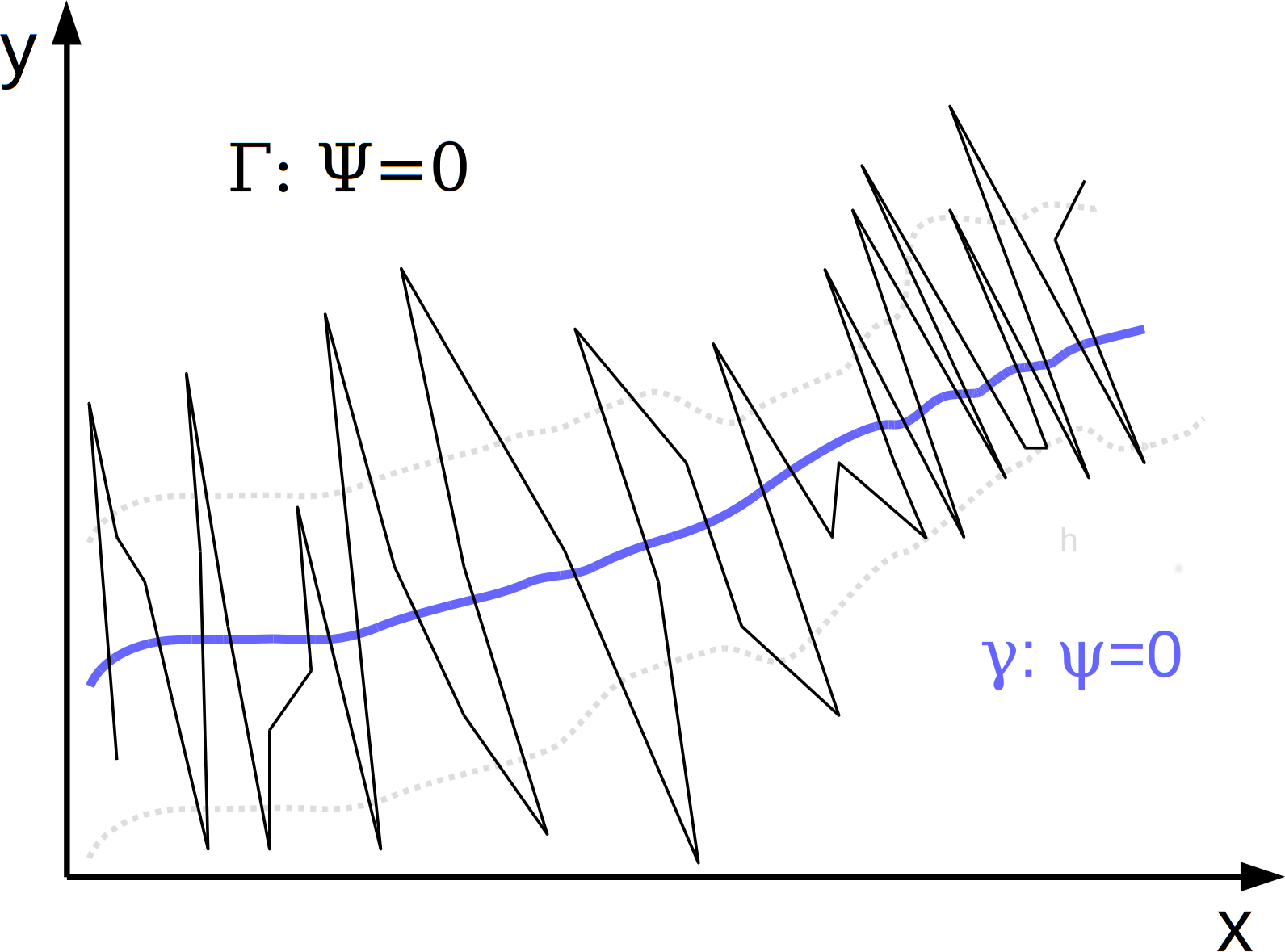}
	\caption{\small{Sketch of one realization of
		        the considered stochastic process,
		        the sharp interface 
		        $\Gamma: \Psixt\!=\!0$ (black-solid line)
		        disturbed by the field of stochastic forces
		        is oscillating
		       	around its expected position $\gamma: \psixt \!=\!0$ (blue-solid line)
                        that defines the regularized interface.
	    	        The intensity
			of  $\Gamma: \Psixt\!=\!0$ oscillations
			can in general be variable in space and/or time.			
		      }}
	\label{fig1}
\end{figure}
\elnm 

In 
the present
work, 
the analogy
between 
turbulence/gas-liquid interface 
and thermal-fluctuations/mesoscopic interface
interactions
is further
exploited.
We note,
when 
the
gas-liquid
interface
is in 
the equilibrium
state,
the
characteristic
length scale $\ephxt\,[m]$ 
governing 
its thickness
is constant
in time 
and space.
In 
the converse
case
the gas-liquid
interface is in 
the non-equilibrium
state.
This scheme
is extended 
to
the case 
of turbulence/gas-liquid
interface
interactions.
The distinction
between equilibrium
and non-equilibrium 
states of 
the mesoscopic
intermittency
region,
allows
the assessment
of the
extent
to which
classical 
sharp/diffusive
interface models
account for 
the stochastic
characteristics
of the
gas-liquid 
interface 
and explains
why they allow
predictions of
topological
changes
governed
by the molecular
effects 
\citep{perumanath19}.
Further
in the
paper,
the physical models
and numerical methods
which allow $\ephxt$
to be variable
in time and space
are proposed
and used
during numerical
solution
of the intermittency
region evolution
equation.

The present
paper
is organized
as follows.
In \Sec{sec2},
the equation governing
the evolution of 
the intermittency region
is derived
from the stochastic
viewpoint
and the conditions
for the intermittency region
equilibrium 
and non-equilibrium
are elucidated.
Afterwards,
it is argued 
why 
the sharp/diffusive models
of the intermittency region 
have
the potential 
to predict
topological 
changes during
break up or 
coalescence
occurring
on a molecular
level.
In \Sec{sec3}
it is shown
that
the stationary
solution of 
the intermittency
region evolution
equation
accounting
for variable
$\ephxt$
is equivalent
to
the minimization
of the 
corresponding
energy
functional.
This result
permits 
the stationary
solution of 
the intermittency
region evolution
equation
to be used as
the local equilibrium
condition.
In \Sec{ssec32},
the local equilibrium
condition
accounting 
for $\ephxt$
is employed
to derive
the mapping
function 
that
accounts
for the
non-equilibrium
effects.
Next,
the generalized
mapping function
is used
during 
numerical solution
of the intermittency
region evolution
equation.
In \Sec{sec4}, 
the numerical
method employed
to integrate 
the intermittency
region evolution
equation
is described 
and results
of the intermittency
region 
evolving due to
$\ephxt$  
are presented
and discussed.
\Sec{sec5}
provides 
the conclusions 
and perspectives
for future work.

\section{Derivation of the intermittency region evolution equation}
\label{sec2}
%
%

Let us assume
the evolution
of mesoscopic
sharp interface
$\Gamma$
is governed by
the  phase indicator
function $\HGPsixt$
transport equation 
\blnm
\be
\pd{H_\Gamma}{t} \!+\! \bW \!\cdot\! \nabla H_\Gamma \!=\! 
\pd{H_\Gamma}{t} \!+\! \deltaGp |\nabla \Psi| \bW \!\cdot\! \bnG   \!=\! 0,
\label{eq1}
\ee
\elnm
where 
$\Psixt\,[m]$
is the signed distance 
function from
the points $\deltaGp\,[1/m]$ 
located
on the two-dimensional
surface $\Psixt\!=\!0$
defining the sharp 
interface $\Gamma$
with the normal vector
$\bnG \!=\! \nabla \Psi/|\nabla \Psi|$,
see solid-black line
in \Fig{fig1}.
$\bW \! \cdot \! \bnG \,[m/s]$
is the stochastic 
velocity field
governing the motion of
$\deltaGp$.
$\HGPsixt$  
represents one, 
instantaneous
realization of 
the stochastic
process generated 
by thermal fluctuations
$\bW' \!=\! \bW \!-\! \av{\bW}\,[m/s]$.
For this reason,
\Eq{eq1}
is of no use
in the continuous
description
of gas/fluid systems.
To derive
its continuum
version the ensemble
averaging
\citep{pope88, mwaclawczyk11, twacl17} 
must be applied to \Eq{eq1}.
It is noticed,
unlike
in the recent work 
\citep{thiesset20},
analysis in the 
present paper
is based on
one-point surface
statistics.

The ensemble average
of the first LHS term
in \Eq{eq1} results in 
\blnm
\be
\pd{}{t}\av{H_\Gamma \lr \Psi \rr} \!=\!
\pd{}{t} \int_{-\infty}^{\infty} H_\Gamma \lr \xi \rr \av{ \delta \! \lr \Psixt - \xi \rr}  d \xi 
\label{eq2}
\ee
\elnm
where $\av{\delta \! \lr \Psixt \!-\! \xi \rr}$
is the ensemble
average $\av{\cdot}$
of the fine 
grained p.d.f.'s $\delta \! \lr \Psixt - \xi \rr$
characterizing each realization
of the stochastic process in
the sample space $\xi\,[m]$.
$\av{\delta \! \lr \Psixt - \xi \rr}$
provides the probability density
that $\xi \!<\! \Psi \lr \bx,t \rr \!<\! \xi + d \xi$. 

The
contributions
to the ensemble
average $\av{\cdot}$ 
from the
second RHS term in \Eq{eq1} 
are non-zero
only if 
the interface $\Gamma$
is present
at point $\bx$ 
and time $t$
where/when the averaging
is carried out.
This issue was
recognized by \citep{pope88}
who proposed
the conditional surface
average $\avG{\cdot}$
to account for the
smearing of the 
interface $\Gamma$ 
due to the 
averaging process.
Its application to 
the second RHS term
in \Eq{eq1} 
leads to
\blnm
\be
\av{\bW \! \cdot \! \nabla H_\Gamma} \!=\! \av{\bW \!\cdot\! \bnG \deltaGp} \!=\! \avG{\bW \! \cdot \! \bnG} \Sigma 
\label{eq3}
\ee
\elnm
where  $\avG{\cdot}$ is 
the surface average 
\blnm
\be
 \avG{\bW \! \cdot \! \bnG} \!=\! \frac{1}{\Sigma} \iint_{\Gamma} 
 \av{\bW \! \cdot \! \bnG 
 \delta \lr \mu     \rr  
 \delta \lr \lambda \rr
 \delta \lr \Psi\!-\!\Psi' \lr \mu,\lambda, t \rr \rr
 A\lr \mu,\lambda, t \rr} d\mu d\lambda
\label{eq4}
\ee
\elnm
and 
$\Sigma \,[1/m]$
is given by 
the formula
\blnm
\be
 \Sigma  
 \!=\! \iint_{\Gamma} 
 \av{ \delta \lr \mu \rr  
 	  \delta \lr \lambda \rr 
 	  \delta \lr \Psi\!-\!\Psi' \lr \mu,\lambda, t \rr\rr
 	  A\lr \mu,\lambda, t \rr} d\mu d\lambda.
\label{eq5}
\ee
\elnm
In equations 
(\ref{eq4}-\ref{eq5})
$\mu,\lambda,\Psi$ 
define
the local,
orthonormal
coordinate
system
of the
infinitesimally
small element
$A\lr \mu,\lambda, t \rr d\mu d\lambda$
where $\Psi$ 
is the coordinate
in the normal direction. 
$\Sigma\,[1/m]$
can be
interpreted as 
the amount of 
the expected 
surface-to-volume
ratio
\citep{pope88};
in the general case
$\Sigma \lr \bx,t \rr$
in \Eq{eq3} is unknown 
and must be closed
by a model.
Using the
decomposition
$\bW \!=\! \av{\bW}\!+\!\bW'$
and Eqs.~(\ref{eq3}-\ref{eq5})
one obtains
\blnm
\be
\avG{\bW \! \cdot \! \bnG} \Sigma = \av{\bW} \avG{\bnG}\Sigma + \avG{\bW' \! \cdot \! \bnG} \Sigma.
\label{eq6}
\ee
\elnm
Next,
the exact 
relations
$\av{\bW' \! \cdot \! \nabla H_\Gamma} \!=\! \avG{\bW' \! \cdot \! \bnG} \Sigma $
and $\avG{\bnG} \Sigma \!=\! \nabla \av{H_\Gamma}$,
see \Eqs{eqA1}{eqA8} respectively,
lead
to 
averaged
\Eq{eq1}
with the RHS term
that must be closed
\blnm
\be
\pd{\alpha}{t} \!+\! \bw \nabla \alpha 
\!=\! - \avG{\bW'\!\cdot\! \bnG}\Sigma
\!=\! - \av{\bW' \! \cdot \! \bnG \deltaGp}
\!=\! - \av{\bW' \!\cdot\! \nabla H_\Gamma},
\label{eq7}
\ee
\elnm
where we have denoted
$\alpha \!=\! \av{ H_\Gamma }$
and $\bw  \!=\! \av{\bW }$.

As it was  
put forward by
\citep{mwaclawczyk11},
the unknown RHS term
in \Eq{eq7} can be closed
by the eddy diffusivity
model
\blnm
\be
 \avG{\bW' \! \cdot \! \bnG} \Sigma  \!=\! - D   \nabla \!\cdot\! \avG{\bnG} \Sigma.
\label{eq8}
\ee
\elnm
Taking the
divergence
of the exact
relation 
$\avG{\bnG} \Sigma \!=\! \nabla \alpha$
leads to
\blnm
\be
  \nabla \!\cdot\! \avG{\bnG} \Sigma = \nabla^2 \alpha \!-\! \avG{\bnG} \!\cdot\! \nabla \Sigma.
\label{eq9}
\ee
\elnm
Substitution 
of \Eq{eq9} and \Eq{eq8}
into \Eq{eq7} results in
\blnm
\be
\pd{\alpha}{t} \!+\! \bw \nabla \alpha 
\!=\! D\nabla^2 \alpha \!-\! D\avG{\bnG} \!\cdot\! \nabla \Sigma
\label{eq10}
\ee
\elnm
where 
the second, 
unclosed
RHS term 
in \Eq{eq10}
was identified
by Wac{\l}awczyk and Oberlack
as
counter gradient
diffusion.
The above equation 
is not 
in the desired
conservative form.
Thus, 
with
the help
of exact relation
$\avG{\bnG} \Sigma \!=\! \nabla \alpha$,
taking into account
the case when $D\lr \bx,t \rr \!=\! C \ephxt$ 
and noting 
the vector normal to 
the regularized interface
$\bng \!=\! \nabla \alpha / |\nabla \alpha| \!=\! \avG{\bnG}/|\avG{\bnG}|$,
\Eq{eq10} is rewritten as
\blnm
\be
\pd{\alpha}{t} \!+\! \bw \nabla \alpha 
\!=\! \nabla \cdot \lr D \nabla \alpha \rr 
-  |\avG{\bnG}| \nabla{\lr D \Sigma \rr}  \!\cdot\! \bng.
\label{eq11}
\ee
\elnm
One notes
that
\Eq{eq11}
accounts
for the variable
characteristic 
length scale $\ephxt$,
however,
it is still
unclosed
due to
the presence
of
the counter
gradient diffusion
term.
The conservative
closure of this
unknown term
\citep{waclawczyk2015}
leads to
the equation
introduced
by \citep{chiu11}
inspired by
the conservative
level-set (CLS) method
\citep{olsson05}
\blnm
\be
\pd{\alpha}{t} \!+\! \nabla \! \cdot \! \lr \bw \alpha \rr \!=\! \nabla \cdot \ls D |\nabla \alpha|\bng \!-\! C \alpha \lr 1-\alpha \rr \bng \rs
\label{eq12}
\ee
\elnm 
where $\bw\,[m/s]$ 
is  velocity of
the regularized
interface $\gamma$,
and in the general 
case
$C \lr \bx,t \rr \,[m/s]$
and 
$D\!=\!C\ephxt\,[m^2/s]$
are velocity and diffusivity scales
characterizing
the intermittency
region, respectively.
The coefficients $C, D$
uniquely specify
the
characteristic
 length $\eph \!\sim\! D/C\,[m]$ 
 and time $\tau_h \! \sim \! \eph/C\,[s]$
scales 
governing 
the solution
of \Eq{eq12}.
We note
the presence
of
the two RHS terms
in \Eq{eq12}
is supported
by the fact that
forces always
occur in pairs.
In \Eq{eq12}
due to the 
presence of
contraction  $C \alpha \lr 1-\alpha \rr \bng$,
diffusion $D|\nabla \alpha| \bng$
is counterbalanced.
The former
term 
was 
identified 
as the first order
approximation
of joint probability 
of creation
of the bond between
particles of 
two different types
\citep{cahn1958}.
The steady
state
solution 
of \Eq{eq12}
with $\eph\!=\! const.$
and
$\bw\!=\!\bu \!=\!0$
is given by
the regularized
Heaviside function 
\blnm
\be
\alpp = \frac{1}{1+\exp{\lr -\psix/\eph \rr}}=
\frac{1}{2} \ls 1+\tanh{\lr \frac{\psix}{2\eph} \rr} \rs
\label{eq13}
\ee
\elnm
and its inverse
function that is
the signed distance 
from the expected position of 
the regularized interface $\gamma$ 
defined by the level-set $\psi \lr \alpha \!=\!1/2 \rr \!=\! 0$
\blnm
\be
\psi \lr \alpha \rr = \eph \ln{\ls \frac{\alpha\lr \psi \rr}
	{1 - \alpha \lr  \psi \rr} \rs}.
\label{eq14}
\ee
\elnm
As noticed
by the 
present author
\citep{twacl15},
\Eqs{eq13}{eq14} 
are known 
to characterize
the cumulative 
distribution $\alpp$,
and quantile $\psia$
functions
of the logistic
distribution.
Additionally,
the gradient 
of $\alpp$ 
given by the formula
 \blnm
 \be
 \nabla \alpha = \frac{\deltaa}{\eph} \nabla \psi,
 \label{eq15}
 \ee
 \elnm 
where $\deltaa/\eph \!=\! \alpha \lr 1\!-\!\alpha \rr/\eph$ is 
the probability density function
of the logistic distribution.
Substitution of \Eq{eq15}
into \Eq{eq12} gives
\blnm
\be
\pd{\alpha}{t} + \bw \nabla \alpha =  \nabla \cdot \ls C  \deltaa \lr |\nabla\psi|-1 \rr \bng \rs,
\label{eq16}
\ee
\elnm
where 
$\bng \!=\! \nabla \alpha /|\nabla \alpha|\!=\! \nabla \psi /|\nabla \psi|$,
and
$\bw$ in \Eq{eq14}
may now be replaced 
by the fluid velocity $\bu$
since in  absence of 
phase changes and/or advection
$\bng \!\cdot\! \bw \!=\! 0$.

In the present work
we separate the advection 
and re-initialization
steps in \Eq{eq16},
which leads to
\blnm
\be
\pd{\alpha}{t} \!+\! \bw \nabla \alpha \!=\!
  \pd{\alpha}{t} \!+\! \frac{\deltaa}{\eph} \bw \!\cdot\! \nabla \psi  \!=\! 0,
\label{eq17}
\ee 
\be
\pd{\alpha}{\tau} \!=\!  \nabla\! \cdot \! \ls C  \deltaa \lr \left |\nabla\psi \right| \!-\! 1 \rr \bng \rs.
\label{eq18}
\ee
\elnm
This form
of \Eq{eq12}
is preferred.
It allows
to consider separately
advection equation \Eq{eq17}
and the model 
of the evolution 
of the intermittency
region using \Eq{eq18}.
Solution 
of \Eq{eq18},
known in the
literature as
the re-initialization
step,
was
shown
to be equivalent
to minimization
of the 
interfacial
energy
functional
containing 
the term 
which accounts
for
the regularized 
interface $\gamma$ 
deformation. 
For this reason,
it was argued 
in the previous
work \citep{twacl17}
that
\Eqs{eq17}{eq18}
with the mapping 
between $\alpp\!-\!\psia$
functions
given by \Eq{eq14}
describe the non-flat,
intermittency
region in
the equilibrium
state as therein
$C\!=\!const.$,
$D\!=\!C\eph\!=\!const.$.
Furthermore,
the statistical
interpretation
of \Eqs{eq17}{eq18}
based on Eqs.~(\ref{eq13})-(\ref{eq15})  
reveals
the relation between
the sharp and diffusive
interface models.

As it was 
mentioned
in \Sec{sec1},
\Eq{eq12} can
be interpreted
as the statistical model
of the mesoscopic
or macroscopic
intermittency region.
The physical
interpretation
depends upon
the character of
the stochastic
force field inducing $\bW$
in \Eq{eq1}
and chosen
time/length
scales.
In the mesoscopic
interpretation,
the deformation
of the sharp
interface
$\Gamma$
is caused by
random,
thermal 
fluctuations.
In the
macroscopic
interpretation,
velocity
$\bW$ in \Eq{eq1}
can be related to
the instantaneous
turbulent 
velocity field.
In the
next
section,
terms under 
which
the intermittency
region is
in the equilibrium
or non-equilibrium
state are
discussed.
%

%
\subsection{Equilibrium and non-equilibrium state of the intermittency region}
\label{ssec21}

%
First, 
the mesoscopic
interface  $\Gamma$
agitated
by the thermal
fluctuations
is considered,
see \Fig{fig1}.
After conditional
averaging
described in \Sec{sec2}
its evolution is
described by \Eq{eq12}.
Let us note
$\alpp$ and $\deltaa$ 
in \Eq{eq12} 
have  
infinite support
as $\eph \!\sim\! \sqrt{k_B T/\sigma} \! > \! 0$
where
$k_B\,[J/K]$ is 
the Boltzman constant,
$T\,[K]$ is
the absolute 
temperature and
$\sigma\,[J/m^2]$ is 
the surface tension
coefficient
\citep{vrij1973,aarts2004}.
$k_BT/V\,[J/m^3]$ 
is the root mean
square measure of
the thermal fluctuations
of molecules
acting to distort
the mesoscopic 
interface $\Gamma$
between two phases
in the infinitesimally
small volume $V$.
The amount
of thermal 
energy in $V$
determines
the kinetic energy 
of molecules.
A surface
tension 
$\sigma/V\,[J/m^5]$
represents
net work
done by
the cohesive
forces between
fluid molecules
per unit area of
$\Gamma$ 
in $V$.
The cohesive 
forces between
fluid molecules 
act to suppress
increases in
interfacial
area of
$\Gamma$.  

In the  
case of 
the turbulence/gas-liquid interface
interactions,
the interface 
$\Gamma$ in \Fig{fig1}
is the sharp
representation
of the gas-liquid,
macroscopic
interface.
By analogy
to the mesoscopic case,
the characteristic 
length scale $\ephxt$
is governed
by the
ratio
of net
turbulent 
kinetic
energy $\rho k\,[J/m^3]$
and
the work 
of forces
generating
turbulent
stresses 
per
unit area 
of the 
interface $\Gamma$
in $V$, $[J/m^5]$.
These
forces
are
acting
to
decrease 
or
increase 
the sharp
interface 
$\Gamma$
interfacial
area.
In the macroscopic
interpretation 
of \Eq{eq12},
the ratio of
work done by
volume/surface
forces 
in the intermittency
region
is
altered by
the work
done by
the gravitational 
and surface tension
forces, 
respectively.
\blnm
\begin{figure}[!ht] \nonumber
	\begin{minipage}{.5\textwidth}
		\centering\includegraphics[width=0.85\textwidth,height=0.65\textwidth,angle=0]{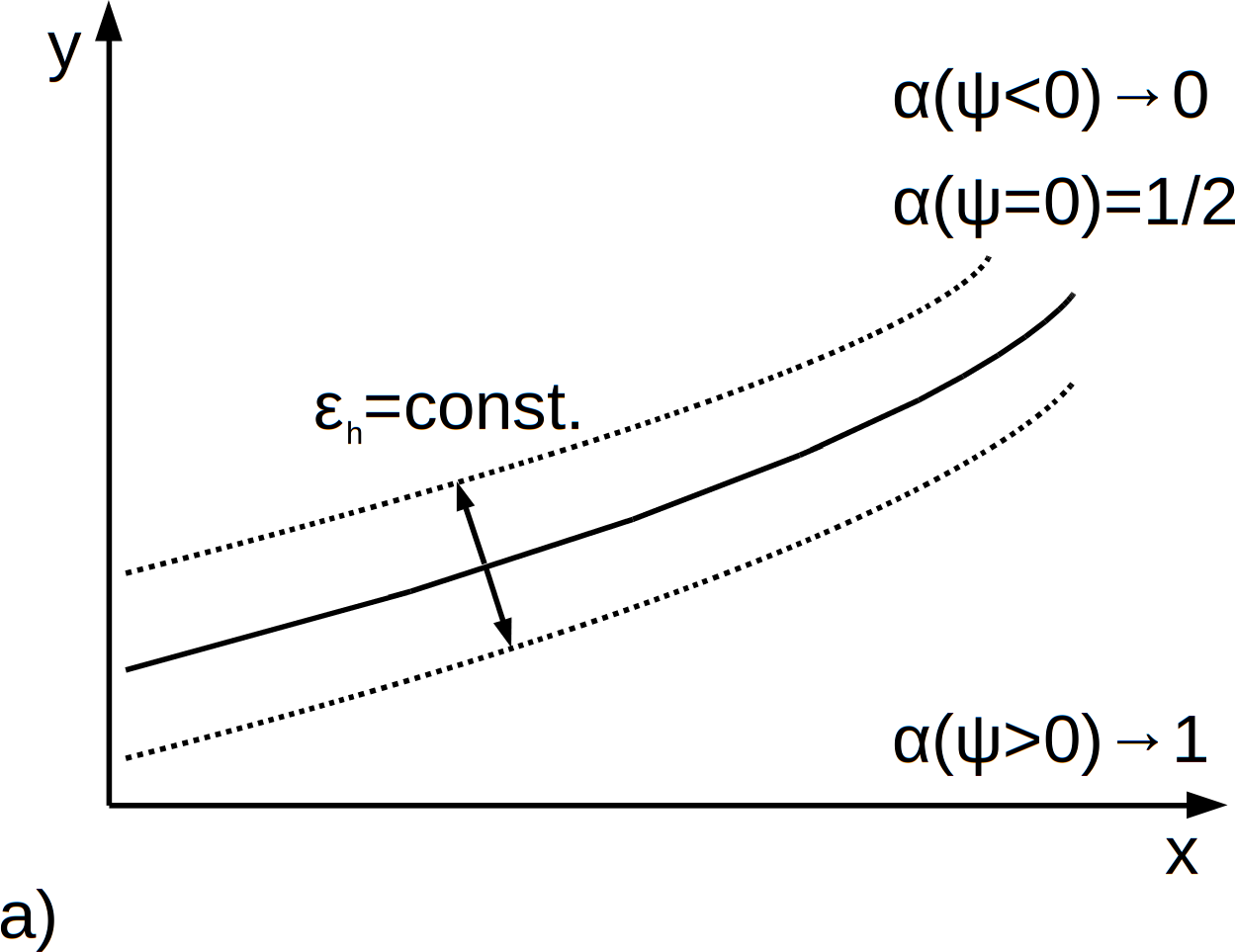}
	\end{minipage}
	\begin{minipage}{.5\textwidth}
		\centering\includegraphics[width=0.85\textwidth,height=0.65\textwidth,angle=0]{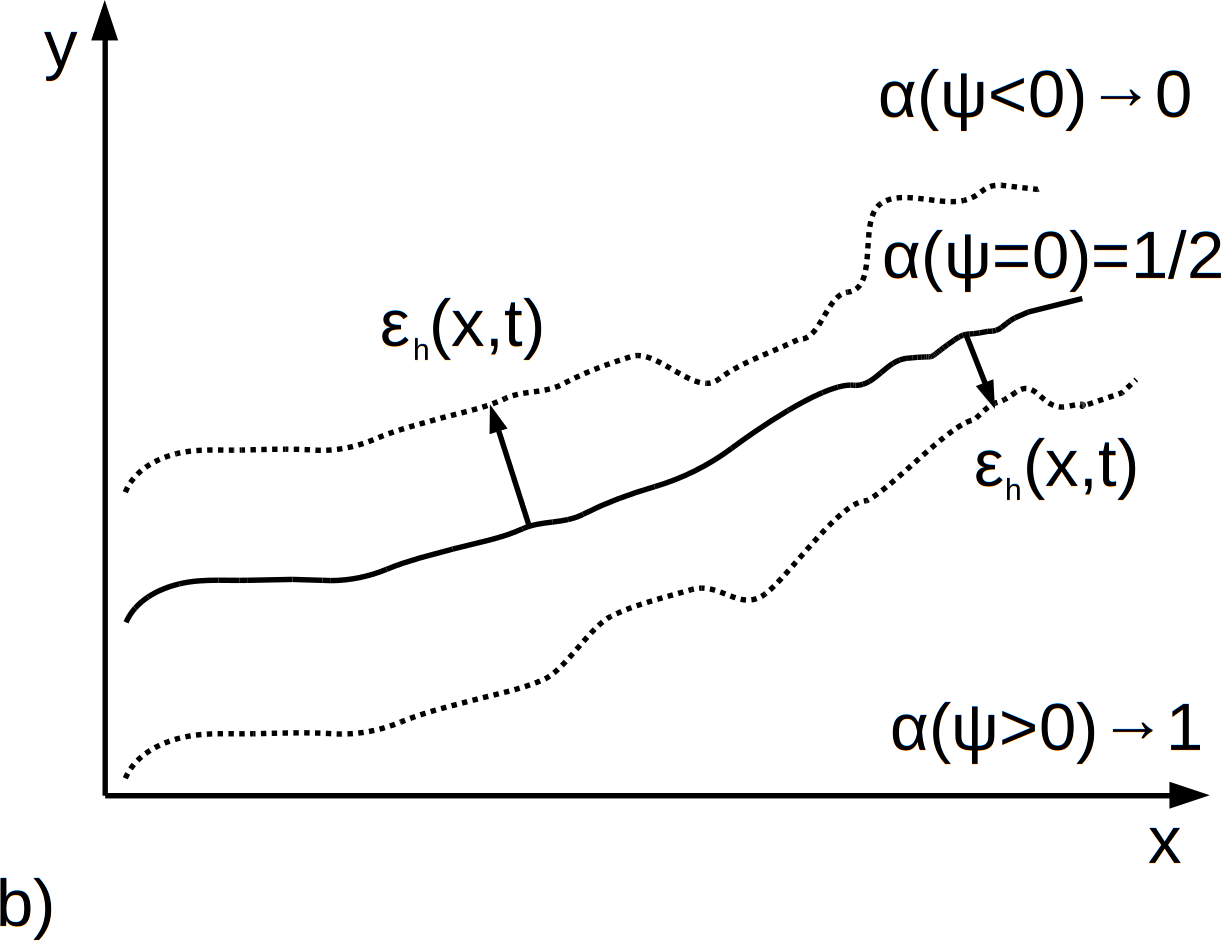}
	\end{minipage}
	\caption{\small{Schematic picture of the intermittency region
			in  (a) equilibrium  $\eph \!=\! const.$  
			and (b) non-equilibrium  $\ephxt$ states.
			The expected position of the interface $\gamma$
			is marked with a black-solid line. }}
	\label{fig2}
\end{figure}
\elnm 

If
the 
ratio
of
work
done 
by the
random
volume
forces
(inducing
fluctuations
of $\Gamma$) 
and surface
forces (per unit area of $\Gamma$)
is constant
in time and space,
then also
$\ephxt\!=\!const.$
and 
the intermittency 
region is 
in the equilibrium
state, see \Fig{fig2}a.
In 
the opposite
case,
the characteristic
length scale $\ephxt$
may change
in time and space,
and for this reason, 
the intermittency
region is 
in the 
non-equilibrium
state, see \Fig{fig2}b.

In what follows,
the statistical 
interpretation of \Eq{eq12}
is used
to localize 
the sharp interface
tracked or captured
in the VOF, SLS
sharp interface models.

\subsection{Stochastic coalescence and sharp/diffusive interface models}
\label{ssec22}
%
%

In  a
series of
molecular
dynamics
simulations
of two droplets
collisions
\citep{perumanath19}
have identified
the characteristic
thermal length scale
$l_T \!\sim\! 2\sqrt{\eph R}$
where
$\eph \!\sim\! \sqrt{k_B T/\sigma}\,[m]$
is  the intermittency
region thickness,
$R\,[m]$ denotes colliding
droplets  radii.
The existence of
the thermal length scale
$l_T\,[m]$ shows that
before the capillary
forces take control
on the droplets coalescence,
molecular and thermal effects
govern this process.
Therefore,
Perumanath et al. 
conclude that
droplet
coalescence is
a stochastic
phenomenon
initiated
by thermal
motion of
fluid particles.
Since $l_T\,[m]$ 
is proportional
to the square root
of the droplet
radius $R\,[m]$,
it is   
expected
that
during topological
changes 
the molecular effects
may influence  
phenomena on the
macroscopic scale
resolved in
fluid dynamics.

In light
of this fact,
classical
sharp interface
models seem 
to overlook
molecular
effects 
(see derivations
the signed-distance
and phase indicator 
function
transport equations
in \citep{osher03, trygg11}, respectively.
This gives rise
to the question
of the extent
to which 
the sharp interface
models are able 
to reconstruct
topological changes 
of the gas-liquid
interface during
coalescence 
and/or break up
events.
First we note
the expected position 
of the regularized 
(macroscopic) interface 
$\gamma$: $\psi \!=\! 0$
is different 
than the instantaneous
position of
the mesoscopic sharp
interface $\Gamma$: 
$\Psi \!=\! 0$
defining one realization
of the stochastic
process, see \Fig{fig1}.
Since in 
the case
of gas-liquid
intermittency region
in the equilibrium
state $0\!<\!\eph\!\ll\!1$
and $\eph\!=\!const.$, 
one can assume
$\eph$ does not
depend on
volume 
and surfaces
forces
doing
work 
in the domain
where
a two-phase
system is
changing its
properties.
Hence,
$\eph \! \rightarrow \! 0$ 
means
$\alpp \rightarrow H_\gamma \lr \psi \!=\! 0 \rr \!=\! 1/2$.
For this
reason,
similar
to the
Gibbs dividing 
surface
neglecting
information about 
$0 \!<\! \eph \!\ll\! 1$,
sharp 
interface models
are valid only
when it is 
assumed
that
the
intermittency
region is
in the equilibrium
state.

The phase indicator 
function $H_\gamma$ 
built on  
the expected position 
of the regularized interface
$\gamma$: $\psixt \!=\! 0$
is different than 
the phase indicator 
function $H_\Gamma$
built on the signed distance
function $\Psixt$,
as $H_\Gamma \lr \Psixt \rr$
is one realization 
of the stochastic 
process governed
by \Eq{eq1}.
$H_\gamma$ 
is the phase indicator function
discretized in VOF methods,
$\psixt$ is the signed 
distance function discretized
in SLS methods.
The level-sets 
$H_\gamma \!=\! 1/2$,
$\psi \!=\! 0$
are two equivalent
geometric,
two-dimensional 
representations
of the expected
position 
of the gas-liquid
regularized
interface $\gamma$
as $0 \!<\! \eph \!\ll\! 1$
must remain 
greater than
zero.
This explains
how
the molecular
effects are 
taken into account
in VOF, SLS type
sharp interface
models. 

Because
the sharp
interface
approximations
are formulated 
in the limit of 
$\eph \! \rightarrow \! 0$,
they do not depend
explicitly
on $\eph$.
Hence, they  
do not allow to
account for
the non-zero
volume/interfacial
energy
ratio
governing
the intermittency
region width,
its 
possible 
variations
and 
consequences
of these, too.
Thus, 
without additional
modeling assumptions
VOF, SLS type
sharp interface models
can not account for
thermal effects
described by \citep{perumanath19}.
However, 
due to 
the fact
VOF, SLS 
interface models
sharply reconstruct
the expected
position of
the regularized
gas-liquid
interface $\gamma$: $H_\gamma \!=\!1/2$, $ \psi \!=\!0 $, respectively,
they are able to 
approximate break 
up and (in most of the cases)
coalescence
processes.

As long as
the intermittency region
remains 
in the equilibrium state
and/or
energy of 
stochastic
fluctuations
is small
and independent
of background
physical
phenomena
the sharp 
interface model
is a good
approximation.
In the opposite
case, some physical
effects may be lost
when using the sharp
interface model
as local
variations
of the volume/surface
forces work ratio 
in the intermittency region
can affect
dynamics
of adjacent
gas-liquid
phases,
for example
through 
the local
modifications
of their
material
properties.
The remaining
part of the present
paper proposes
how
the description
given
by Eqs.~$(\ref{eq17},\ref{eq18},\ref{eq14})$
can be extended
to model 
the intermittency
region in 
the non-equilibrium
state.

%
\section{Modeling of non-equilibrium effects in the intermittency region}
\label{sec3}
%
As it has been
explained in 
the previous 
sections,
the motivation
for generalized
numerical solution
of Eqs.~$(\ref{eq17},\ref{eq18},\ref{eq14})$
comes from 
the need 
to account
for the case 
when 
the ratio
of work
done
by the volume/surface
forces
governing
$\eph$
varies
in space
and time.
Generalization
of the equilibrium
model
can also be 
justified
from a
thermodynamic
perspective.
The  intermittency 
region between
two weakly miscible
phases is an
open system
that may not
be in 
the equilibrium
state as
it is
perpetually
exchanging
energy
with 
neighboring
phases.

During 
previous
analytical 
considerations 
and
numerical
experiments  
it has been assumed
that
$C\!=\!const.$
and $\eph \!\sim \! \Delta x$,
see \Fig{fig2}a.
The main 
subject 
of the present
section
is to extend 
the analytical
model
and
numerical
solution
of the set 
of partial differential
algebraic 
Eqs.~(\ref{eq17}, \ref{eq18}, \ref{eq14})
to the case when
$\ephxt$ is variable
as it is schematically
depicted in \Fig{fig2}b.
The case
when the 
characteristic
time scale $\tau_h \! \sim \! \ephxt/ C \lr \bx,t \rr$ 
is variable as well
is left 
for future
studies.

\subsection{Minimization of free energy functional with variable characteristic length scale}
\label{ssec31}
In the previous
work of the present 
author \citep{twacl17}
it has been shown
the Helmholtz free
energy functional
defining 
the energy
of the two-phase
system
\blnm
\be
  F \ls \alpha \rs \!=\! \int_V \sigma \ls \eph |\nabla \alpha|^2
                     + \frac{f\lr \alpha \rr}{\eph} 
                     + k \lr \alpha  \rr\rs dV
\label{eq19}
\ee
\elnm
where $\sigma\,[J/m^2]$
is a known constant and
$f \lr \alpha \rr \!=\! \alpha^2 \lr 1-\alpha \rr^2\,[-]$,
has to contain 
term $k \lr \alpha  \rr$  
accounting for the energy
of the regularized
interface $\gamma$
deformation.
Its presence
in \Eq{eq19} 
is required 
to guarantee 
the equilibrium
state 
of the non-flat
regularized 
interface $\gamma$
by setting 
$\pdFl{F}{\alpha}\!=\!0$.
From 
the equilibrium 
condition given 
by the stationary 
solution to \Eq{eq12}
with $C\!=\!const.$, $\eph\!=\!const.$
it was shown
$k \lr \alpha \rr$ in \Eq{eq19}
satisfies the relation
\blnm
\be
 \int_V \pd{k \lr \alpha \rr}{\alpha} dV \deltaup{\alpha}
 \!=\!
 \int_V 2 \alpha \lr 1\!-\!\alpha \rr \nabla \!\cdot\! \bng dV \deltaup{\alpha}.
\label{eq20}
\ee
\elnm
As $\pdFl{k}{\alpha}$ 
in \Eq{eq20}
does not depend
explicitly on $\ephxt$
the above relation
will be also 
used herein.

Next
it is proven
the functional
derivative
of  \Eq{eq19}
with the variable
characteristic
length scale $\ephxt$
leads
to the stationary 
solution 
of \Eq{eq12}
accounting
for
the non-equilibrium
effects.
The RHS of
stationary
\Eq{eq12}
leads 
to re-initialization
equation 
in the non-conservative
form 
\blnm
\be
 \pd{\alpha}{\tau} 
 \!=\! \eph \nabla^2 \alpha 
 \!+\! \nabla \eph \! \cdot \! \nabla \alpha 
 \!-\! \frac{\alpha \lr 1\!-\!\alpha \rr}{\eph} \ls \lr 1-2\alpha \rr \!+\! \eph \nabla \!\cdot\! \bng \rs
\label{eq21}
\ee
\elnm
where we set $C\!=\!1\,[m/s]$ for 
clarity.
Calculation 
of the functional
derivative 
of \Eq{eq19} 
with $\ephxt$
is carried out
in \ref{appB}.
The minimization
condition given
by \Eq{eqB6} is 
the same
as the RHS
of \Eq{eq21},
therefore
\blnm
\be
\pdF{F}{\alpha} \!=\! \pd{\alpha}{\tau} \!=\! 0.
\label{eq22}
\ee
\elnm
The stationary solution
to \Eq{eq12} 
or the steady state
solution of
the corresponding 
re-initialization equation
in pseudo-time $\tau$
would minimize 
the functional (\ref{eq19})
with the variable 
characteristic length
scale $\ephxt$.
The additional 
term $\nabla \eph \!\cdot\! \nabla \alpha$
forces changes 
of the function
$\alpha$ shape.
When $\eph \!=\! const.$
the equilibrium solution
given by \Eqs{eq13}{eq14}
is recovered. 
In the following
section,
the mapping
function
used 
during 
numerical
solution
of \Eqs{eq17}{eq18}
with variable
$\ephxt$
is derived.

\subsection{Modification of the mapping procedure}
\label{ssec32}
%
In the present
section it is
proposed
how to use the
re-initialization 
equation in 
the form of \Eq{eq18}
taking into account
variable $\ephxt$.
The equilibrium
condition
obtained 
from  
the stationary
solution 
to \Eq{eq12}
reads
\blnm
\be
\nabla \alpha \!=\! |\nabla \alpha| \bng \!=\! \frac{\alpha \lr 1\!-\!\alpha \rr}{\ephxt} \bng. 
\label{eq23}
\ee
\elnm 
\Eq{eq23} is 
formulated
in the direction $\bng$
normal to the  regularized
interface $\gamma$,
hence,
it may
be rewritten as
\blnm
\be
\pd{\alpha}{\psi} \left| \pd{\psi}{\bx} \right| \!=\! \frac{1}{\ephxt}  \alpha \lr 1 \!-\! \alpha \rr,
\label{eq24}
\ee
\elnm
where it 
is assumed
$\pdl{\alpha}{\psi} \! > \! 0$
meaning
$\alpp$ 
is expected 
to be 
the cumulative
distribution
function
with infinite 
support
analogously
to \Eq{eq15}.
Next, we assume
$|\nabla \psi| \!\equiv\! 1$
in \Eqs{eq23}{eq24}.
As a result,
substitution  
of \Eq{eq23}
into \Eq{eq12} with
$D \lr \bx,t \rr\!=\!C \ephxt$
let us derive 
\Eq{eq18}.
The assumption
$|\nabla \psi| \!\equiv\! 1$
means
the signed distance
function $ \psix $
spans 
the space where
surface
averaged oscillations
of the sharp interface 
$\Gamma$ 
take place.
On average,
these
oscillations occur
only in  the direction
$\bng$ normal 
to the expected
position $\psi\!=\!0$
of the regularized
interface $\gamma$.
The above 
interpretation
explains
the difference
between 
$\psia$
and  
$\Psixt$
signed distance
function fields.
$\Psixt$ is
exclusively 
the signed 
distance
from points
$\deltaGp$
located at the
sharp interface 
$\Gamma$ defined
by the level-set
$\Psixt \!=\! 0$.

Further,
it is noticed
at each point 
$\lr \bx,t \rr$ 
of the field
$\ephxt$
the signed 
distance
function
$\psixt$ is given.
Hence, 
the knowledge of
the field $\psixt$
gives
$\lr \bx,t \rr$
and thus
$\ephxt$.
Therefore, 
we introduce
$\epsilon_h^{\psi} \lr \bx,t \rr$
denoting 
$\ephxt$ 
determined 
using $\psixt$.
This 
let us
to integrate
\Eq{eq24}
in the local 
coordinate
system attached
to the 
regularized
interface $\gamma$.
As $\gamma$
is defined 
by $\psixt\!=\!0$,
$\psixt$ is
the normal coordinate
with the origin at
$\psixt\!=\!0$
of this local 
system.
At each
fixed
point 
of given
$\alpp,\,\psia,\,\ephxt$
fields
this
integration
reads
\blnm
\be
\int^{1/2}_{\alpp} \frac{d\alpha'}{\alpha' \lr 1\!-\!\alpha' \rr} = \int^{0}_{\psia} \frac{d \psi'}{\ephppxt}.
\label{eq25}
\ee
\elnm
The integration
(\ref{eq25})
is performed
from the
arbitrary point
located at the signed-distance
from the regularized interface
$\alpp\!-\!\psia$
to the expected
position of the
regularized
interface $\psi \lr \alpha \!=\!1/2 \rr \!=\! 0$.
One notes
the LHS integration 
in \Eq{eq25}
does not
assume
or result in
any specific form/shape
of the function $\alpp$.

To recover 
the equilibrium
solution when
$\ephxt \!=\! const.$
it is necessary  
to preserve the mapping
between $\alpp\!-\!\psia$,
see \Eq{eq14}.
For this reason,
it is more convenient
to reformulate 
the RHS integral
in \Eq{eq25} using
variable
substitution 
as follows 
\blnm
\be
\int^{0}_{\psia} \frac{d \psi'}{\ephppxt} 
\!=\! \psia \int^0_1 \frac{dt'}{\ephtpxt} \!=\! \psia \Ip
\label{eq26}
\ee
\elnm
where $t' \! \in \!  [ 0,1 ]$ 
is the  parameter
such that $\psi' \!=\! t' \psi$
and $d \psi' \!=\! dt' \psi$,
furthermore
$\Ip$ is used to denote
the integral on the RHS
of \Eq{eq26}.
After integration
of \Eq{eq24} with \Eq{eq26}
one obtains 
\blnm
\be
\psia  = \frac{1}{\Ip} ln \ls \frac{\alpp}{1-\alpp} \rs.
\label{eq27}
\ee
\elnm
At
the given,
arbitrary
point $\lr \bx,t \rr$, 
the
signed
distance $\psixt$
has the known value.
For this reason,
at the point $(\bx,t)$
the integral
$\Ip \!=\! const.$
and thus
the inverse 
relation 
is also true
\blnm
\be
\alpp \!=\! \frac{1}{1+\exp{ \lr - \psia \Ip \rr }}.
\label{eq28} 
\ee
\elnm
The only difference
between 
\Eqs{eq13}{eq14} 
and \Eqs{eq28}{eq27}
is the latter 
take into account
variation of $\ephxt$
in the sense
of the local
equilibrium condition
given by \Eq{eq24}.
When
the field
$\ephxt \!=\! const.$,
\Eq{eq25} and \Eq{eq26}
reduce to
the equilibrium
solution,
which is guaranteed
by the
definition of $\Ip$.
Thus, 
the mapping given 
by \Eq{eq27} or
the form of $\alpp$
given by \Eq{eq28}
can be employed during
numerical solution
of the system given 
by \Eqs{eq17}{eq18}
to model
how
the 
$\ephxt$
field
is affecting 
changes of
the
cumulative 
distribution
function
$0 \!<\! \alpp \!<\! 1$
profile.

\section{Numerical solution} 
\label{sec4} 
This
section
introduces 
a numerical 
method
for the
exact and 
approximate
solutions 
of 
the intermittency
region
evolution
equation
with  
the structured
grid solver.
First, 
a
one-dimensional 
study is carried
out showing
how 
the re-initialization
equation (\ref{eq18})
with the modified
mapping 
procedure
defined 
by \Eq{eq27}
can be
used 
to reconstruct
the intermittency
region in 
the non-equilibrium
state.
Afterwards,
the coupled
solution
is compared
with 
the semi-analytical
approach 
using \Eq{eq18}
where
$\eph \!=\!const.$
and \Eq{eq28}
is accounting
for variable
$\ephxt$.
Finally,
the semi-analytical
solution
is used
in two-dimensional
studies 
without
and with
advection
to reconstruct
more
complex 
behavior
of 
the 
intermittency
region.
Details of 
discretization
and  numerical 
solution of \Eqs{eq17}{eq18}
using the mapping given by \Eq{eq14}
where $\eph \!=\! const.$
are  described 
in \citep{twacl15, twacl17}.
In \ref{appD}, 
minor modifications
to these schemes
required to take
into account
variable $\ephxt$
are described.

\subsection{Approximation of the local equilibrium condition}
\label{ssec41}
%
The main 
problem during
numerical solution
of the set  
of algebraic, partial differential 
equations  
(\ref{eq17}, \ref{eq18}, \ref{eq27})
or (\ref{eq17}, \ref{eq18}, \ref{eq14})
where $\eph\!=\!const.$ and \Eq{eq28}
is accounting
for $\ephxt$,
is approximation
of the integral 
in the local
equilibrium 
condition 
given by \Eq{eq26}. 
Namely,
one
needs to find
the quadrature
$I \lr \psi \rr$
for integral
\blnm
\be
 \Ip \!=\! \int^0_1 \frac{dt'}{\ephtpxt } \approx I \lr \psi \rr   
\label{eq29}
\ee
\elnm
keeping in mind
the parameter $t'$
is changing along
the signed distance
function $\psia$,
from 
the local position
on the computational
grid at $t'\!=\!1$
to the interface at $t'\!=\!0$.
As function 
$\psia$
is also
the solution
to \Eq{eq18},
one can use 
this to formulate
an effective 
numerical
integration
procedure.
\blnm
\begin{figure}[!ht] \nonumber
	\centering\includegraphics[width=0.5\textwidth,height=0.5\textwidth,angle=0]{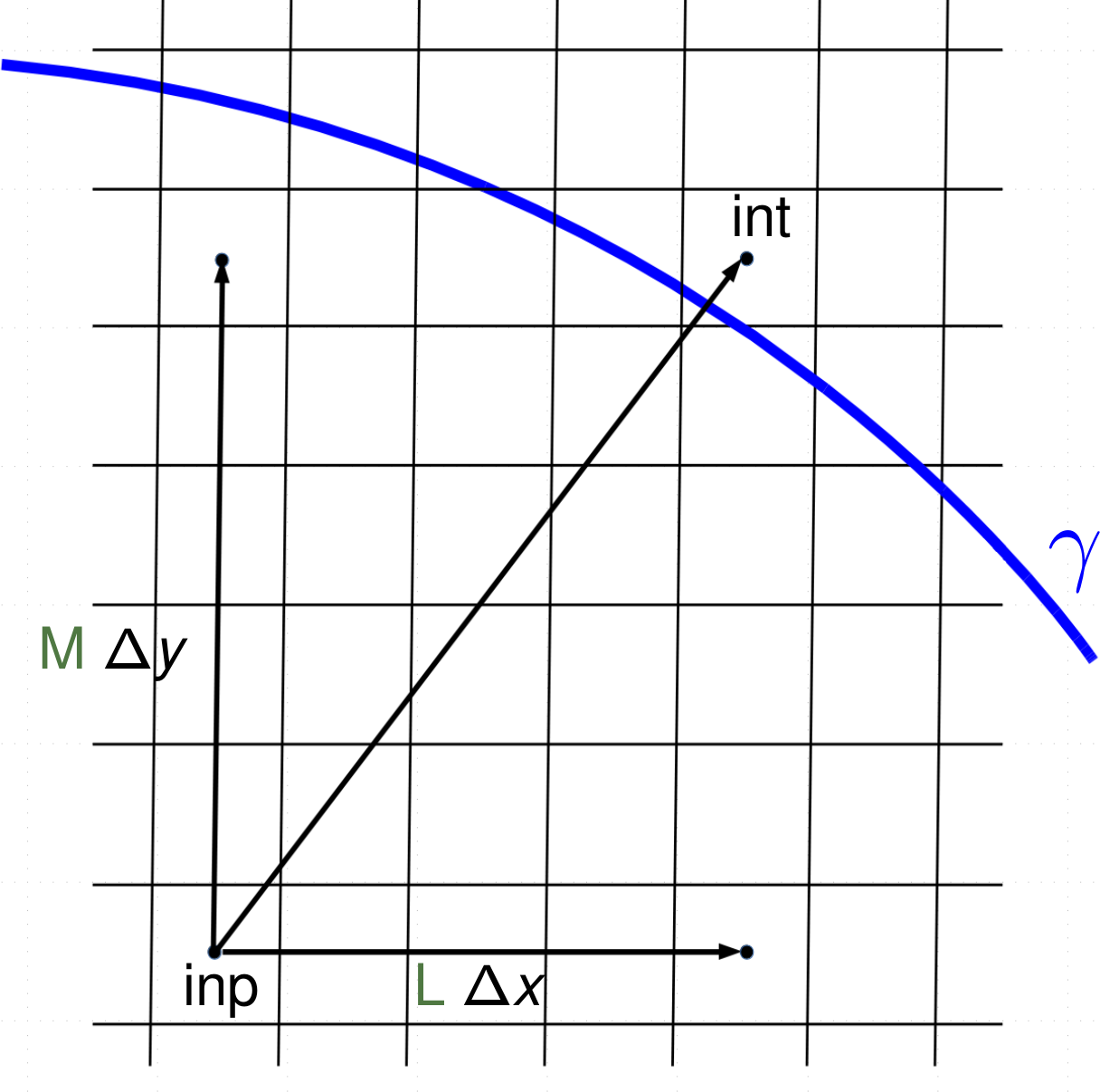}
	\caption{\small{Determination of the interface $\gamma$ position on 
			        structured grid using the signed distance function $\psia$
			        (not shown).
			         Schematic presentation of the points at $t'\!=\!0$ \texttt{(int)}
			        and $t' \!=\! 1$ \texttt{(inp)} required to compute
			        the trapezoidal quadrature, see \Eq{eq30}.
		             }}
	\label{fig3}
\end{figure}
\elnm

Let us
note
if the 
two point,
first-order accurate
quadrature is used,
the discussed
problem 
is reduced to 
finding index
\texttt{(int)}
of the control
volume 
with 
the interface $\gamma$ 
during
loop over
all grid 
points  
with the
index \texttt{(inp)}. 
The 
sketch of 
this procedure
in the two-dimensional
case
for the one pair
of control volumes
\texttt{(inp)}
and \texttt{(int)}
is depicted
in \Fig{fig3}.
The simplest 
quadrature
taking into 
account the two
point information
required to approximate
\Eq{eq29}
is given
by the first-order 
accurate trapezoidal
rule
\blnm
\be
  I_T \lr \psi \rr \!=\! \frac{1}{2} \ls \frac{1}{ \eph \texttt{(int)} } + \frac{1}{\eph \texttt{(inp)}} \rs.
\label{eq30}
\ee
\elnm
To obtain
a higher order 
of accuracy,
the third-order
accurate Simpson
rule can 
be used to
approximate
\Eq{eq29} as well,
it  reads 
\blnm
\be
I_S \lr \psi \rr \!=\! \frac{1}{6} \ls \frac{1}{ \eph \texttt{(int)} } 
                                     + \frac{4}{ \eph \texttt{(inm)} }
                                     + \frac{1}{ \eph \texttt{(inp)}} \rs,
\label{eq31}
\ee
\elnm 
where $\texttt{(inm)}$
denotes control volume
in the center between
$\texttt{(inp)}$,
$\texttt{(int)}$.
In what follows
it is compared how
approximations of 
the integral (\ref{eq29})
given
by \Eqs{eq30}{eq31}
affect
the obtained
solutions.
Introduction
of an even higher
order of accuracy
in the approximation
of \Eq{eq29}
requires
considering
additional
control
volumes   
in-between
the local
position on
the mesh \texttt{(inp)} and 
the expected position
of the interface $\gamma$
\texttt{(int)}.
When
the number
of control volumes
between
\texttt{(inp)} and \texttt{(int)}
is smaller
than
the quadrature
stencil,
the higher-order
quadrature
has
to be
replaced
by 
the appropriate
lower-order
quadrature
or interpolation
of $\ephxt$.

As the present
results are obtained
using
the structured grid solver,
computation of
\Eq{eq30} or \Eq{eq31}
is straightforward.
Knowing
the local position
at the grid \texttt{(inp)}
and the value of 
the signed distance
function
in this
cell $\psi \lr inp \rr$
one needs 
to project
it on $(x,y)$
directions
to obtain: 
$\psi_x \!=\! -\psi \lr inp \rr n_{\gamma,x}$,
$\psi_y \!=\! -\psi \lr inp \rr n_{\gamma,y}$.
Next,  
compute
constants $\texttt{(L,M)}$ 
(see \Fig{fig3})
where $L \! \approx \! N\!I\!N\!T \lr  \psi_x/\Delta x \rr$
and $M \! \approx \! N\!I\!N\!T \lr \psi_y/\Delta y \rr$,
and finally determine 
the index of the cell 
containing interface \texttt{(int)},
on structured
grid 
$int \!=\! inp \!+\! L \!\cdot\! N\!J \!+\! M$, 
where $N\!J$ is the number of grid cells
in $j$ direction, $N\!I\!N\!T$ is 
the intrinsic function
returning nearest 
integer.
Point 
\texttt{(inm)}
in \Eq{eq31} 
is obtained
in a similar
way taking
$\psi_x^m \!=\! - 0.5 \psi \lr inp \rr n_{\gamma,x}$,
$\psi_y^m \!=\! - 0.5 \psi \lr inp \rr n_{\gamma,y}$
and then 
computing
$\texttt{($L^m$,$M^m$)}$.
If the stencil
where \Eq{eq31}
is computed
is smaller 
than three 
control volumes,
the $\eph \texttt{(inm)}$
value is 
obtained
as $\eph \texttt{(inm)}\!=\!(\eph \texttt{(inp)}\!+\!\eph \texttt{(int)})/2$. 
Due to introduction
of $\ephppxt$ in \Eq{eq25},
in one and 
three dimensional
cases this procedure
can be easily 
adopted by taking
into account one less
or one more
spatial
direction
to compute \texttt{(int)} 
and/or \texttt{(inm)}. 
%

\subsection{Evolution of one-dimensional cumulative distribution function}
\label{ssec42}
%

To
compare 
the exact
and approximate
semi-analytical
solutions
and estimate 
numerical 
accuracy 
of the  procedure
introduced
in Section \ref{ssec41} 
the evolution of
the one-dimensional 
$\alpp$ profile
disturbed by 
variable 
$\ephxt$ is 
studied
on 
three,
gradually
refined
grids
$m_i \!=\! 2^{5+i} \!\times\! 2^{5+i},\, i\!=\!1,2,3$.
The characteristic
length scale
$\ephxt$ is
predefined 
as the step $\ephxtS$
or bell $\ephxtB$
shaped disturbance,
see \Eqs{eqC1}{eqC2} 
respectively;
only \Eq{eq18}
is solved, 
as $\bw\!=\!\bu\!=\!0$
advection is neglected.
The number of physical
time steps is set to 
$N_t=72$, $\Delta t \!=\! 10^{-3}\,[s]$.
The number 
of re-initialization
time steps, 
if not stated otherwise,
is 
set to $N_\tau \!=\! 256$
with the size
$\Delta \tau \!=\! \ephb/2$ 
to guarantee the steady
state solution of  
\Eq{eq18} after 
each time iteration
$it$.
The minimum (base)
thickness of the
interface is 
set to 
$\ephb\!=\!\Delta x_i,\,i=1,2,3$.
The discretization
of \Eq{eq18} is the
same as in \citep{twacl17},
the only modification
accounting for $\ephxt$
is introduced to the 
constrained interpolation
used to approximate
$\deltaa \!=\! \alpha \lr 1\!-\!\alpha \rr$
in \Eq{eq18},
see \Eq{eqD5}.

\begin{figure}[!ht] \nonumber
	\begin{minipage}{.5\textwidth}
		\includegraphics[width=1.\textwidth,height=.75\textwidth,angle=0]{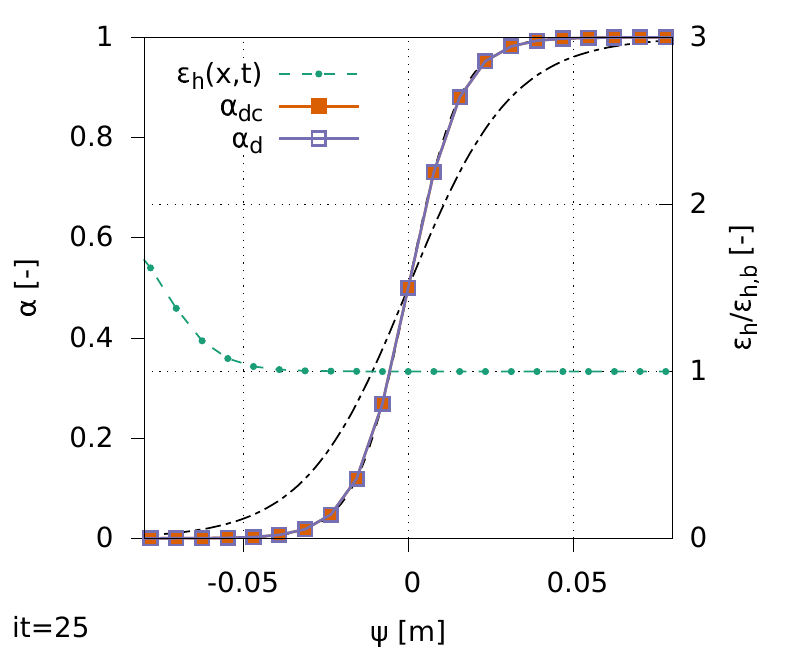}
	\end{minipage}
	\begin{minipage}{.5\textwidth}
		\includegraphics[width=1.\textwidth,height=.75\textwidth,angle=0]{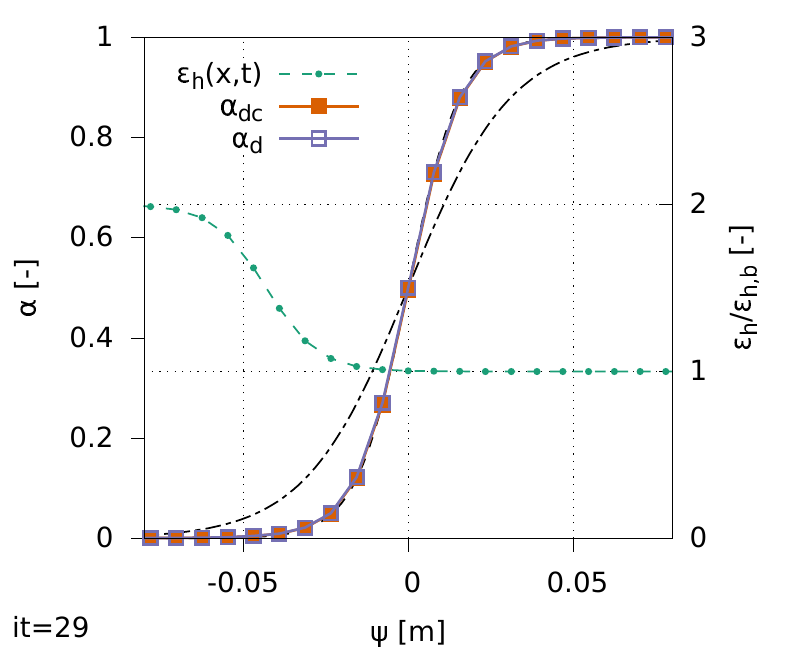}
	\end{minipage}
	\begin{minipage}{.5\textwidth}
		\includegraphics[width=1.\textwidth,height=.75\textwidth,angle=0]{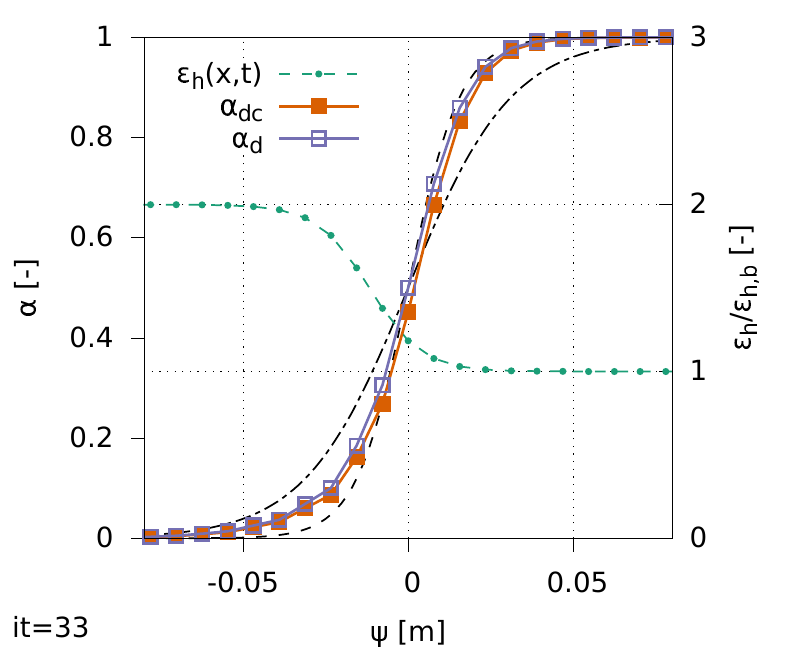}
	\end{minipage}
	\begin{minipage}{.5\textwidth}
		\includegraphics[width=1.\textwidth,height=.75\textwidth,angle=0]{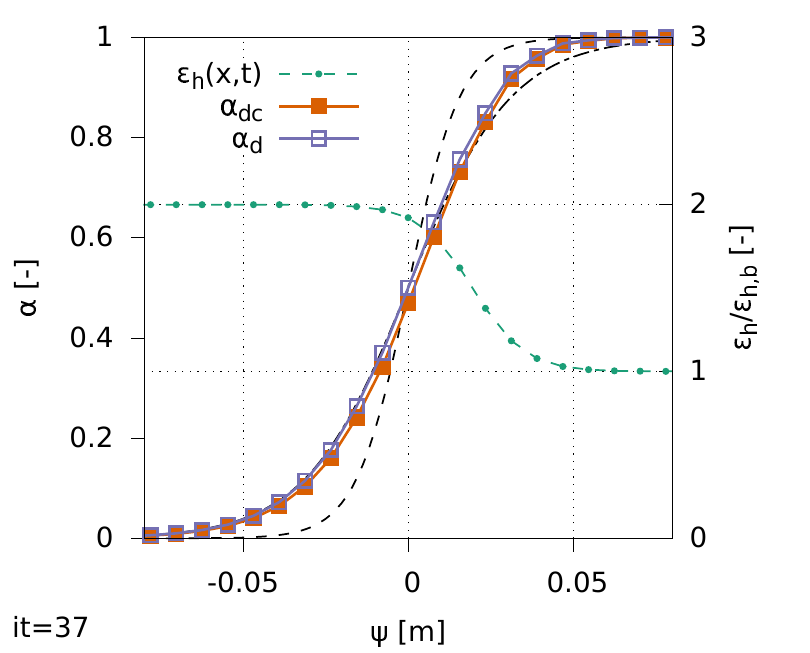}
	\end{minipage}
	\begin{minipage}{.5\textwidth}
		\includegraphics[width=1.\textwidth,height=.75\textwidth,angle=0]{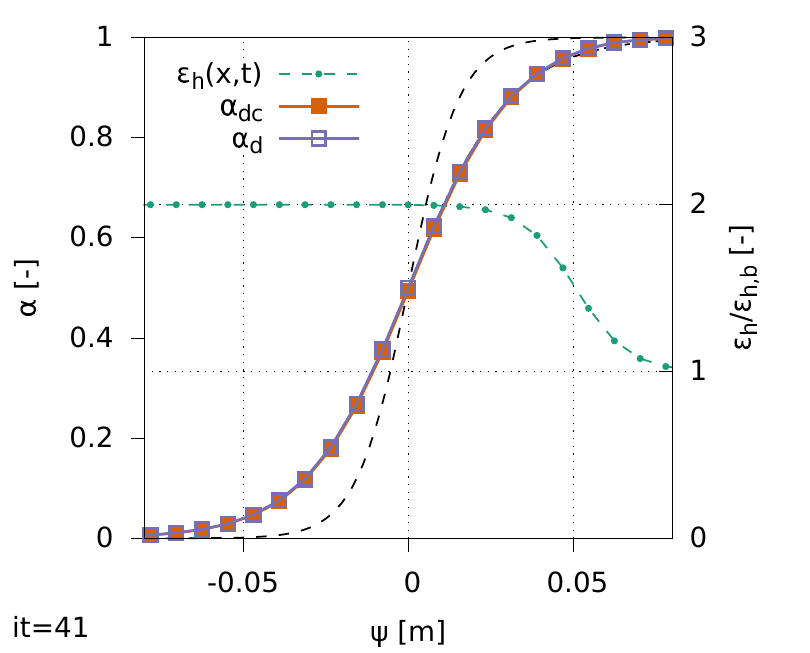}
	\end{minipage}
	\begin{minipage}{.5\textwidth}
		\includegraphics[width=1.\textwidth,height=.75\textwidth,angle=0]{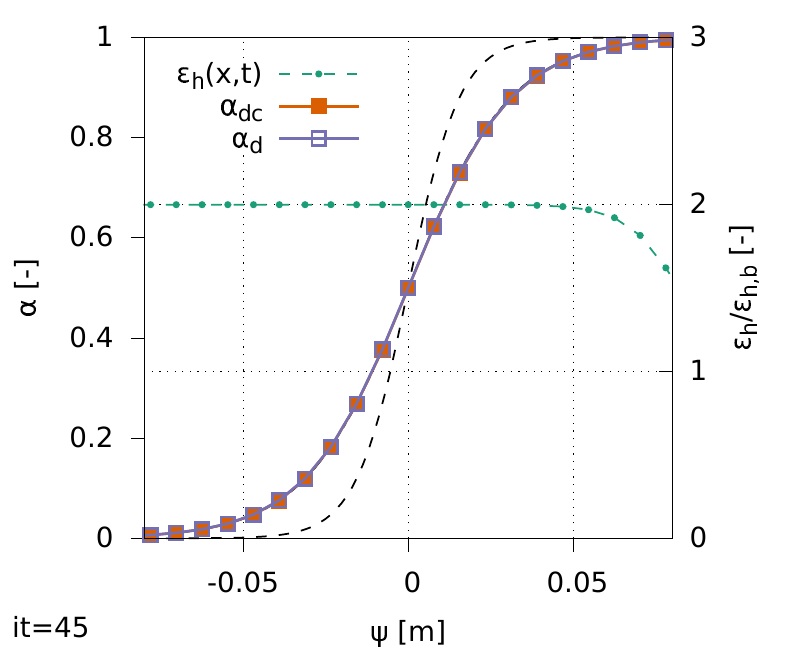}
	\end{minipage}
	\caption{\small{Comparison of 
		       	$\alpp$ 
		       	affected by 
		        the step shaped 
			variation $\ephxtS$ defined by \Eq{eqC1}
			shown after (left to right)
			equal time intervals; $\alpha_{dc}$ denotes the coupled
			and $\alpha_d$ semi-analytical solution, respectively.
			Black-dashed lines mark the analytical
		        profiles of $\alpp$ where $\eph\!=\!\ephb$ or $\eph\!=\!2\ephb$, see \Eq{eq13}. }}
	\label{fig4}
\end{figure}

\Figs{fig4}{fig5}
present,
respectively,
the evolution of
$\ephxtS$, $\ephxtB$
profiles
and corresponding
variations of 
$\alpp$
at equal 
time intervals
on the mesh $m_2$.
The results
presented therein
are obtained after
each time iteration
$it$ at the end
of re-initialization
process, see \Fig{m2fig6}.
The black-dashed
lines in \Figs{fig4}{fig5} represent
the analytical
$\alpp$ profiles 
obtained  
using \Eq{eq13}
with 
$\eph\!=\!\ephb$ (dashed line), 
and $\eph\!=\!2\ephb$ (dashed-dotted line).
\blnm
\elnm
Therein,
$\alpha_{dc}$
denotes 
solutions
obtained using 
the direct coupling 
of \Eqs{eq18}{eq27},
$\alpha_d$
denotes
an approximate, semi-analytical
solution 
obtained using 
\Eq{eq18} where
$\eph \!=\! \ephb$
and \Eq{eq28}
accounting
for $\ephxt$.
In both cases
the third-order
accurate
Simpson 
rule (\ref{eq31})
is used to 
approximate $\Ip$
integral in \Eq{eq26}.

The convergence
space 
of the re-initialization
equation (\ref{eq18})
for the coupled 
$\alpha_{dc}$  cases
is presented
in \Figs{m1fig6}{m3fig6}
on the meshes $m_i\!=\!2^{5+i}\!\times\!2^{5+i},\,i=1,2,3$
respectively.
Therein,
the $L_{1,\tau}$ 
norm defined
by \Eq{eqE1} 
characterizing
the numerical
solution
of \Eq{eq18}
in times $t,\,\tau$
is presented.
The top row
shows convergence
of \Eq{eq18}
obtained using
the first-order
accurate 
quadrature (\ref{eq30}),
the bottom
row using the third-order
accurate quadrature (\ref{eq31}).
\begin{figure}[!ht] \nonumber
	\begin{minipage}{.5\textwidth}
		\includegraphics[width=1.\textwidth,height=.75\textwidth,angle=0]{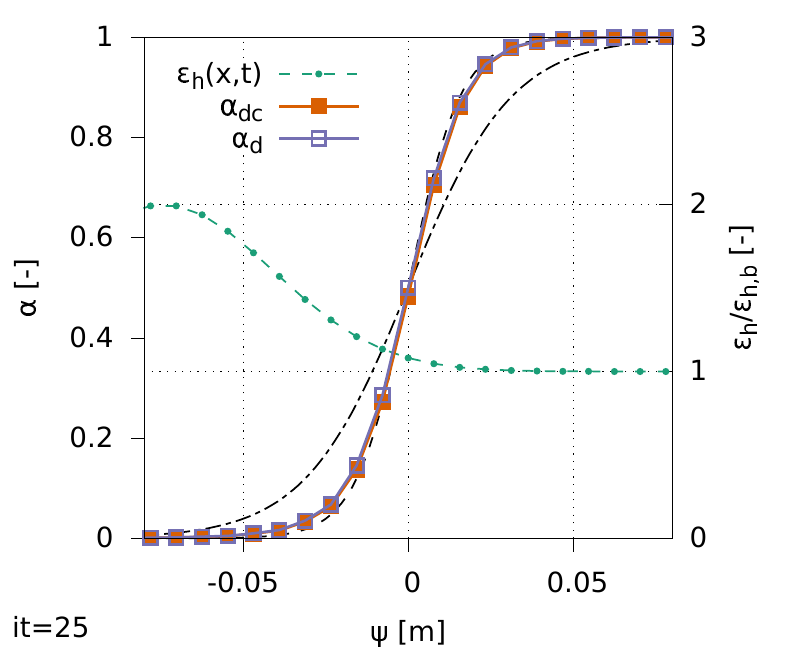}
	\end{minipage}
	\begin{minipage}{.5\textwidth}
		\includegraphics[width=1.\textwidth,height=.75\textwidth,angle=0]{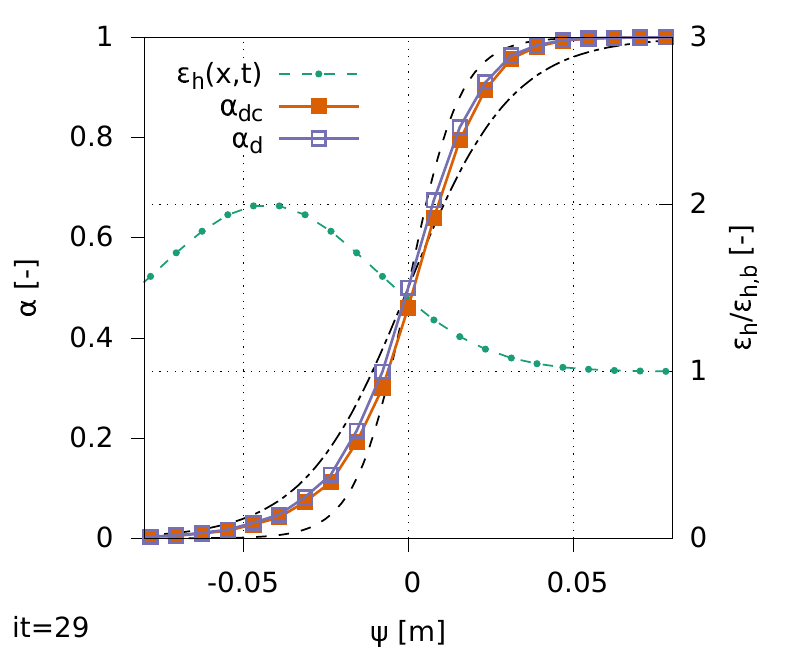}
	\end{minipage}
	\begin{minipage}{.5\textwidth}
		\includegraphics[width=1.\textwidth,height=.75\textwidth,angle=0]{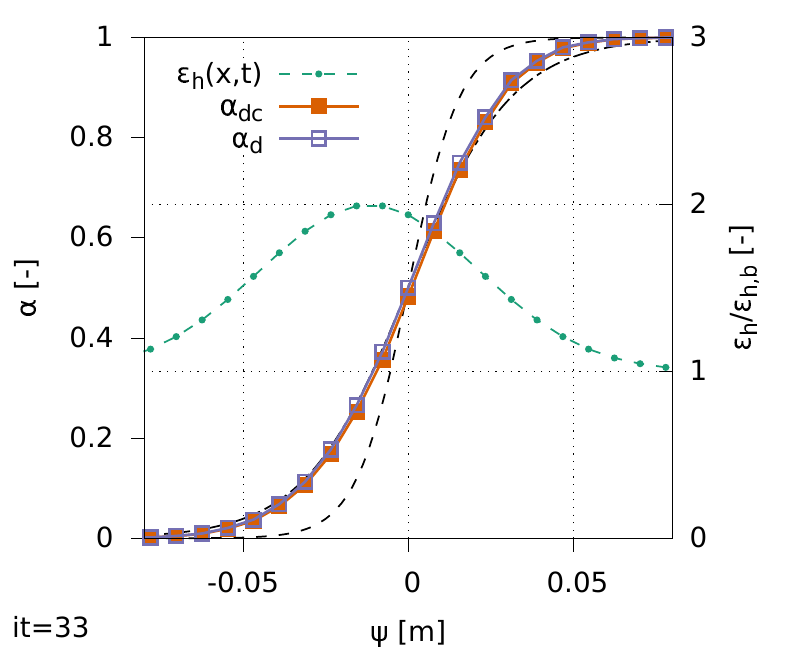}
	\end{minipage}
	\begin{minipage}{.5\textwidth}
		\includegraphics[width=1.\textwidth,height=.75\textwidth,angle=0]{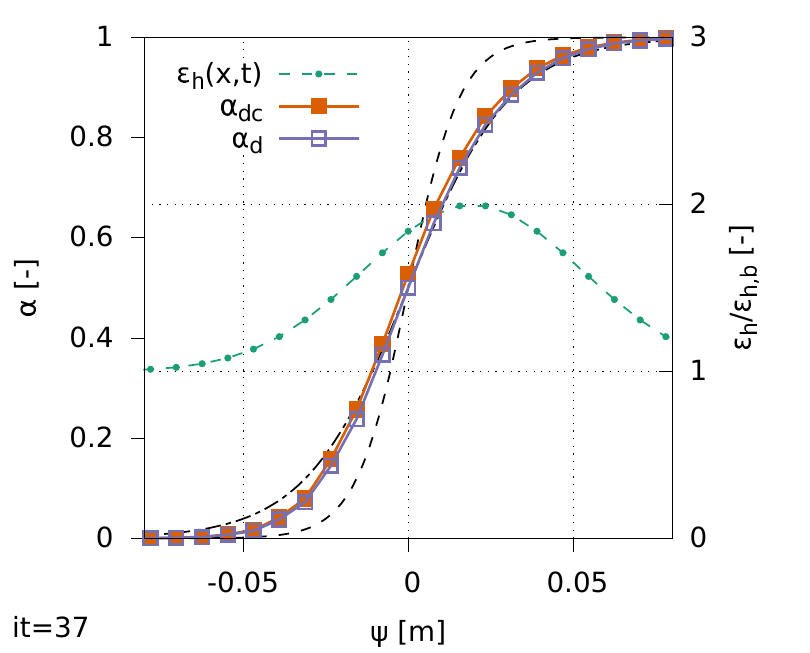}
	\end{minipage}
	\begin{minipage}{.5\textwidth}
		\includegraphics[width=1.\textwidth,height=.75\textwidth,angle=0]{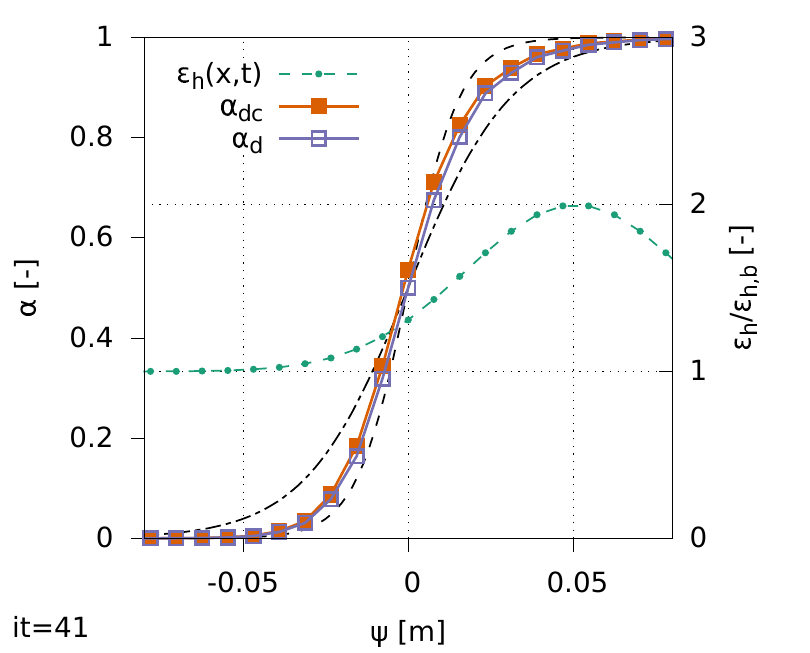}
	\end{minipage}
	\begin{minipage}{.5\textwidth}
		\includegraphics[width=1.\textwidth,height=.75\textwidth,angle=0]{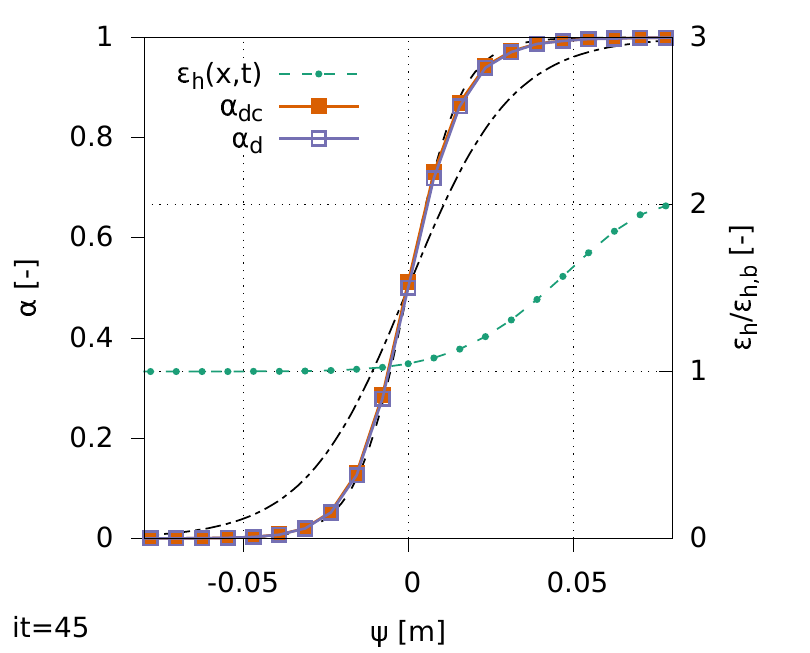}
	\end{minipage}
	\caption{\small{Comparison of 
		$\alpp$ 
		affected by 
		the bell shaped 
		variation $\ephxtB$ defined by \Eq{eqC2},
		shown after (left to right)
		equal time intervals; $\alpha_{dc}$ denotes the coupled
		and $\alpha_d$ semi-analytical solution, respectively.
		Black-dashed lines mark the analytical
		profiles of $\alpp$ where $\eph\!=\!\ephb$ or $\eph\!=\!2\ephb$, see \Eq{eq13}. }}
	\label{fig5}
\end{figure}
Although in 
all cases
the convergence of
the numerical solution
is obtained,
one observes 
the variation 
of $\alpp$
caused by $\ephxt$
strongly
affects
the numerical
solution
of \Eq{eq18}.
In order 
to avoid
this dependence,  
the approach
where
$\eph \!=\! \ephb \!=\! const.$
in \Eq{eq18} 
and $\ephxt$
variation 
is modeled
using \Eq{eq28}
is introduced. 
In this case,
the $L_{1,\tau}$
norm 
remains almost 
constant $\sim 10^{-16}$
for all steps
$t,\tau$ \citep{twacl15};
$\psi$ field
obtained during
the solution of
\Eq{eq18}
where $\eph\!=\!\ephb$  
is treated
as the carrier
function
for $\alpp$
given by \Eq{eq28}.
\blnm
\begin{figure}[!ht] \nonumber
	\begin{minipage}{.49\textwidth}
		\subfloat[]{\centering\includegraphics[width=.75\textwidth,height=1.\textwidth,angle=-90]{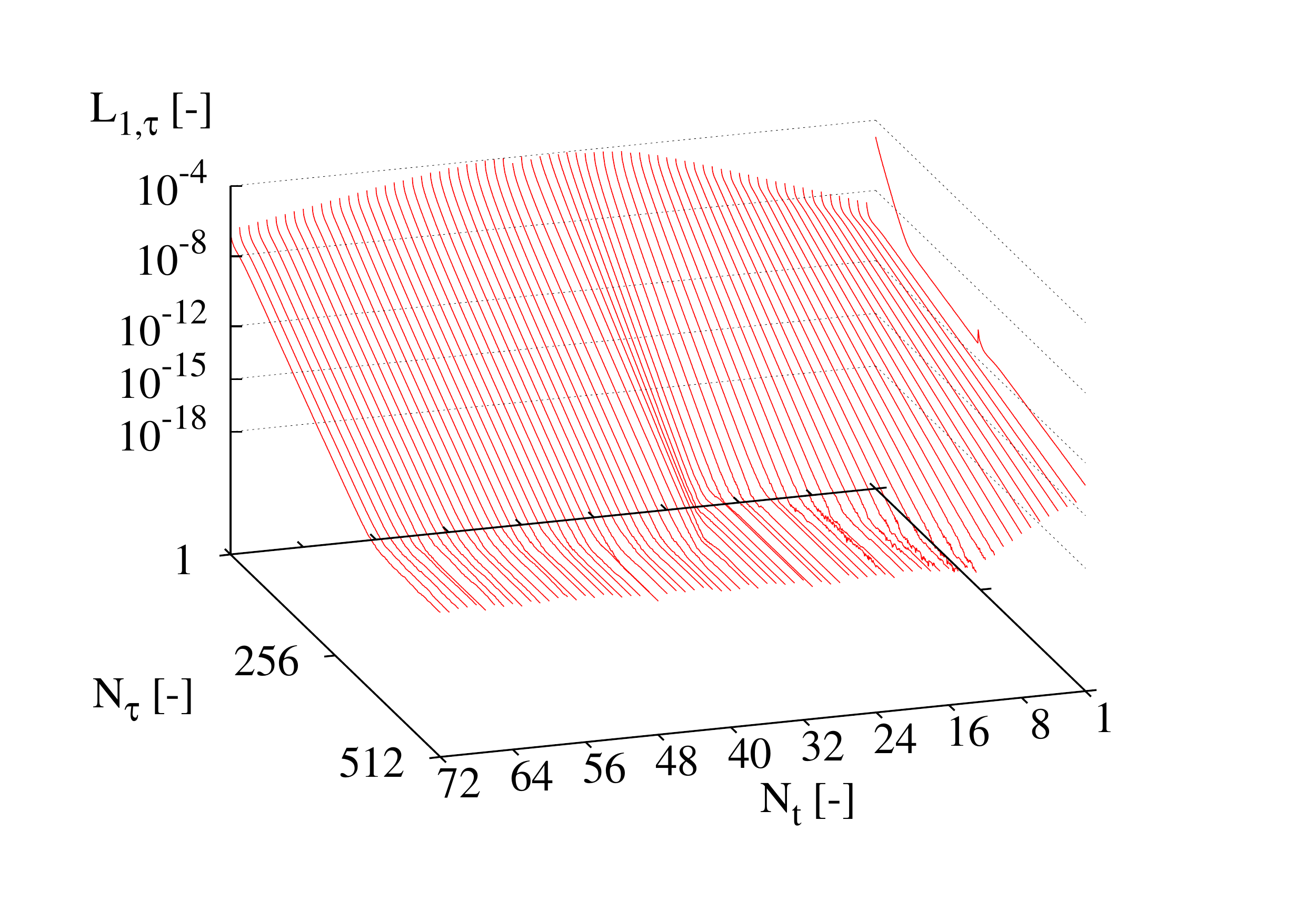}}
	\end{minipage}%
	\begin{minipage}{.49\textwidth}
		\subfloat[]{\centering\includegraphics[width=.75\textwidth,height=1.\textwidth,angle=-90]{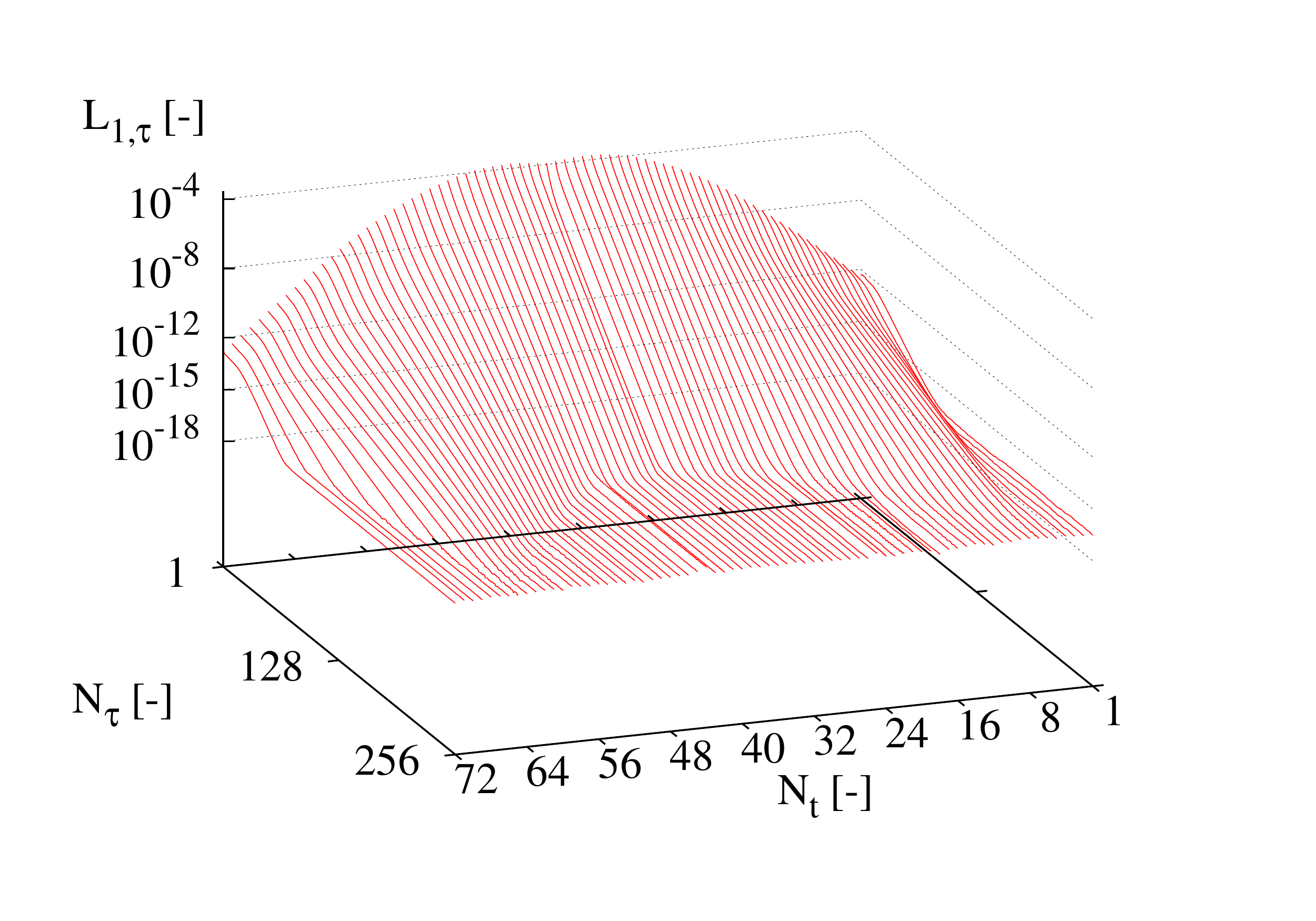}}
	\end{minipage}
	\begin{minipage}{.49\textwidth}
		\subfloat[]{\centering\includegraphics[width=.75\textwidth,height=1.\textwidth,angle=-90]{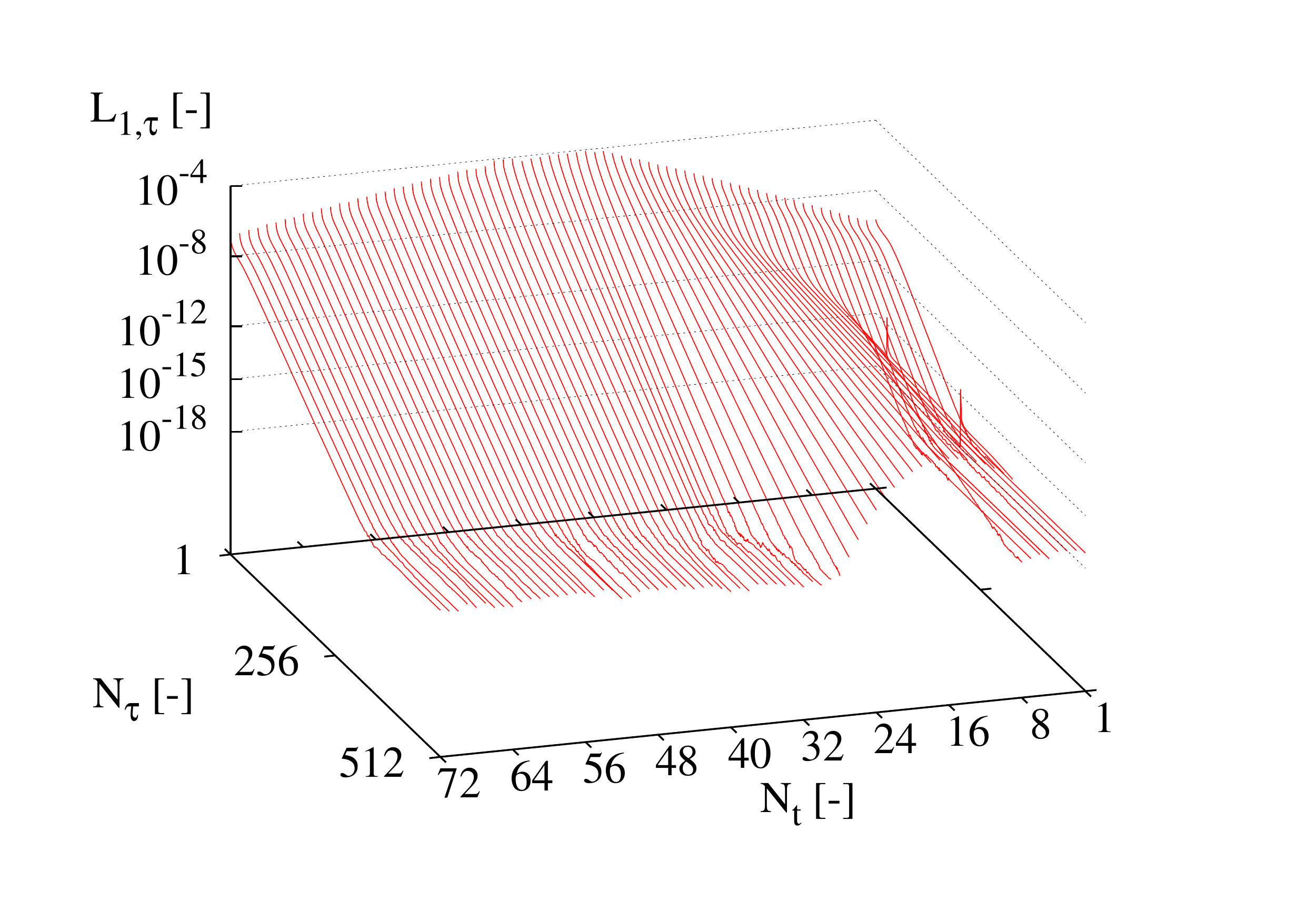}}
	\end{minipage}
	\begin{minipage}{.49\textwidth}
    	\subfloat[]{\centering\includegraphics[width=.75\textwidth,height=1.\textwidth,angle=-90]{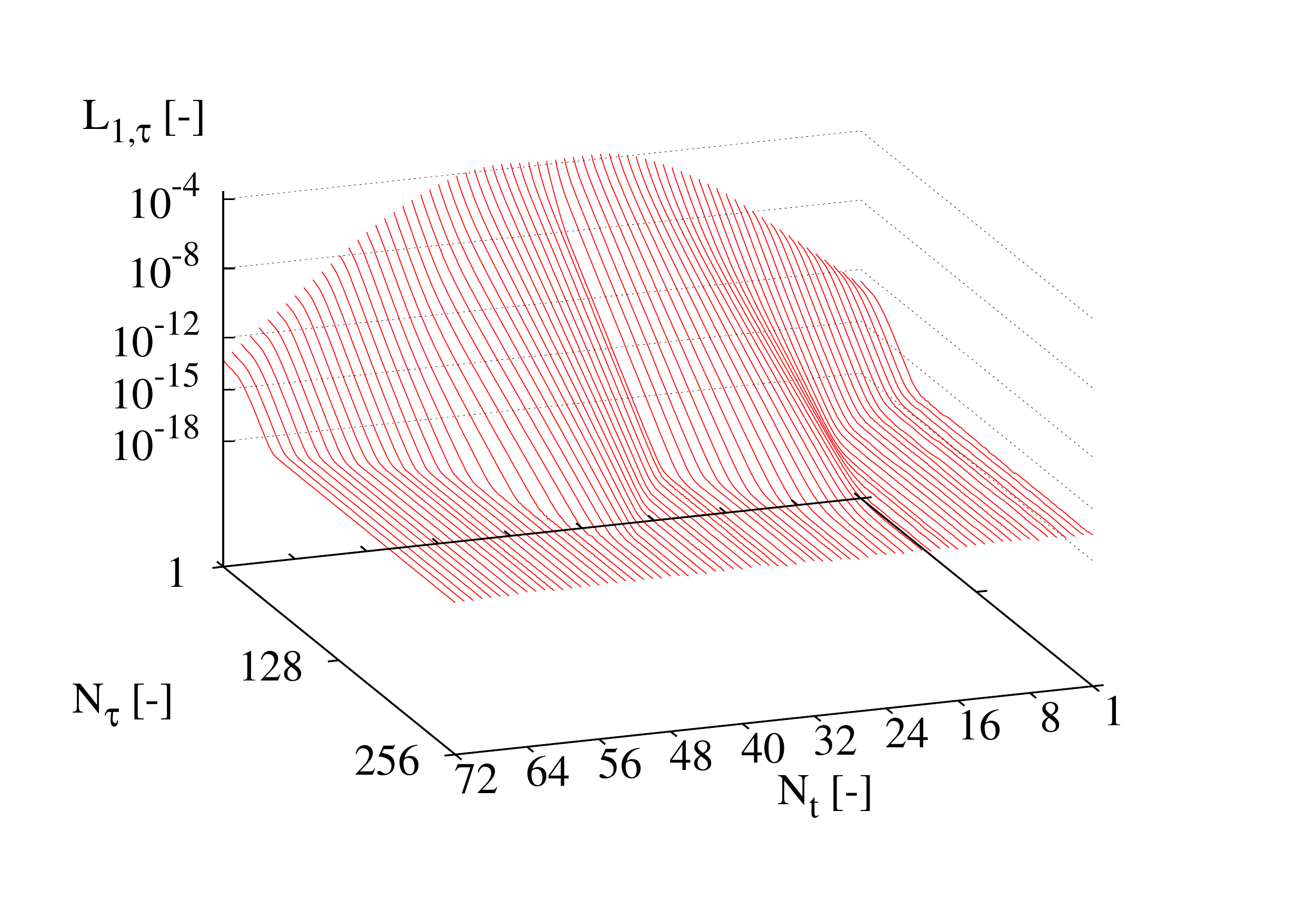}}
    \end{minipage}
	\caption{\small{Convergence space of $L_{1,\tau}$ norm given by \Eq{eqE1},
		            during solution of \Eq{eq18} and \Eq{eq27} on the mesh $m_1$.
					The one-dimensional $\alpp$ profile is altered 
					by $\ephxtS$ (left column)
					or $\ephxtB$ (right column)
					see \Figs{fig4}{fig5}, respectively.
		            The results are
	       		    obtained with $N_t\!=\!72$
		            time and $N_\tau \!=\! 512$ (a,c) or $N_\tau \!=\!256 $ (b,d)  
		     	    re-initialization steps;
					the trapezoidal (top row) 
					or Simpson (bottom row) rules
					are employed to approximate integral in  \Eq{eq26}.
					}}
 	\label{m1fig6}
\end{figure}
\elnm 
\blnm
\begin{figure}[!ht] \nonumber
	\begin{minipage}{.49\textwidth}
		\subfloat[]{\centering\includegraphics[width=.75\textwidth,height=1.\textwidth,angle=-90]{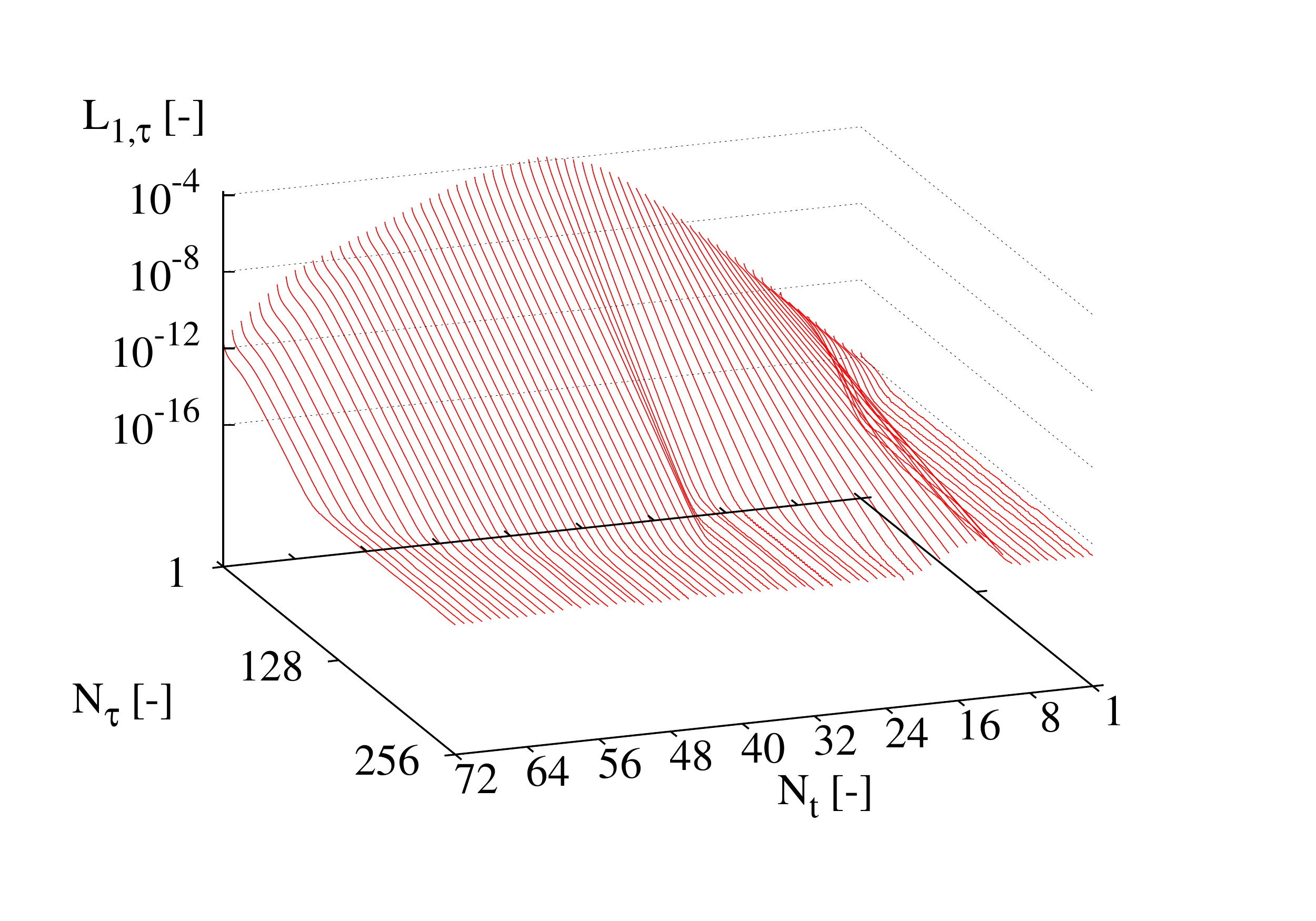}}
	\end{minipage}%
	\begin{minipage}{.49\textwidth}
		\subfloat[]{\centering\includegraphics[width=.75\textwidth,height=1.\textwidth,angle=-90]{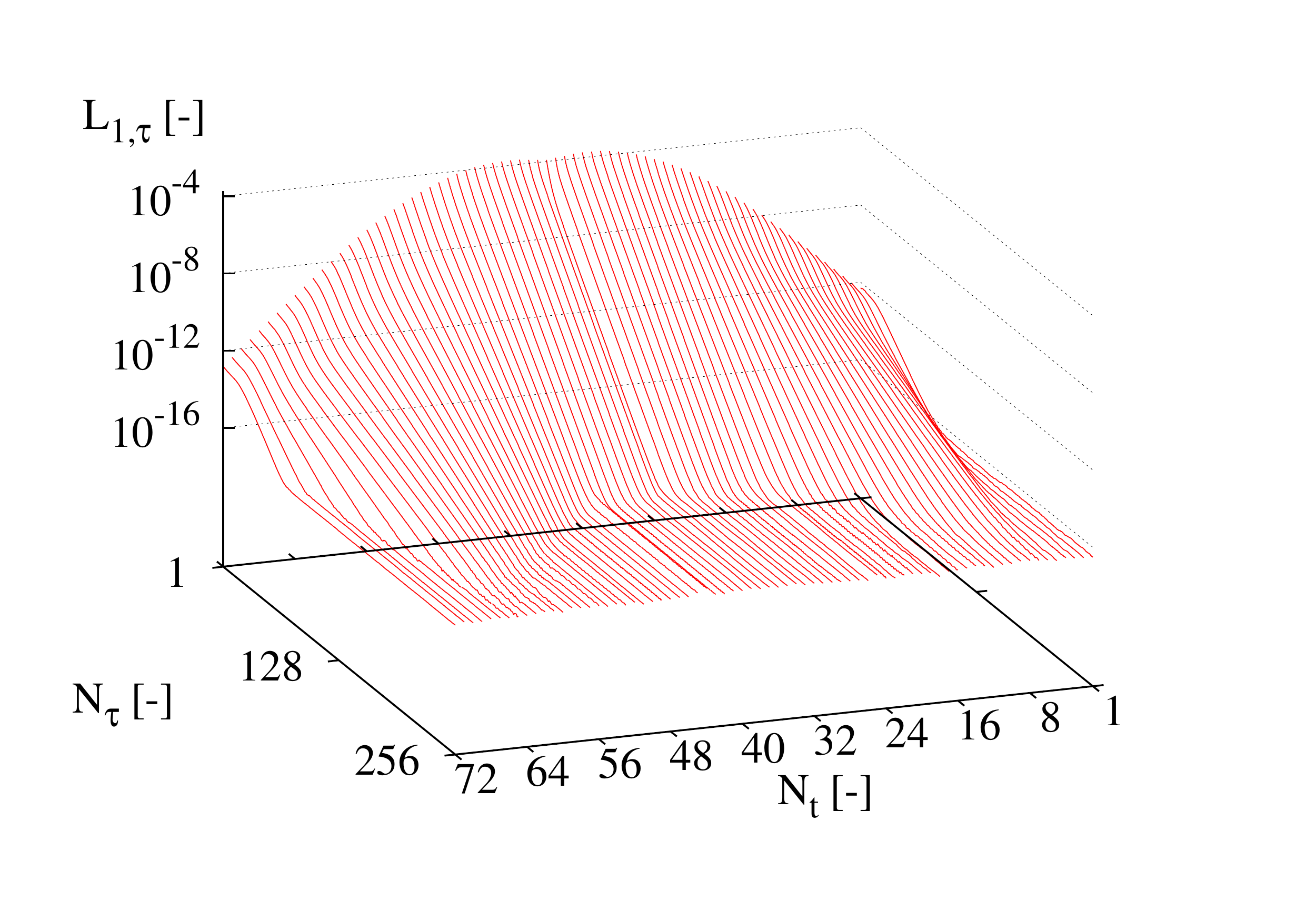}}
	\end{minipage}
	\begin{minipage}{.49\textwidth}
		\subfloat[]{\centering\includegraphics[width=.75\textwidth,height=1.\textwidth,angle=-90]{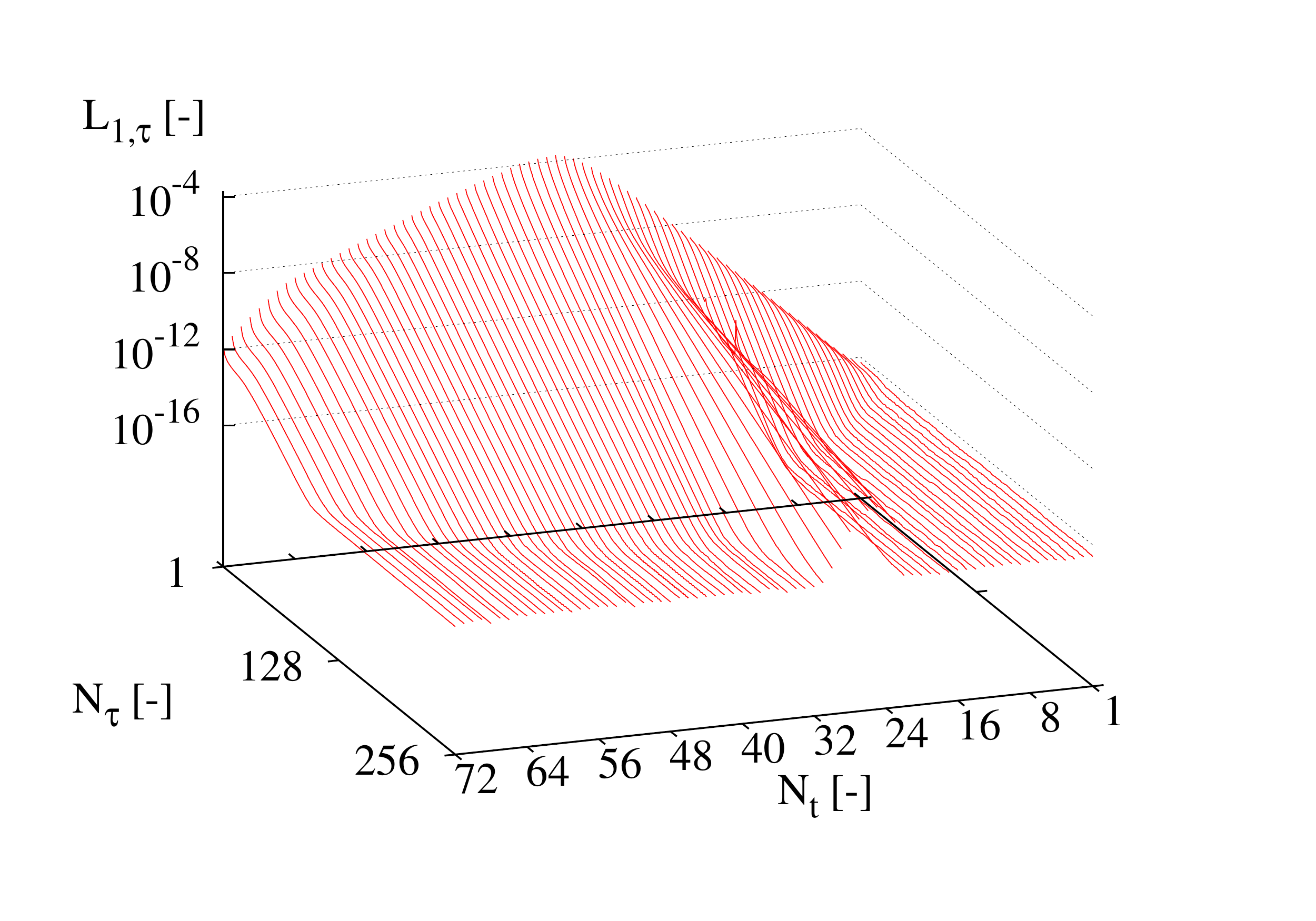}}
	\end{minipage}
	\begin{minipage}{.49\textwidth}
		\subfloat[]{\centering\includegraphics[width=.75\textwidth,height=1.\textwidth,angle=-90]{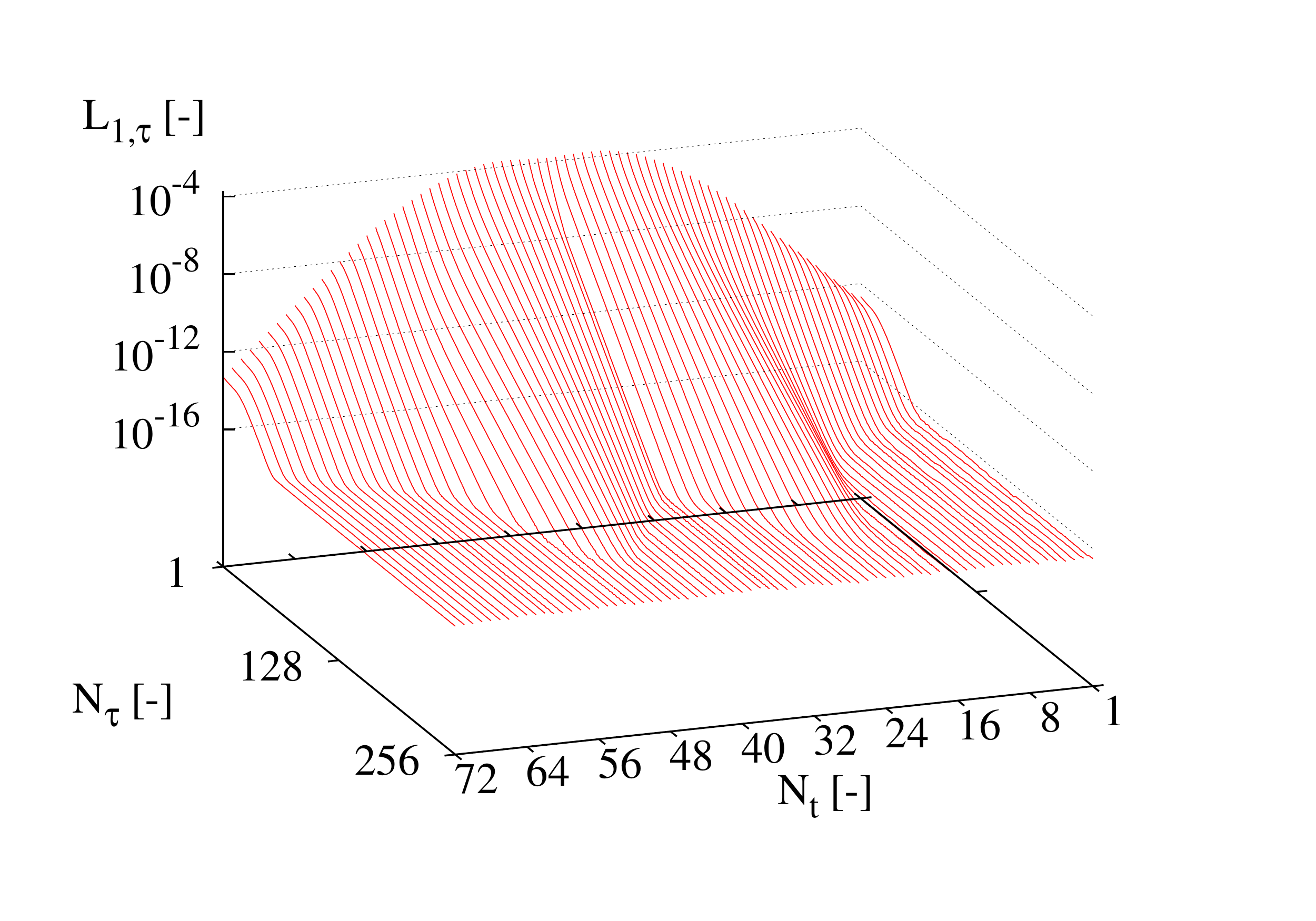}}
	\end{minipage}
	\caption{\small{Convergence space of $L_{1,\tau}$ norm given by \Eq{eqE1},
					during solution of \Eq{eq18} and \Eq{eq27} on the mesh $m_2$.
					The one-dimensional $\alpp$ profile is altered 
					by variable $\ephxtS$ (left column)
					or $\ephxtB$ (right column).
					The results are
					obtained with $N_t\!=\!72$
					time and $N_\tau \!=\!256 $ 
					re-initialization steps;
				    the trapezoidal (top row) 
					or Simpson (bottom row) rules
					are employed to approximate integral in  \Eq{eq26}.
			 }}
	\label{m2fig6}
\end{figure}
\elnm 
\blnm
\begin{figure}[!ht] \nonumber
	\begin{minipage}{.49\textwidth}
		\subfloat[]{\centering\includegraphics[width=.75\textwidth,height=1.\textwidth,angle=-90]{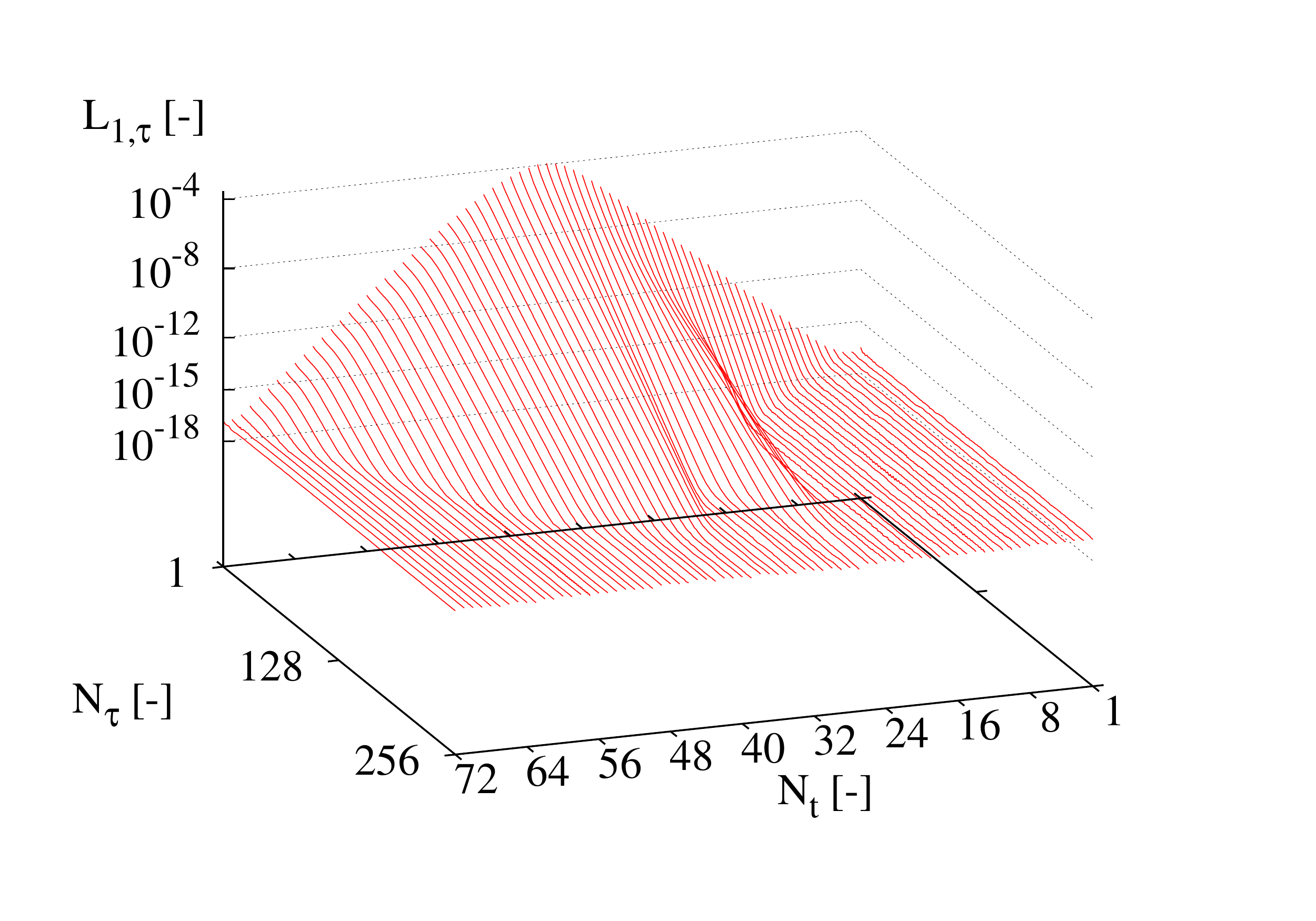}}
	\end{minipage}%
	\begin{minipage}{.49\textwidth}
		\subfloat[]{\centering\includegraphics[width=.75\textwidth,height=1.\textwidth,angle=-90]{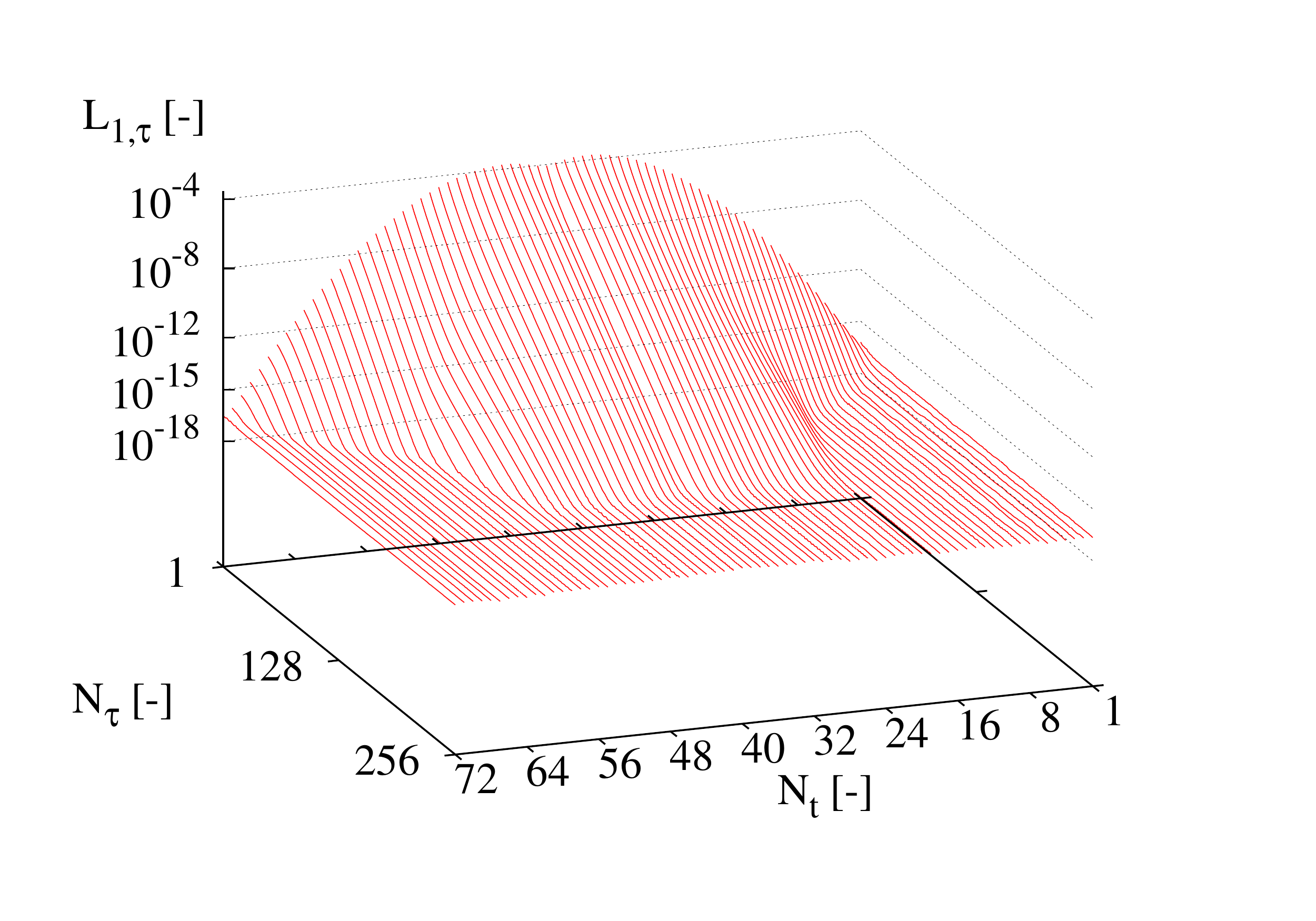}}
	\end{minipage}
	\begin{minipage}{.49\textwidth}
		\subfloat[]{\centering\includegraphics[width=.75\textwidth,height=1.\textwidth,angle=-90]{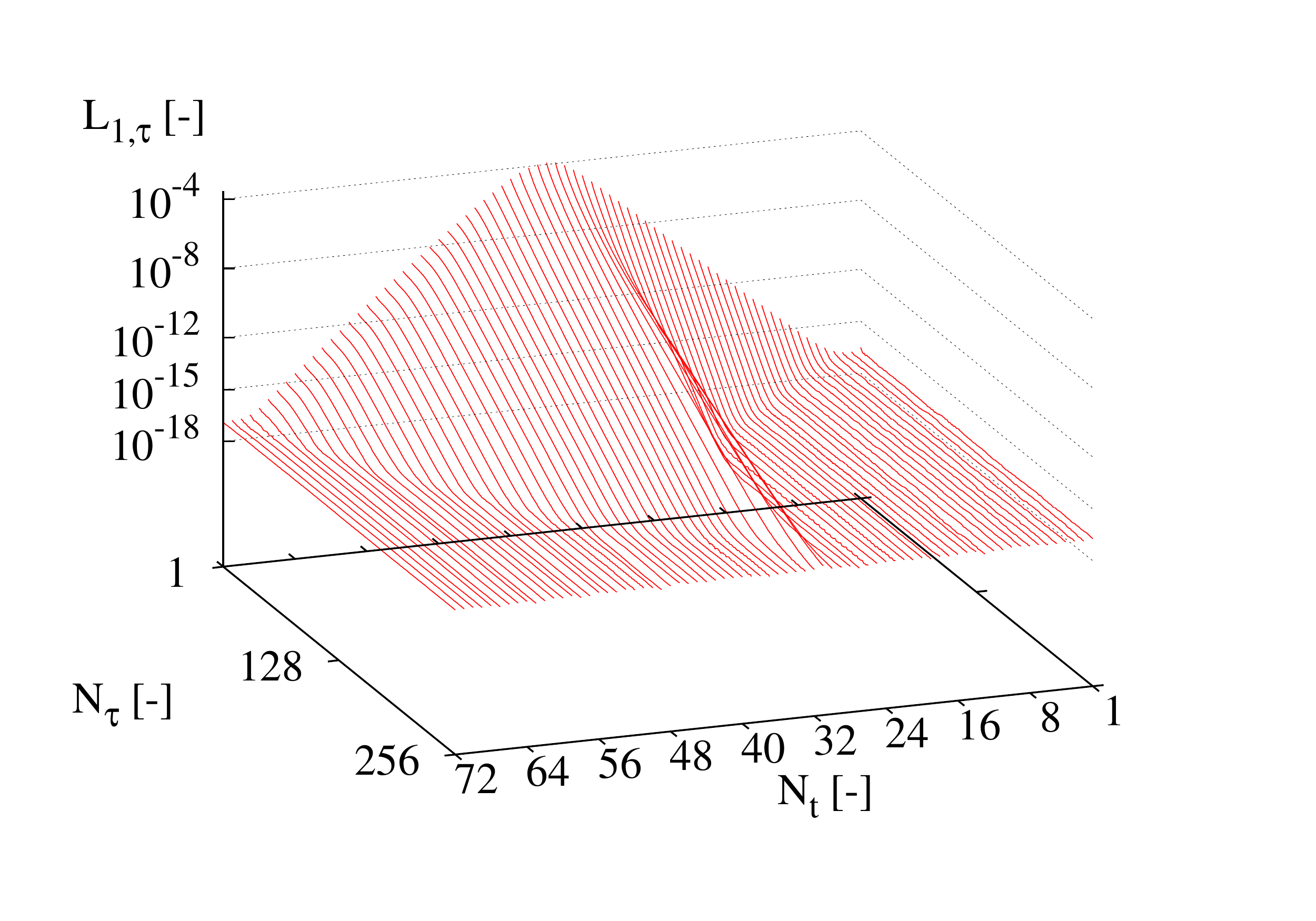}}
	\end{minipage}
	\begin{minipage}{.49\textwidth}
		\subfloat[]{\centering\includegraphics[width=.75\textwidth,height=1.\textwidth,angle=-90]{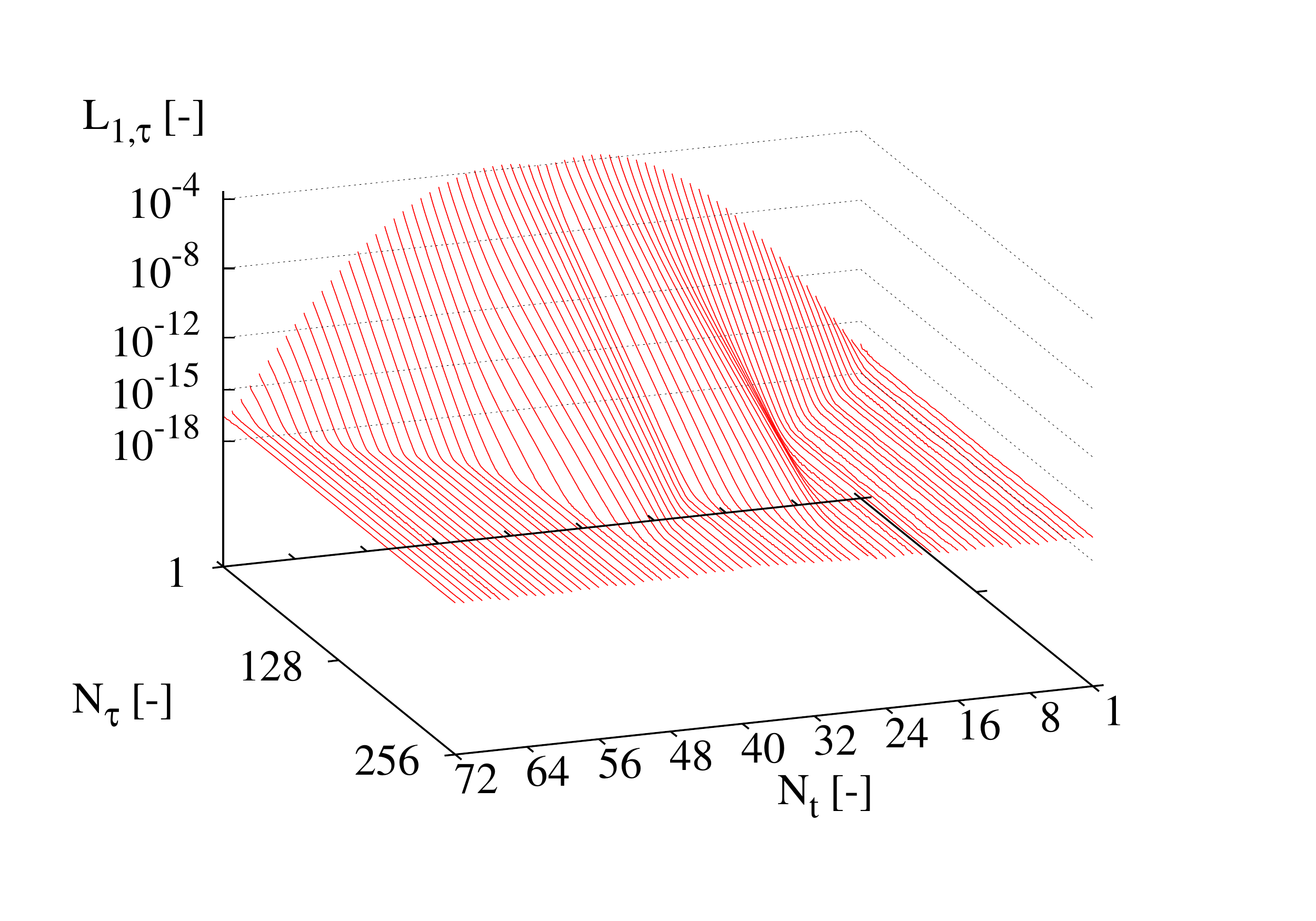}}
	\end{minipage}
	\caption{\small{Convergence space of $L_{1,\tau}$ norm given by \Eq{eqE1},
			during solution of \Eq{eq18} and \Eq{eq27} on the mesh $m_3$.
			The one-dimensional $\alpp$ profile is altered 
			by variable $\ephxtS$ (left column)
			or $\ephxtB$ (right column).
			The results are
			obtained with $N_t\!=\!72$
			time and $N_\tau \!=\!256 $ 
			re-initialization steps;
			the trapezoidal (top row) 
			or Simpson (bottom row) rules
			are employed to approximate integral in  \Eq{eq26}.
		    }}
	\label{m3fig6}
\end{figure}
\elnm 
\blnm
\begin{figure}[!ht] \nonumber
	\begin{minipage}{.5\textwidth}
		\includegraphics[width=.75\textwidth,height=1.\textwidth,angle=-90]{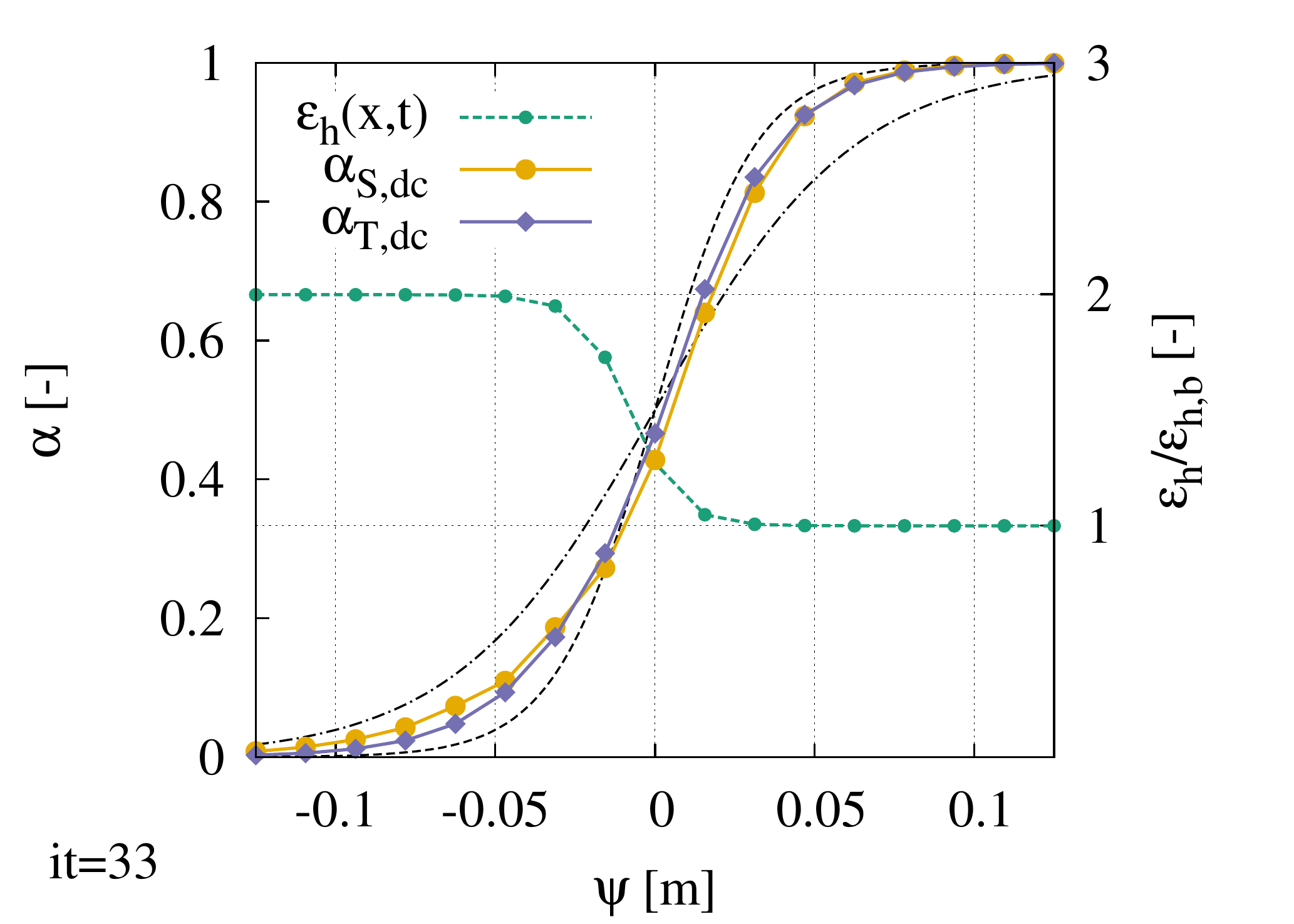}
	\end{minipage}
	\begin{minipage}{.5\textwidth}
		\includegraphics[width=.75\textwidth,height=1.\textwidth,angle=-90]{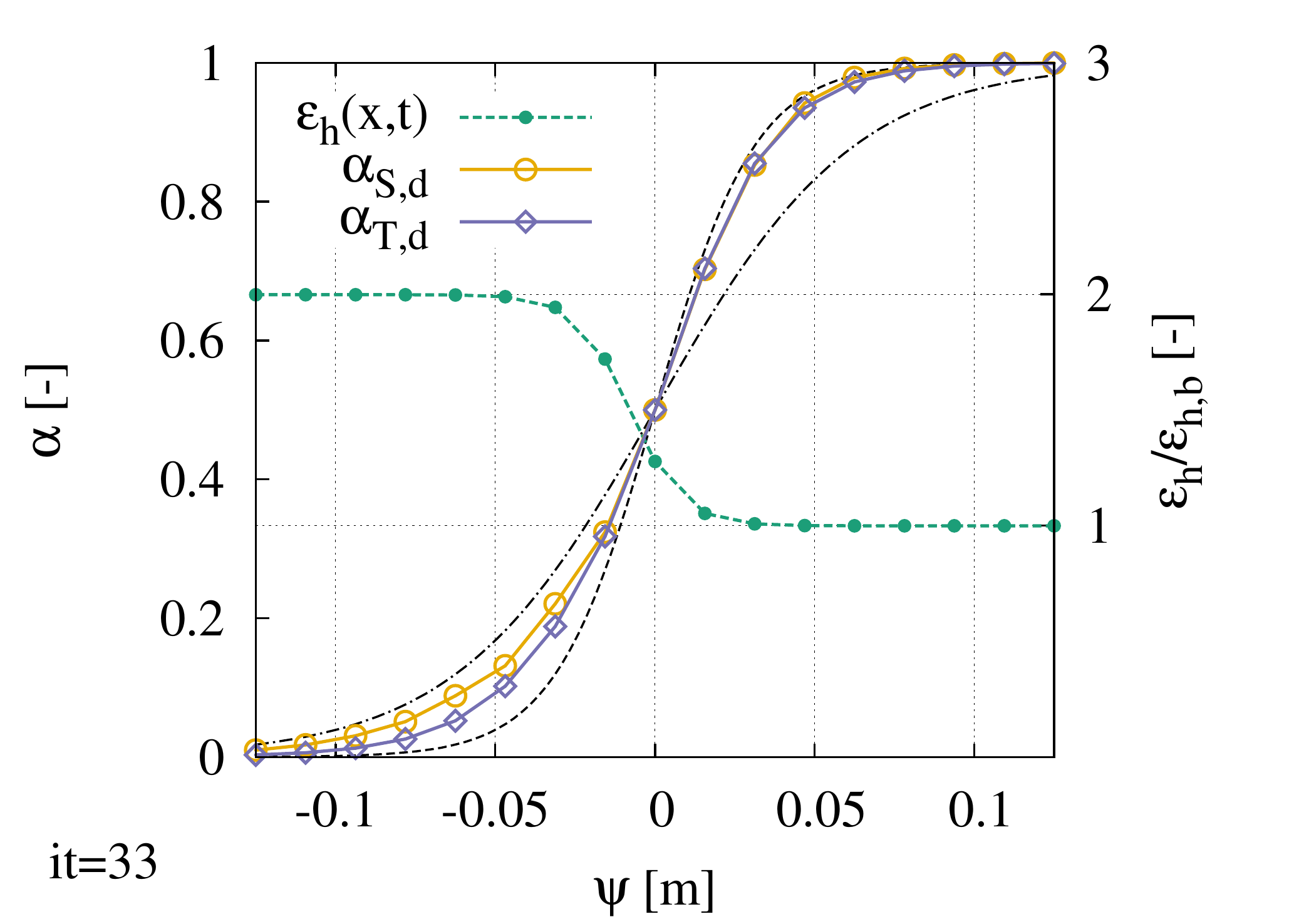}
	\end{minipage}
	\begin{minipage}{.5\textwidth}
		\includegraphics[width=.75\textwidth,height=1.\textwidth,angle=-90]{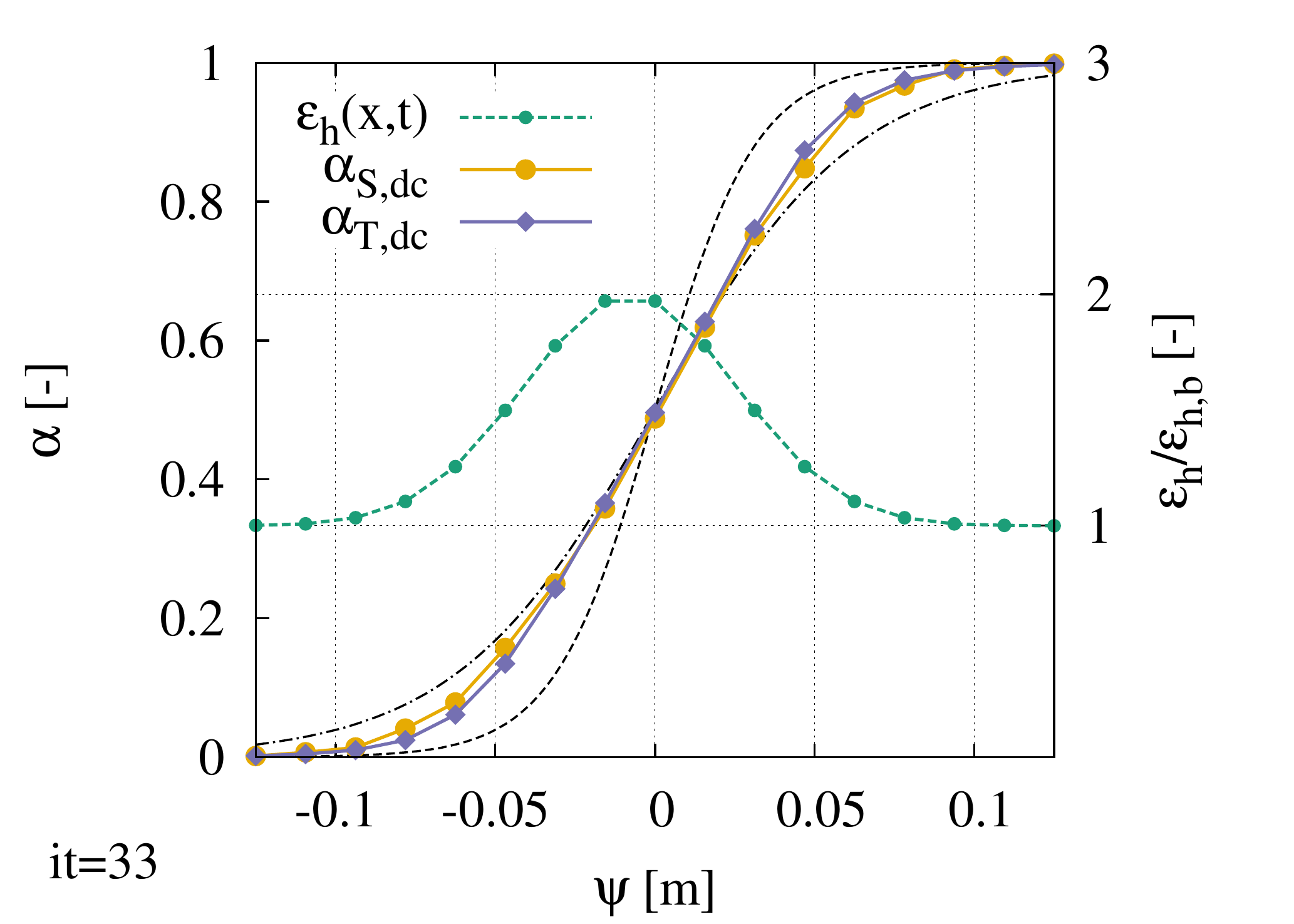}
	\end{minipage}
	\begin{minipage}{.5\textwidth}
		\includegraphics[width=.75\textwidth,height=1.\textwidth,angle=-90]{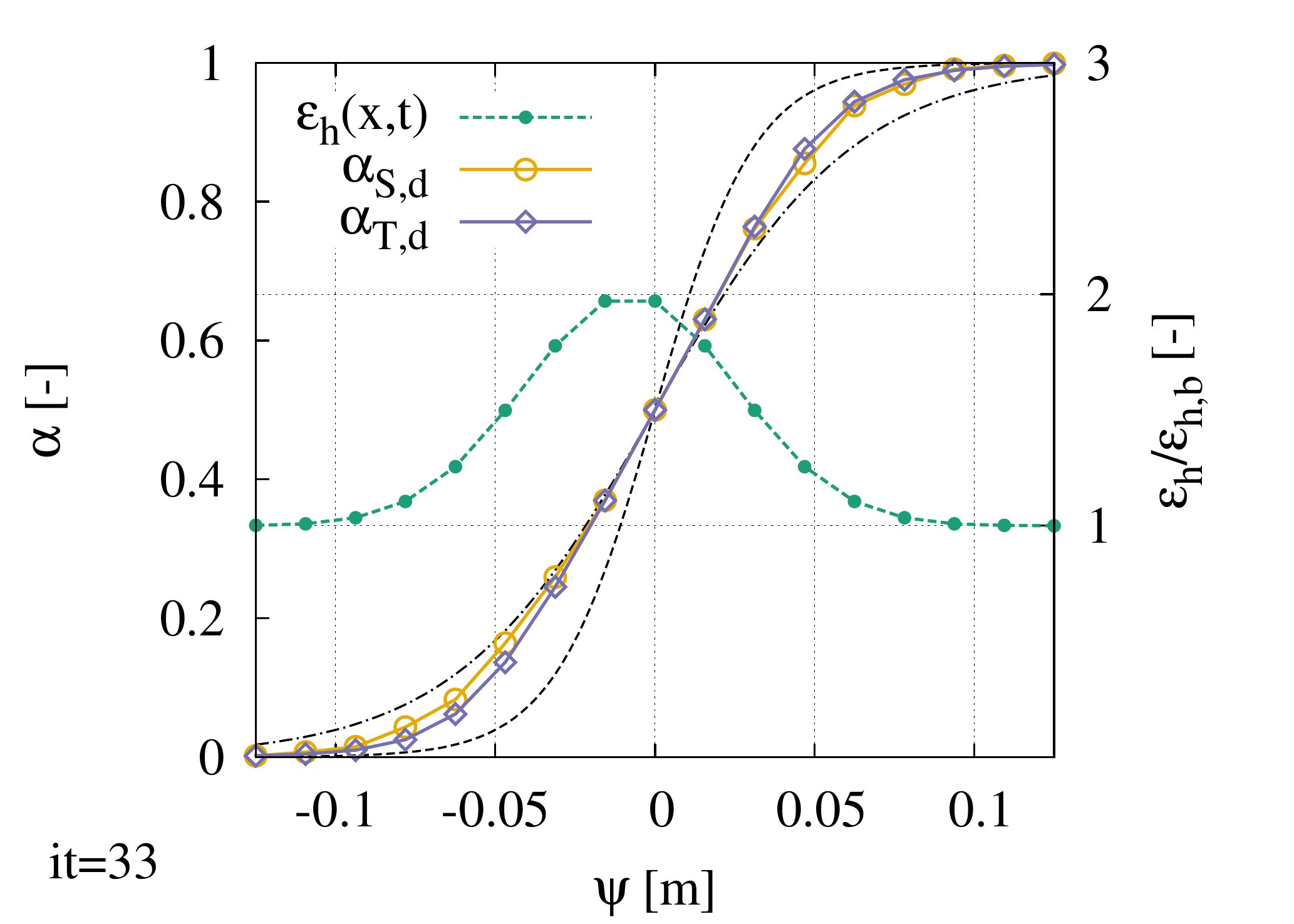}
	\end{minipage}
	\caption{\small{Comparison of 
			$\alpp$ profiles
			affected by 
			the step (top row) 
			or bell (bottom row) shaped 
			variation of the $\ephxt$ field.
			Results are
			predicted using the trapezoidal ($\alpha_T$) 
			or Simpson rules ($\alpha_S$) in the case
			of direct  ($\alpha_{dc}$)
			or semi-analytical ($\alpha_{d}$) solutions
			at two selected time moments $it \!\cdot\! \Delta t$ using the grid $m_1$.
			Black dashed lines mark the analytical solution
			given by \Eq{eq13} where $\eph\!=\!\ephb$ or $\eph\!=\!2\ephb$,
			respectively.
		       }}
	\label{fig7}
\end{figure}
\elnm

The 
differences
between
$\alpha_{dc}$
and $\alpha_d$
profiles observed
in \Figs{fig4}{fig5}
are the consequence
of differences
in $\psi$ fields
obtained 
during
the coupled
and semi-analytical
solutions.
During 
the direct
coupling, 
$\psia$
is 
part of
the numerical solution
and hence 
the expected position
of the regularized 
interface $\psi \!=\!0$
can change its location.
In the semi-analytical 
case, $\eph \!=\! \ephb \!=\!const.$,
for this reason
the position $\psi \!=\! 0$
is not affected
by variations
of the characteristic
length scale field
$\ephxt$.
The latter
approach
simplifies
numerical
solution of 
\Eq{eq18},
but still
the semi-analytical
solution
$\alpha_d$
closely mimics
the exact one 
$\alpha_{dc}$,
compare
results 
in \Figs{fig4}{fig5}.

\Fig{fig7}
illustrates
how 
the order of
accuracy 
of the quadrature
used in \Eq{eq26}
affects 
the numerical
results
obtained on
the grid $m_1$.
Therein
it can be
observed
the quadrature 
selection
has 
an notable
but small
impact
on the 
obtained 
results.
Differences 
between fully
coupled solutions
reconstructed using
the first-order
trapezoidal 
$\alpha_{T,dc}$
or third-order
accurate
Simpson
$\alpha_{S,dc}$
rules (solid symbols)
are almost
the same as
differences
between 
analogical
semi-analytical
solutions:
$\alpha_{T,d}$ 
and $\alpha_{S,d}$
(hollow symbols),
compare 
results
in \Fig{fig7}.
The deviations
between 
the 
results
obtained
using
first- or third-order
accurate quadrature
 are
most pronounced
in the regions
where the slope change
of the $\ephxt$ profile
is significant.
This is expected
in view of
the definition
of the order of
accuracy 
of the both integration
rules.

The impact 
of $\ephxt$
on $\alpp$ 
studied in
\Figs{fig4}{fig5}
can be summarized
as follows.
As it is anticipated,
variations
of $\ephxt$
affect
the shape of 
the cumulative distribution
function $\alpp$.
In the case
of asymmetric,
step shaped
$\ephS$ 
this ultimately
leads to
the increase
of the
width of 
the intermittency
region, 
see \Fig{fig4}.
One observes
the 
$\alpp$
profile
approaches
the equilibrium,
analytical solution
given by   
\Eq{eq13} with $\eph\!=\!2\ephb$,
see \Fig{fig4} $it\!=\!45$. 
We note that
in the 
present numerical 
procedure the width
of the intermittency region
is constrained 
by $\ephb$, hence, 
the variation of $\eph$ 
can not result
in the intermittency
region being thinner 
than $\ephb$.
In \Fig{fig4},
the approximate
solution $\alpha_d$
is reacting slightly  
faster on $\ephS$
variation
than  $\alpha_{dc}$.
However,
$\alpha_d$ follows 
the direct solution $\alpha_{dc}$
very closely.
The 
solutions
$\alpha_{dc}$, $\alpha_d$
are bounded by 
the analytical $\alpp$ profiles
(black-dashed lines) with
the extreme
values of
$\eph\!=\!\ephb$ and $\eph\!=\!2\ephb$. 

The same
conclusions
can be drawn
from
the results
presented
in \Fig{fig5}.
During
the non-symmetric
changes
of $\alpp$ 
both,
the exact and
approximate
solutions display
similar behavior,
initially
resulting 
in
the increase
and then
decrease of 
the intermittency
region
width.
After the peak
of  $\ephxtB$ 
passes
the expected position
of the interface at $\psi\!=\!0$ 
the analytical
$\alpp$ profile with
$\eph\!=\!\ephb$ is
recovered, see \Fig{fig5}, $it\!=\!45$.
It is noted,
the return 
of $\alpp$
to the equilibrium
state is guaranteed
by design
of the mapping
function \Eq{eq27}
and quadrature 
in \Eq{eq29}. 

\begin{figure}[!ht] \nonumber
	\begin{minipage}{.5\textwidth}
		\subfloat[]{\includegraphics[width=.7\textwidth,height=1.\textwidth,angle=-90]{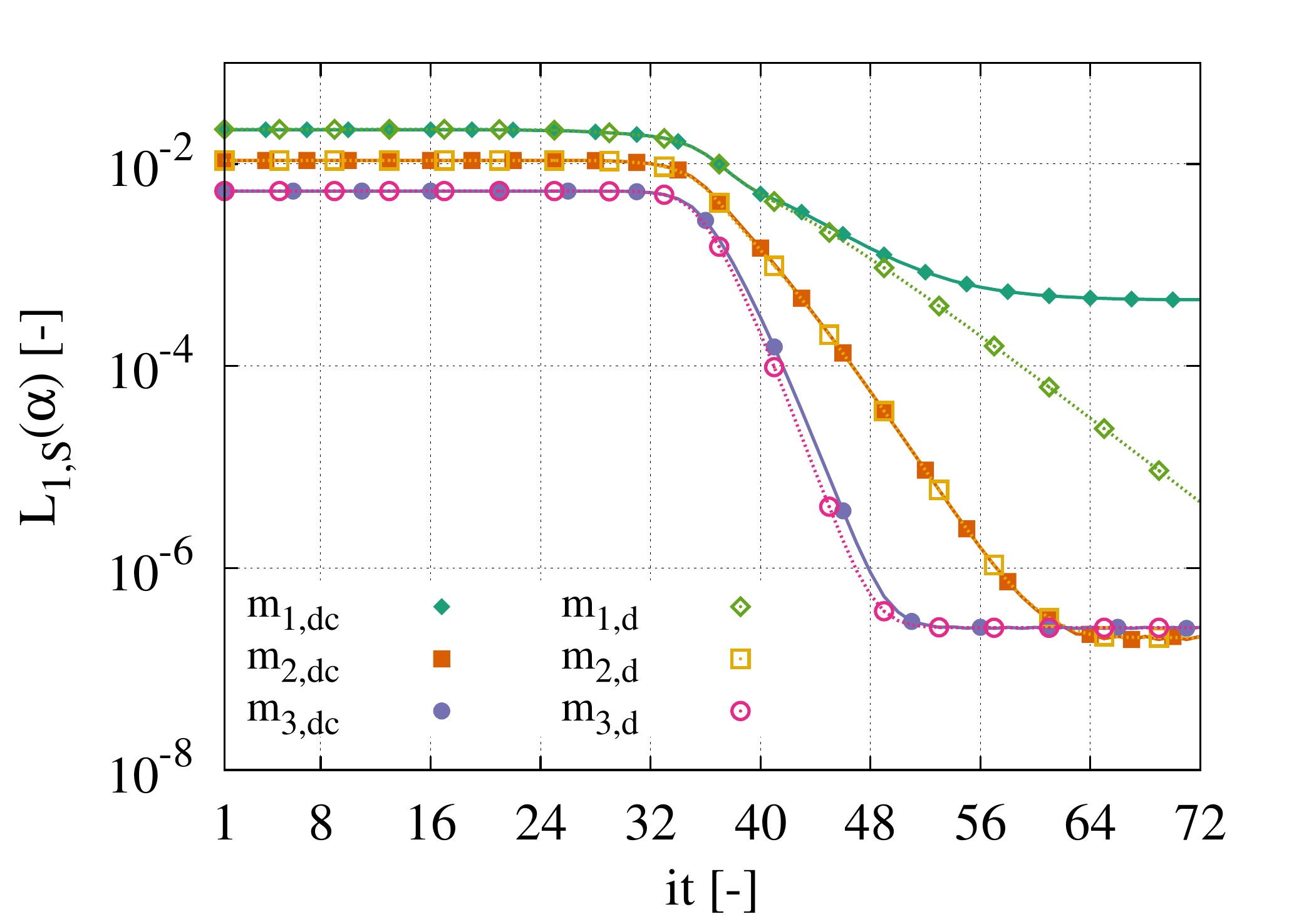}}
	\end{minipage}
	\begin{minipage}{.5\textwidth}
		\subfloat[]{\includegraphics[width=.7\textwidth,height=1.\textwidth,angle=-90]{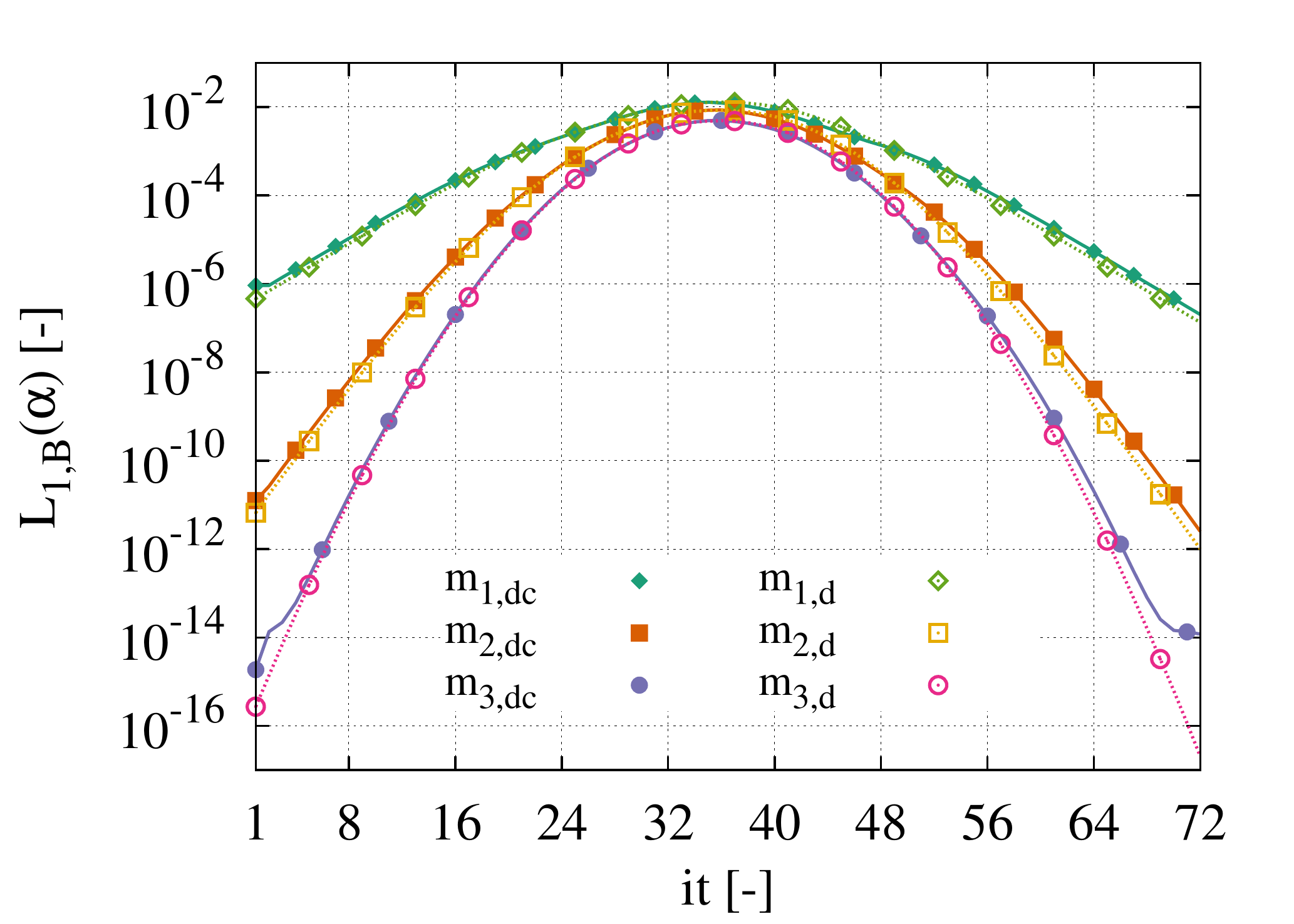}}
	\end{minipage}	
	\begin{minipage}{.5\textwidth}
		\subfloat[]{\includegraphics[width=.7\textwidth,height=1.\textwidth,angle=-90]{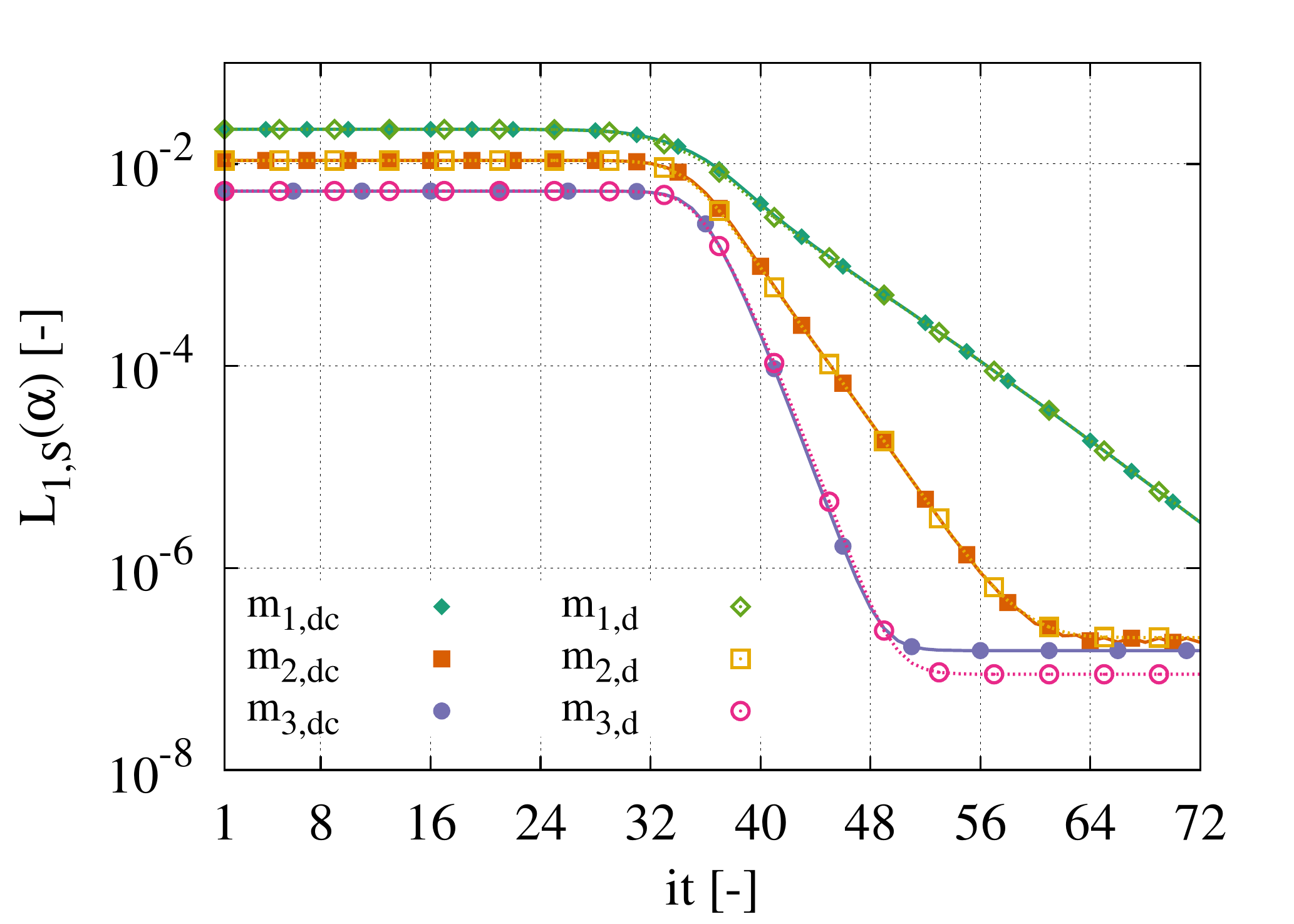}}
	\end{minipage}
	\begin{minipage}{.5\textwidth}
		\subfloat[]{\includegraphics[width=.7\textwidth,height=1.\textwidth,angle=-90]{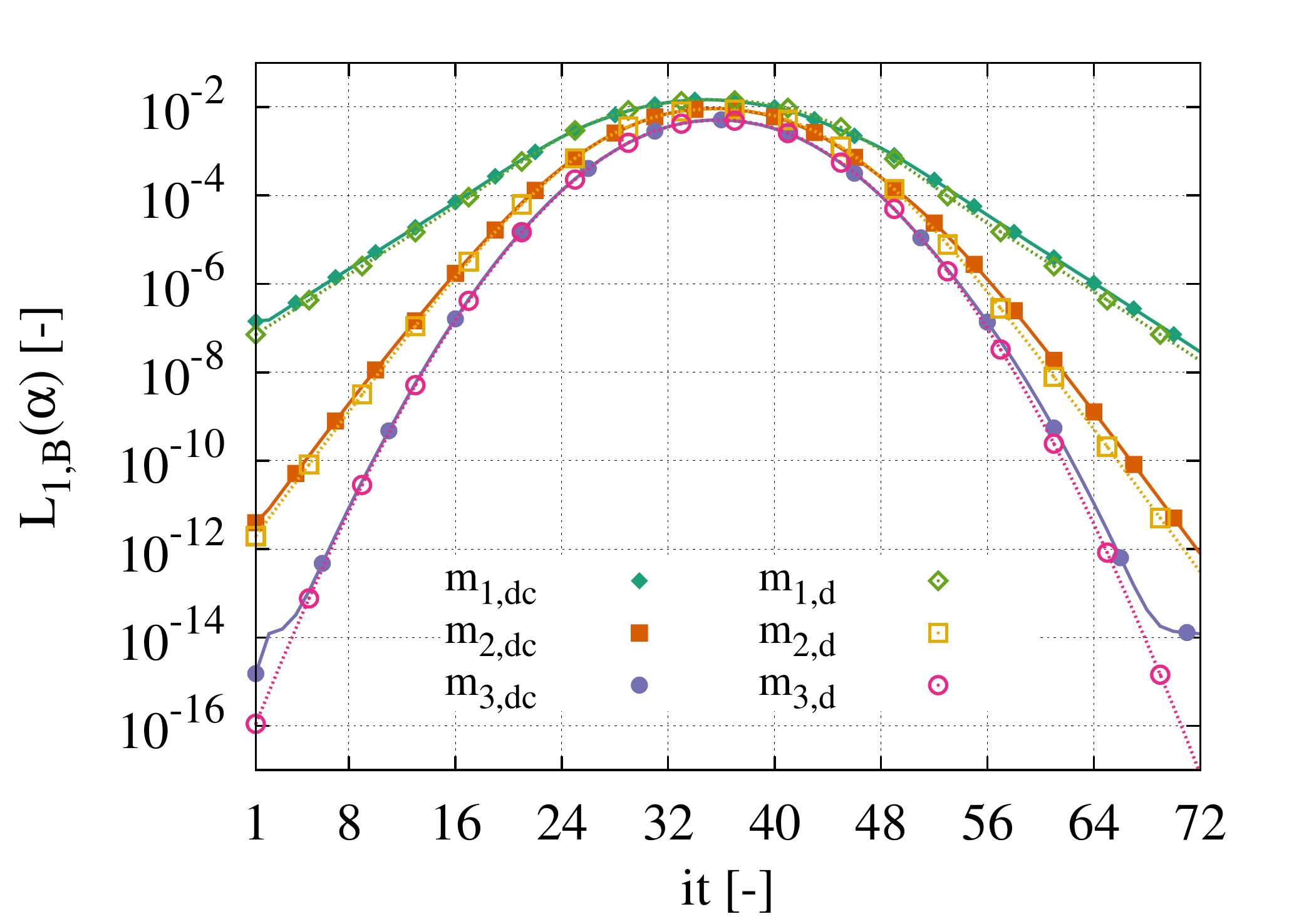}}
	\end{minipage}
	\caption{\small{Temporal  evolution 
			of the $L_{1,S} \lr \alpha \rr$ and
			$L_{1,B} \lr \alpha \rr$ norms
			defined by \Eq{eqE2}
			in physical time $it \!\cdot\! \Delta t$.
			The error of the coupled solution (dc) (\Eqs{eq18}{eq27}) 
			is marked with the solid symbols, 
			the error of the semi-analytical solution (d) 
			(\Eq{eq28})  is marked with 
			the hollow symbols;
			$\eph\!=\!\ephS$ (left column) 
			or $\eph\!=\!\ephB$ (right column).
			The integral 
			in \Eq{eq29} is approximated 
			using the trapezoidal (top row) 
			or Simpson (bottom row) rules, 
			respectively.
	}}
	\label{nfig10}
\end{figure}

The history
of  
convergence  
of the numerical
error 
on the three
gradually 
refined grids
$m_i,\,i\!=\!1,2,3$ 
is presented
in \Fig{nfig10}.
It shows 
$L_{1,S}\lr \alpha \rr$
and $L_{1,B}\lr \alpha \rr$ norms
defined by \Eq{eqE2}
plotted after
each time iteration
$it\!=\!1,\ldots,72$.
We note that
the differences
in the numerical
error
of the semi-analytical
solutions (hollow symbols
in \Fig{nfig10}(a,c)
and  \Fig{nfig10}(b,d))
are only caused  
by the errors
of the quadrature
used to 
approximate
\Eq{eq26}.
\blnm
\elnm 

Convergence of
the solution 
is obtained
on each grid $m_i,\, i\!=\!1,2,3$
what 
is confirmed
by the results
depicted in \Figs{m1fig6}{m3fig6}.
When 
the Simpson rule
is used to approximate
the integral in \Eq{eq26}
the level
of the numerical errors
is  lower
in comparison
with 
the trapezoidal
rule,
compare the
results 
in the bottom and
top row 
in \Fig{nfig10}.
The errors
of the 
coupled $L_{1,S}^{dc}$, $L_{1,B}^{dc}$   (solid symbols)
and semi-analytical $L_{1,S}^{d}$, $L_{1,B}^{d}$ (hollow symbols)
solutions 
display 
similar behavior;
the error of
the coupled solution
is always higher than 
the error of
the respective
semi-analytical
solution, 
see \Fig{nfig10}(a,b,c,d).
One notes
that in the case
$\eph\!=\!\ephS$
when 
the trapezoidal
quadrature is 
used on the
grid $m_1$ (see \Fig{nfig10}(a)) 
the coupled and
semi-analytical
solutions are
different opposite
to the case
when the Simpson
rule is employed,
compare $m_{1,dc}$, $m_{1,d}$
in \Fig{nfig10}(a,c). 
In the case
$\eph\!=\!\ephB$ 
the differences
between $L_{1,B}^{dc}$ (solid symbols)
and $L_{1,B}^{d}$ (hollow symbols)
can be spotted
only on the grid $m_3$
at
the beginning
$it \!<\! 4$
and at the end
of simulation $it \!>\! 68$,
see $m_{3,dc}$, $m_{3,d}$ in \Fig{nfig10}(b,d).
\blnm
\begin{figure}[!ht] \nonumber
	\begin{minipage}{.5\textwidth}
		\subfloat[]{\centering\includegraphics[width=0.9\textwidth,height=0.95\textwidth,angle=-90]{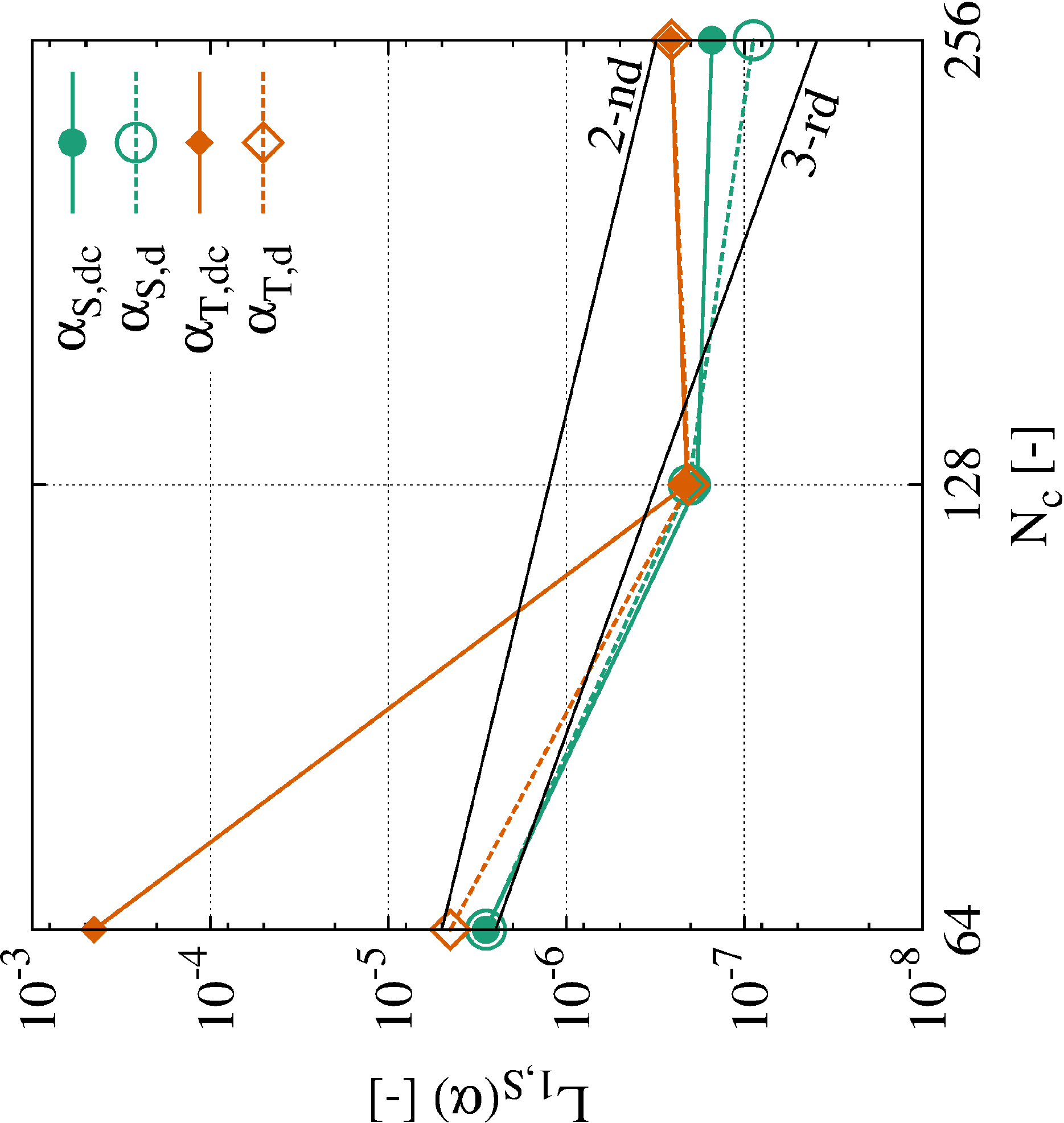}}
	\end{minipage}
	\begin{minipage}{.5\textwidth}
		\subfloat[]{\centering\includegraphics[width=0.9\textwidth,height=0.95\textwidth,angle=-90]{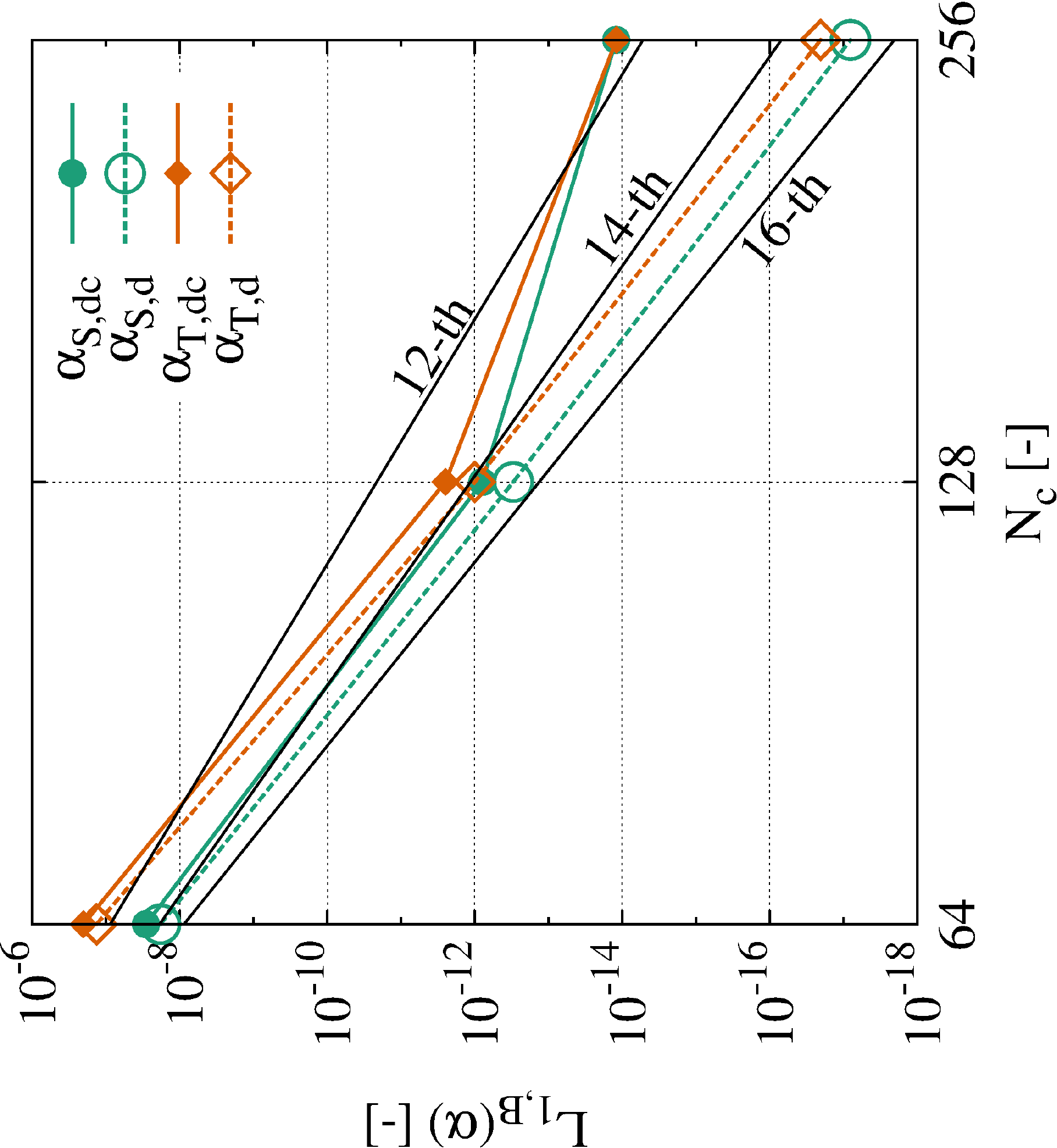}}
	\end{minipage}
	\caption{\small{Spatial convergence rates of 
			the numerical errors $L_{1,S}, L_{1,B}$ defined by \Eq{eqE2}
			at $it=72$, see \Fig{nfig10}; (a) $\eph\!=\!\ephS$,  (b) $\eph\!=\!\ephB$.
			Figure shows the convergence rate of the coupled $\alpha_{dc}$ (solid symbols) 
			and semi-analytical $\alpha_d$ (hollow symbols) solution.
			The intergral in \Eq{eq26} is discretized
			using the trapezoidal ($T$) or Simpson ($S$) rules.
			}}
	\label{nfig11}
\end{figure}
\elnm 

To stabilize
the numerical
solution of \Eqs{eq18}{eq27}
on the grid $m_1$ 
when
$\eph\!=\!\ephS$
it was necessary
to reduce the 
time step size
$\Delta \tau$
and
increase 
the number 
of re-initialization
steps $N_\tau$, see 
results 
in \Fig{m1fig6}(a,c).
In 
this case 
$N_\tau \!=\! 512$ and $\Delta \tau \!=\! 0.25 \Delta x$,
for both integration rules
used.

Some instabilities
in convergence
of the numerical 
solutions
on the grid $m_1$, $m_2$ 
can be seen 
in \Fig{m1fig6}(a,c) and \Fig{m2fig6}(c).
However,
they are 
present only
in the case $\eph\!=\!\ephS$
and vanish
with increasing 
spatial resolution,
compare results
in Figs. \ref{m2fig6}(a,c)-\ref{m3fig6}(a,c).
It is 
emphasized 
that the 
algorithm
employed in the
present work
to reconstruct
solution of
\Eqs{eq18}{eq27}
does not use 
flux limiters 
or non-oscillatory
schemes in order
to stabilize
convergence
of the numerical
solution.

On the 
grid $m_1$
when
$\eph\!=\!\ephS$,
the first- or third-order
accuracy 
is not sufficient
to obtain the 
truncation 
error level
after all
physical time
steps $it$
after predefined
$N_\tau\!=\!512$
re-initialization
steps, see \Fig{m1fig6}(a,c).
For this reason,
the norm $L_{1,S}\lr \alpha \rr$
(see $m_1$ results in \Fig{nfig10}(a,c))
does not converge 
to the stationary 
state, too.
In particular,
this is
visible in 
the semi-analytical
solution, see hollow symbols
in \Fig{nfig10}(a,c). 
This might 
be caused
by not fully
converged computations
at the time iterations
 $1 \!<\! it \!<\! 16$  (see \Fig{m1fig6}(a))
and $8 \!<\! it \!<\! 32$ (see \Fig{m1fig6}(c)) .
We note similar 
behavior of the numerical
error is observed
when grid $m_2$ is used,
see \Fig{m2fig6}(c) $12 \!<\! it \!<\! 24$.
However
therein,
the 
error level 
is approximately
five orders 
of the magnitude
lower than in
\Fig{m1fig6}(a,c)
and the convergence
to the stationary 
state is achieved
with both integration
rules,
see  results $m_{2,dc}$, $m_{2,d}$
in \Fig{nfig10}(a,c).

Interestingly,
in the case of 
$\eph\!=\!\ephB$
the problem described
above does 
not show up,
see 
the right 
column 
in \Fig{nfig10}.
Both quadratures
in the coupled (solid symbols)
and semi-analytical (hollow symbols)
solutions
predict similar
evolution 
of the numerical 
error $L_{1,B} \lr \alpha \rr$,
see \Fig{nfig10}(b,d).
A cursory explanation
for this result
is that
in the present study
spatial dimensions
of 
$\ephxt$ are constant
on all used 
grids $m_i$
(count the number of grid points
 in $\ephxt$ profile
 in \Fig{fig7}, $it\!=\!33$ and
 in  \Fig{fig4}, $it\!=\!33$).
Thus on 
the grid 
$m_1$
the step profile $\ephxtS$
is represented
by four grid
points 
and
this impairs
solution
of \Eq{eq27}.
This reckoning
is confirmed by
the convergence 
rates of 
the spatial
discretization
error presented
in \Fig{nfig11}.
Therein,
the 
$L_{1,S} \lr \alpha \rr$
and
$L_{1,B} \lr \alpha \rr$ 
norms
depicted
in \Fig{nfig10}
are 
plotted at 
the time moment
$t\!=\!72 \!\cdot\! \Delta t$.
One notes
that
the spatial convergence rate
is strongly affected 
by the choice
of the
shape of
$\ephxt$
profile
and
the order
of accuracy 
of the
quadrature
has little 
impact on
the 
$L_{1,S}\lr \alpha \rr$, 
$L_{1,B}\lr \alpha \rr$
norms
convergence
rates, 
compare solid
and hollow
symbols 
in \Fig{nfig11}(a,b).

\Fig{nfig11}(a)
shows
that
in the case $\eph\!=\!\ephS$
the  convergence rate
of the coupled (dc) and
semi-analytical (d)
solution 
is second-order
accurate for 
both 
quadratures
used.
As the semi-analytical
solution introduces 
lower discretization
error the level
of $L_{1,S}$ norm for
$\alpha_{S,d}$, $\alpha_{T,d}$
is lower than
in the coupled case,
see \Fig{nfig11}(a).
When $\alpp$
is affected by
$\ephB$ the
high-order of
the convergence rate
of all numerical
solutions 
is obtained.
The first-,
and third-order
accurate
quadrature  
reconstruct
the numerical solution
with the same
convergence rate.
The error level
of the coupled
solution is higher
than the error level
of the semi-analytical
solution, see \Fig{nfig11}(b).

At first,
high-orders 
of the convergence
rate presented
in \Fig{nfig11}
appear to be
unexpected.
However, 
in the present study
$\ephb \!=\! \Delta x_i$, $i\!=\!1,2,3$
this guarantees
the resolution 
in the one-dimensional case
is four times higher
than in the previous
papers of the present
author \citep{twacl15,twacl17}
where the convergence
of \Eq{eq18} was
investigated.
Moreover, 
we note 
herein
solution 
of \Eq{eq18} 
in time $\tau$
is carried out
until 
the level 
of the truncation
error is achieved in
the double precision
computations,
see \Figs{m1fig6}{m3fig6}.
In aforementioned works
the tests with advection
were carried out using
$N_\tau\!=\!4$  (\cite{twacl17} see Fig.~7,~9)
or $1 <N_\tau < 16$ (\cite{twacl15} see Fig.~31)
re-initialization
steps.
Yet other argument
supporting results
in \Fig{nfig11}(b) 
is 
the present 
method is based on
the semi-analytical solution
of the set of partial differential
algebraic equations (\ref{eq17}, \ref{eq18}, \ref{eq27}),
see \Sec{ssec41} and \ref{appD}. 
This is the main 
difference between
the present semi-numerical 
approach
and purely numerical
techniques 
(VOF, SLS methods)
known 
in the 
literature.
The results
presented
in \Figs{fig4}{fig5}
and convergence
study presented
in \Figs{nfig10}{nfig11}
confirm 
the
semi-analytical
solution
$\alpha_d$
is providing 
close 
estimation 
of  
the coupled
problem
$\alpha_{dc}$. 
In 
the next section
this
solution 
is used 
to model
variation 
of the 
intermittency
region
around
circular drop
without and with
advection.
As the 
third-order
accurate
Simpson quadrature (\ref{eq31})
is more sensitive
to variations
of $\ephxt$
it is used
in  all  the
following
numerical 
tests.

\subsection{Two-dimensional semi-analytical solution}
\label{ssec43}
\begin{figure}[!hbt] \nonumber
\begin{minipage}{.94\textwidth}
	\begin{minipage}{.33\textwidth}
		\centering\includegraphics[width=.95\textwidth,height=.95\textwidth,angle=0]{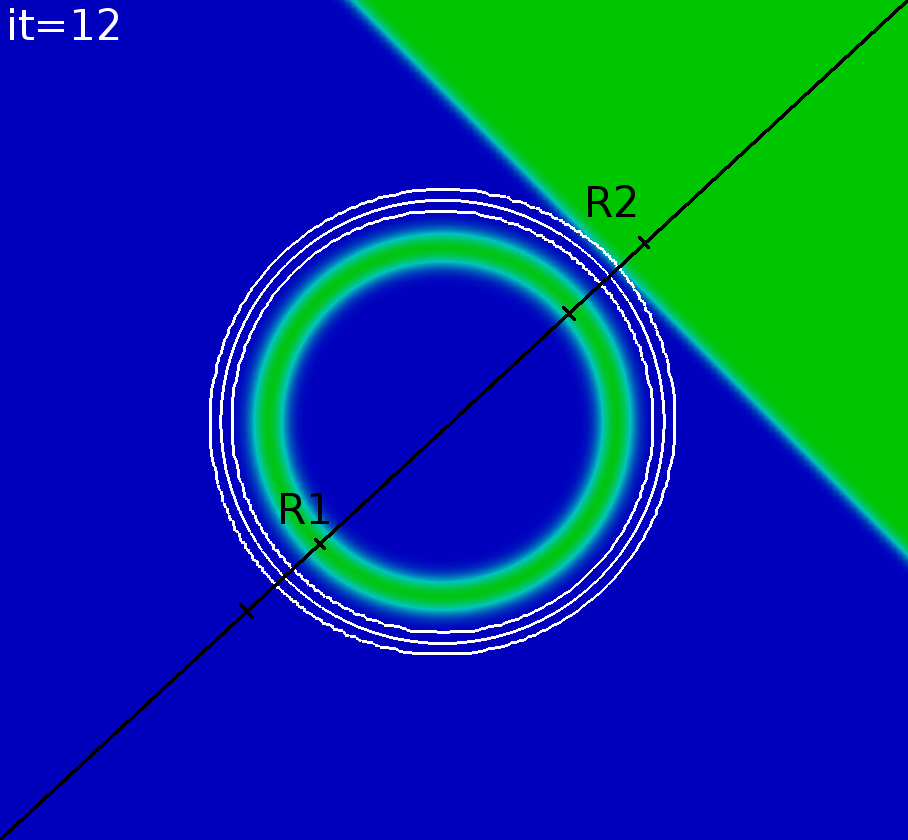}
	\end{minipage}%
	\begin{minipage}{.33\textwidth}
		\centering\includegraphics[width=.95\textwidth,height=.95\textwidth,angle=0]{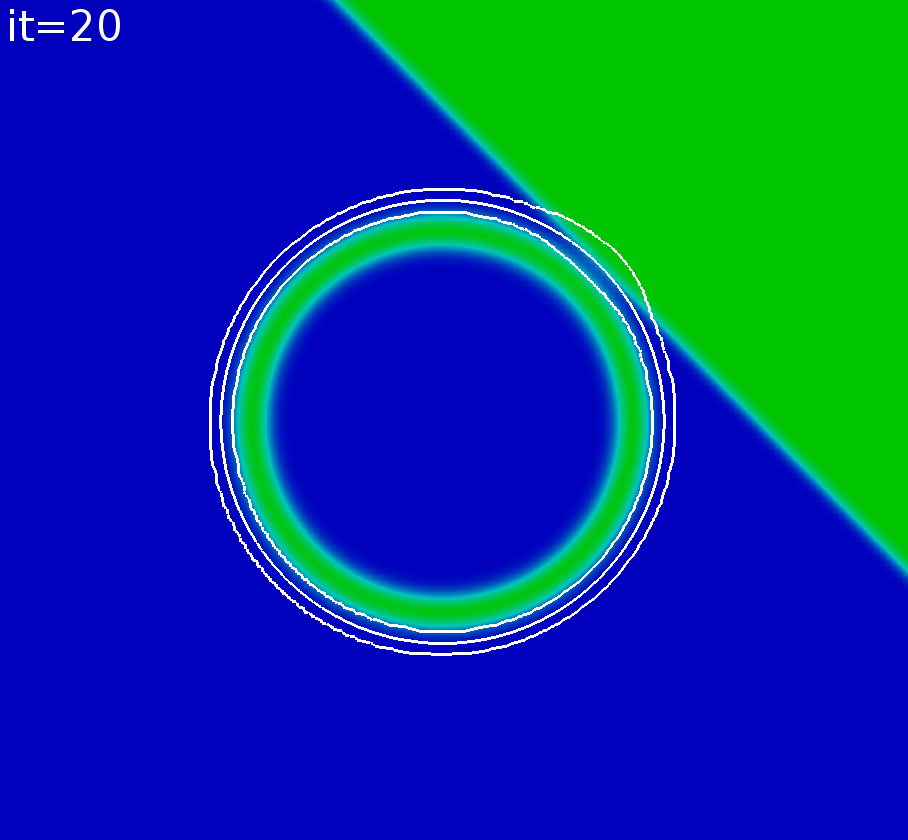}
	\end{minipage}%
	\begin{minipage}{.33\textwidth}
		\centering\includegraphics[width=.95\textwidth,height=.95\textwidth,angle=0]{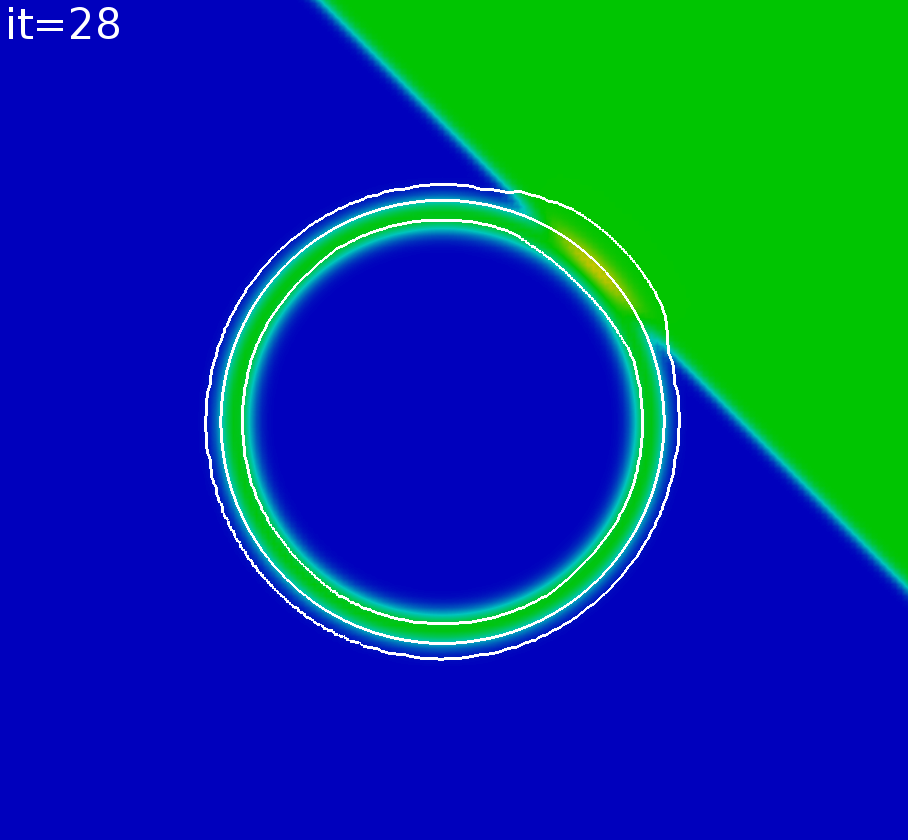}
	\end{minipage}      
	\begin{minipage}{.33\textwidth}
		\centering\includegraphics[width=.95\textwidth,height=.95\textwidth,angle=0]{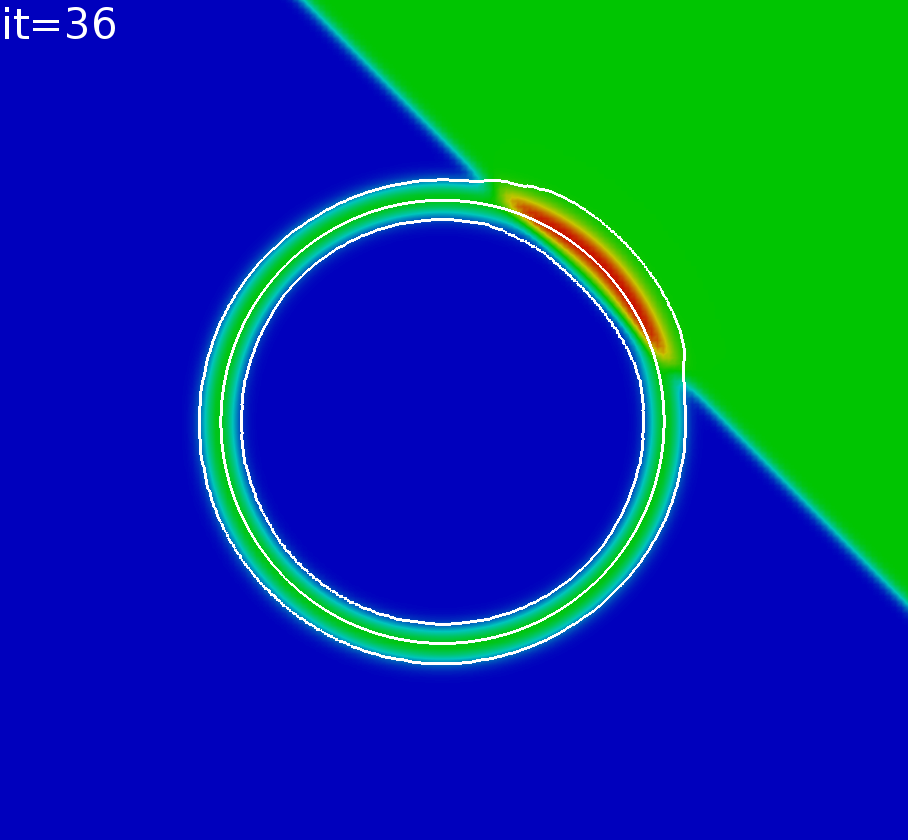}
	\end{minipage}%
	\begin{minipage}{.33\textwidth}
		\centering\includegraphics[width=.95\textwidth,height=.95\textwidth,angle=0]{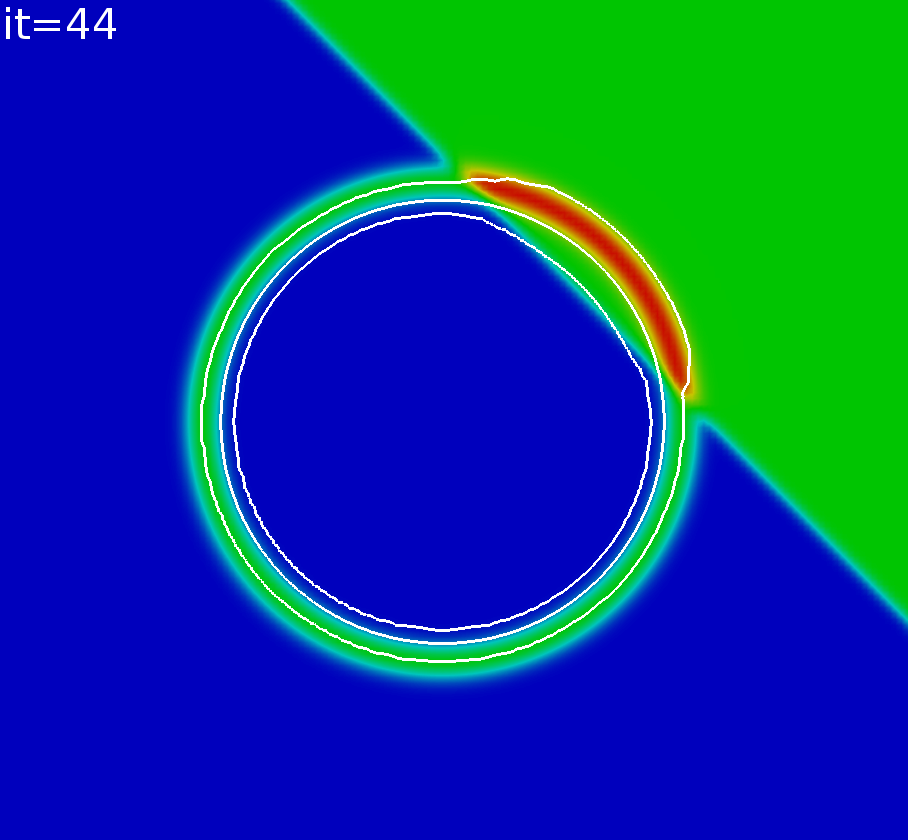}
	\end{minipage}%
	\begin{minipage}{.33\textwidth}
		\centering\includegraphics[width=.95\textwidth,height=.95\textwidth,angle=0]{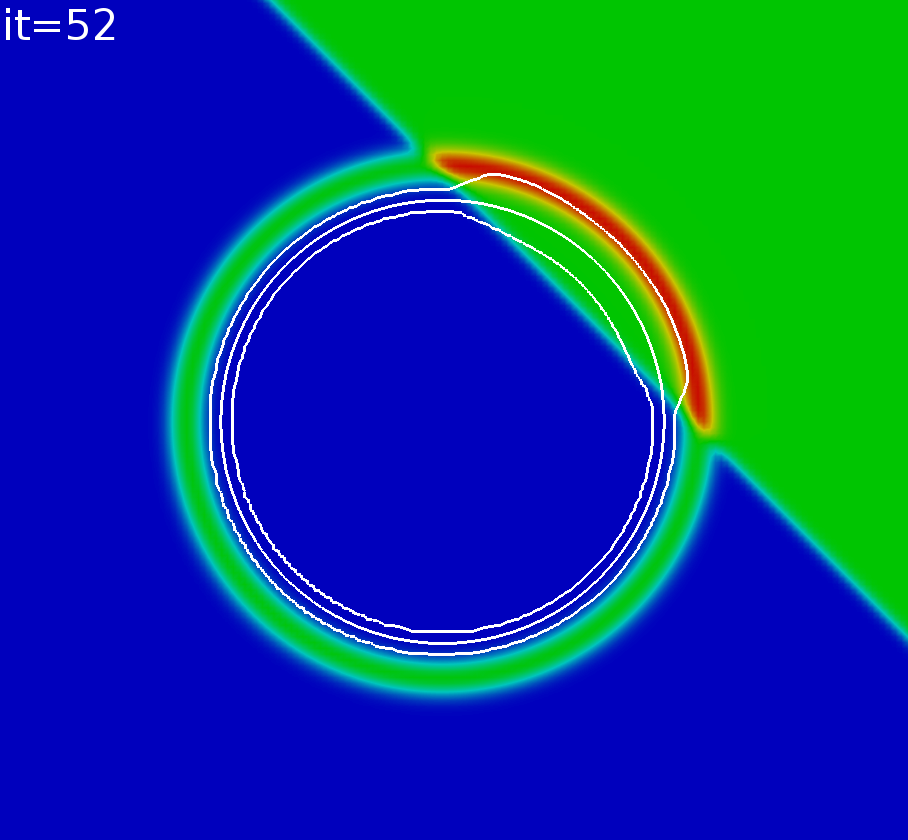}
	\end{minipage}
\end{minipage}
\hspace*{-0.1575cm}
\begin{minipage}{.0575\textwidth}
\includegraphics[width=1.\textwidth,height=8.\textwidth,angle=0]{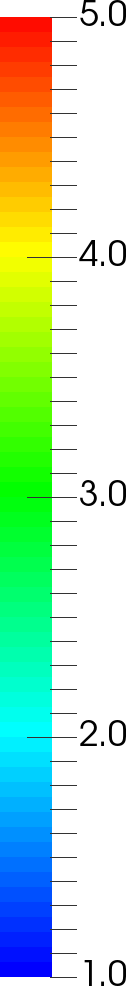}
\end{minipage}
    
	\caption{\small{Evolution of the intermittency region affected by
	 	        the variable, defined by \Eq{eqC3} field of 
	 	        $\epsilon_{h,b} \!\le\! \ephxt \!\le\! 5\epsilon_{h,b}$
		        (colors);
			all figures show contours $\alpha(\psi \!=\! -4\ephb),\,
			\alpha(\psi \!=\! 0),\,\alpha(\psi \!=\!4\ephb)$.
	}}
	\label{fig8}
\end{figure}

%
To assess
how the
numerical method
introduced in 
\Sec{ssec41}
works in
the two-dimensional
case, 
the resting, 
circular drop
centered at 
the point $\lr 0.5,0.5 \rr$
with radius $R_B\!=\!0.15\,[m]$
surrounded by
the intermittency 
region  and disturbed
by the $\ephB\!\le\!\ephxt \!\le\! 5\ephB$
field
is studied,
see \Fig{fig8}.
\blnm
\elnm 
In this test,
$\ephxt$ is
evolving
according
to \Eq{eqC3}.
The
problem
is solved 
in a two-dimensional
unit square 
box $[0,1] \!\times\! [0,1]$
discretized with
$2^8 \!\times\! 2^8$
control volumes;
the base width
of the intermittency
region is set 
to $\ephb \!=\! \sqrt{2}\Delta x/4\,[m]$,
the physical 
time step size
is $\Delta t \!=\!10^{-3}\,[s]$,
fictitious 
time step size
$\Delta \tau \!=\!\ephb/2\,[s]$,
four re-initialization 
steps $N_\tau\!=\!4$ 
per $\Delta t$
are used.
Only \Eq{eq18}
with $\ephb\!=\!const.$
and \Eq{eq28} taking
into account
$\ephxt$ are
solved as
$\bw\!=\!\bu\!=\!0$.
\blnm
\begin{figure}[!ht] \nonumber
	\begin{minipage}{.5\textwidth}
		\subfloat[]{\includegraphics[width=1.\textwidth,height=.75\textwidth,angle=0]{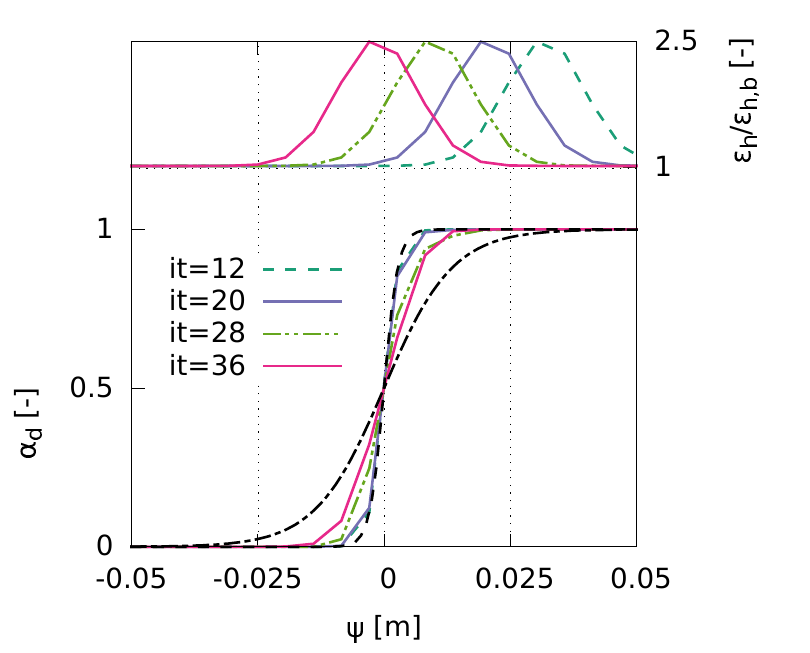}}
	\end{minipage}
	\begin{minipage}{.5\textwidth}
		\subfloat[]{\includegraphics[width=1.\textwidth,height=.75\textwidth,angle=0]{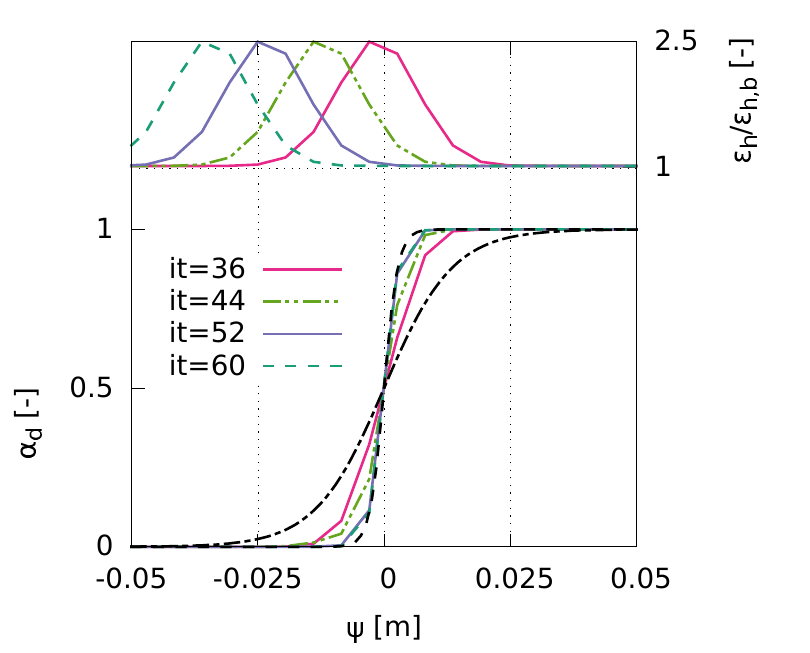}}
	\end{minipage}
	\begin{minipage}{.5\textwidth}
	    \subfloat[]{\includegraphics[width=1.\textwidth,height=.75\textwidth,angle=0]{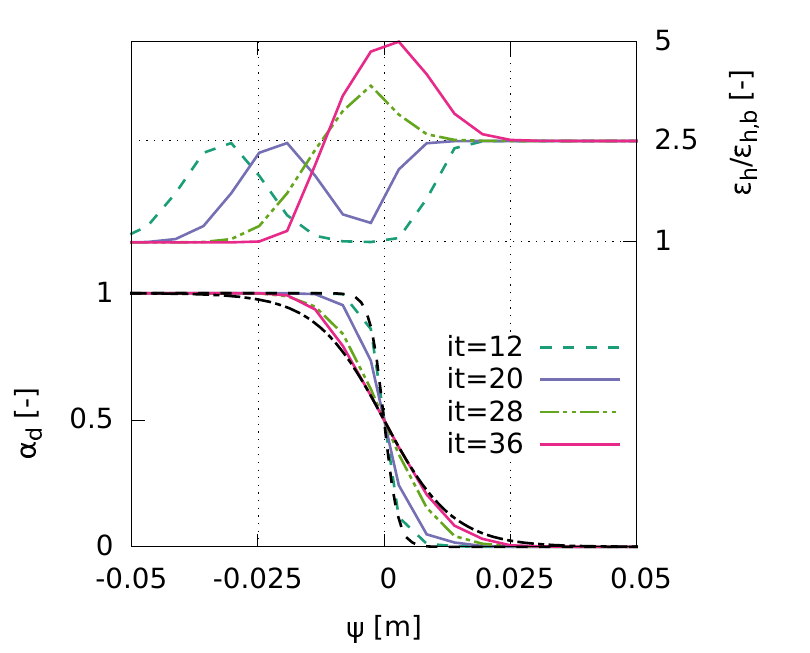}}
\end{minipage}
\begin{minipage}{.5\textwidth}
	   \subfloat[]{\includegraphics[width=1.\textwidth,height=.75\textwidth,angle=0]{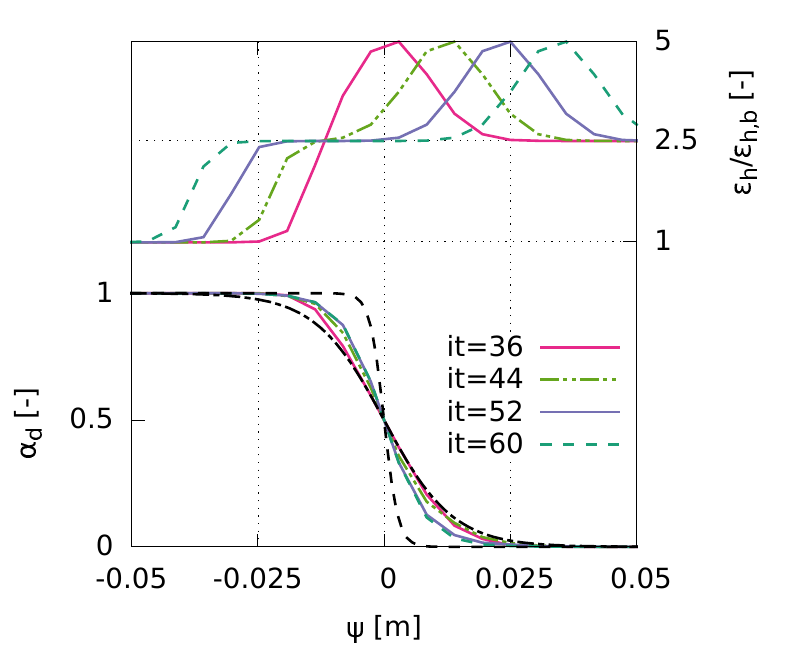}}
\end{minipage}
	\caption{\small{Comparison of $\alpp$ profiles 
			affected by the $\ephxt$ field (see \Fig{fig8})
			with the analytical solutions.
			Diagrams present
			the $\ephxt$ and
			$\alpha_{S,d}$ profiles drawn
			along the region $R1$ (a),(b) and $R2$ (c),(d) 
			of the diagonal (see \Fig{fig8}, $it\!=\!12$) 
			at seven time moments $it$.
			Black lines mark the analytical
			profiles of $\alpp$ where $\eph\!=\!\ephb$ (dashed line) and $\eph\!=\!5\ephb$ 
			(dashed-dotted line) obtained using \Eq{eq13}.
	}}
	\label{fig9}
\end{figure}
\elnm

The evolution
of the intermittency
region due to
variable $\ephxt$
is illustrated 
in \Figs{fig8}{fig9}.
\Fig{fig8}
displays
variation 
of the 
$\ephxt$
field and
its impact
on $\alpp$
illustrated
using contours
$\alpha(\psi \!=\! -4\eph),\,
\alpha(\psi \!=\! 0),\,\alpha(\psi \!=\!4\eph)$.
The variation 
of $\ephxt$
in the subsequent
time moments $it \!\cdot\! \Delta t$
leads
first 
to an increase
and later a decrease
of the intermittency
region width
(similarly to the one-dimensional
predictions in \Fig{fig5}).
Details of
this process 
can be observed
in \Fig{fig9}.
Therein,
$\ephxt$ and $\alpp$
profiles are
drawn
along
the parts
$R1$, $R2$ 
of the diagonal
across
the computational
domain (see \Fig{fig8}, $it\!=\!12$)
at seven different
time moments $it \! \cdot \! \Delta t$.
\Fig{fig9}a,b 
is illustrating variations
in the region $R1$
and
\Fig{fig9}c,d 
in the region $R2$. 

In \Fig{fig9}(a),
the impact of 
$\ephxt$ increasing
in time
on $\alpp$ is
presented.
It can be
observed 
the profile 
of $\alpp$
converges to
analytical
solution given
by \Eq{eq13}
with   $\eph\!=\!2.5\ephb$
(not shown in \Fig{fig9}
for clarity of presentation).
After the bell
shaped disturbance
(moving to the left)
passes 
$\psi\!=\!0$
(near $it\!=\!36$)
the process is
reversed and 
at $it\!=\!60$
the equilibrium
solution with
$\eph\!=\!\ephb$
is reconstructed,
see \Fig{fig9}b.
Variation 
of $\ephxt$
in the region
$R2$ is more 
complex,
see \Fig{fig9}c,d.
Therein,
the step (moving to the left)
and bell (moving to the right) 
shaped
disturbances
interfere,
leading
to
increased
$\alpp$
profile width
close to the one 
obtained when
$\eph\!=\!5\ephb$
in \Eq{eq13}, 
see \Fig{fig9}c, $it\!=\!36$. 
In the subsequent
time moments,
the bell  
and step 
disturbances 
pass $\psi\!=\!0$
and
the $\alpp$ profile
approaches
the equilibrium
solution 
where $\ephb\!=\!2.5\ephb$,
see \Fig{fig9}d $it\!>\!36$.

One notes
that
during 
the evolution
in times $t$, $\tau$
the $\alpp$ profile
remains bounded 
between two
extreme solutions
obtained with
$\eph\!=\!\ephb$ (black dashed line)
and $\eph\!=\!5\ephb$ (black dashed-dotted line),
see \Fig{fig9}.
In the
two- or
three-dimensional 
cases  
the integration
(\ref{eq25})
is carried
out along 
the normal
coordinate
$\psi$ 
in the local
system
attached to 
the each point
of regularized
interface $\gamma$.
Thus,
as in 
the one-dimensional
case,
reduction to 
the equilibrium
solution in 
the points
where $\eph\!=\!const.$
is guaranteed by
the design of the quadrature
(\ref{eq31}).
%

\subsection{Two-dimensional semi-analytical solution with advection}
\label{ssec44}
%
In this
section, 
the
semi-analytical
approach
described
and verified
in \Secs{ssec41}{ssec43}
is used
to reconstruct 
the behavior of
the intermittency 
region surrounding  
a two-dimensional
circular drop with 
the radius $R\!=\!0.15\,[m]$ 
initially located
at the point $(0.5,0.35)$ 
and advected in
the divergence-free,
constant,   
circular
velocity field
$\bu \!=\! (u_1,u_2)=V_0/L \lr y\!-\!0.5,0.5\!-\!x \rr$
where $V_0\!=\!1\,[m/s]$ and $L\!=\!1\,[m]$.
The size of
the computational
domain, number of
control volumes
and settings
of the solver of \Eqs{eq17}{eq18}
are the same 
as described
in \Sec{ssec43}.
\blnm
\begin{figure}[!ht] \nonumber
	\begin{minipage}{.32\textwidth}
		\centering\includegraphics[width=.825\textwidth,height=.1\textwidth,angle=0]{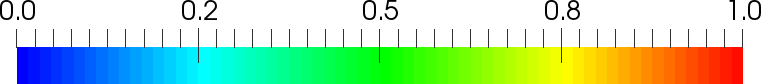}
	\end{minipage}%
	\begin{minipage}{.32\textwidth}
		\centering\includegraphics[width=.825\textwidth,height=.1\textwidth,angle=0]{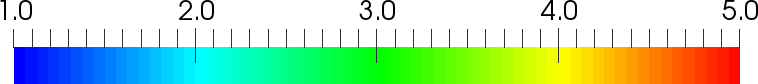}
	\end{minipage}%
	\begin{minipage}{.32\textwidth}
	\end{minipage}
	\begin{minipage}{.32\textwidth}
		\centering\includegraphics[width=.825\textwidth,height=.825\textwidth,angle=0]{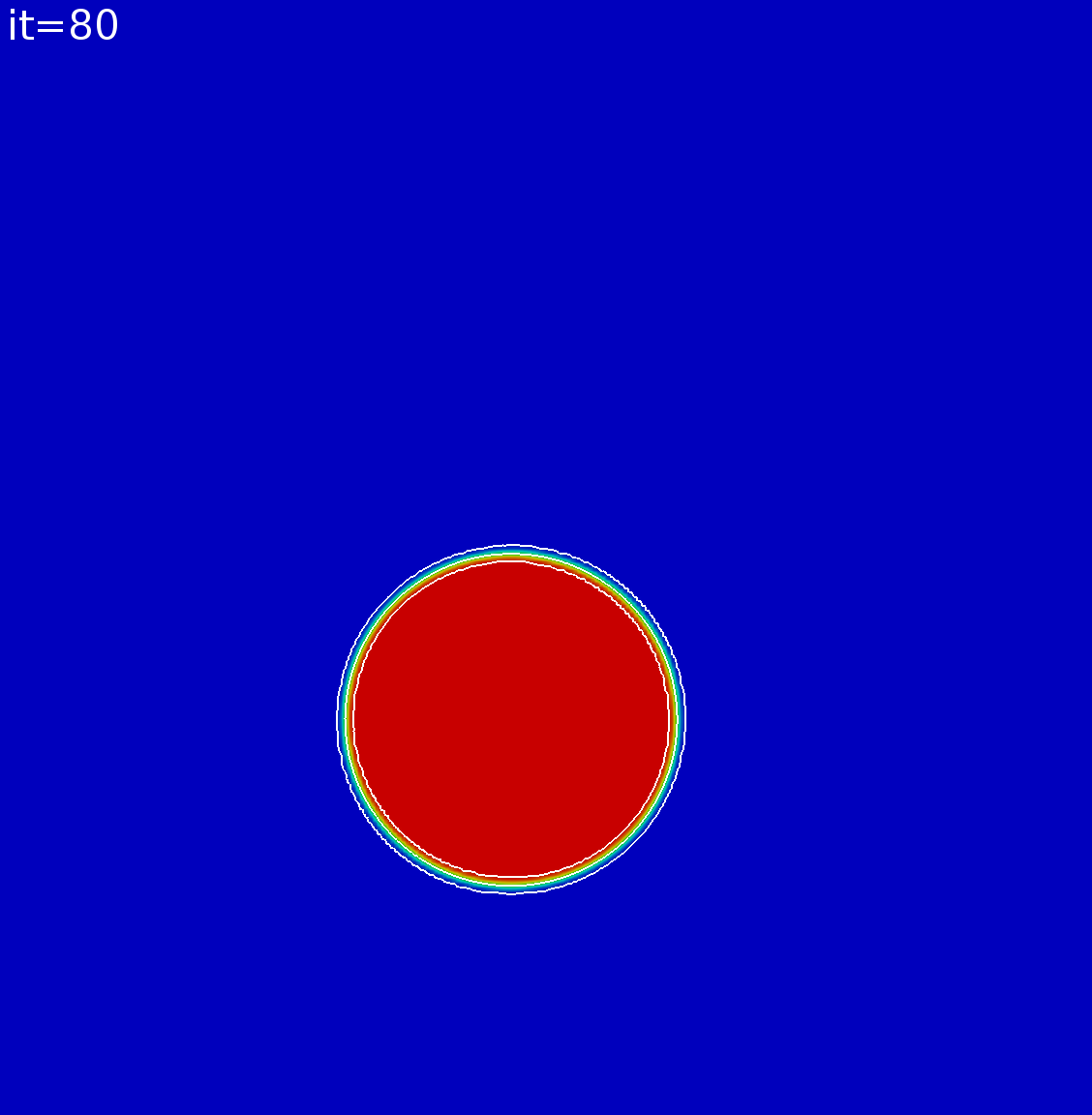}
	\end{minipage}%
	\begin{minipage}{.32\textwidth}
		\centering\includegraphics[width=.825\textwidth,height=.825\textwidth,angle=0]{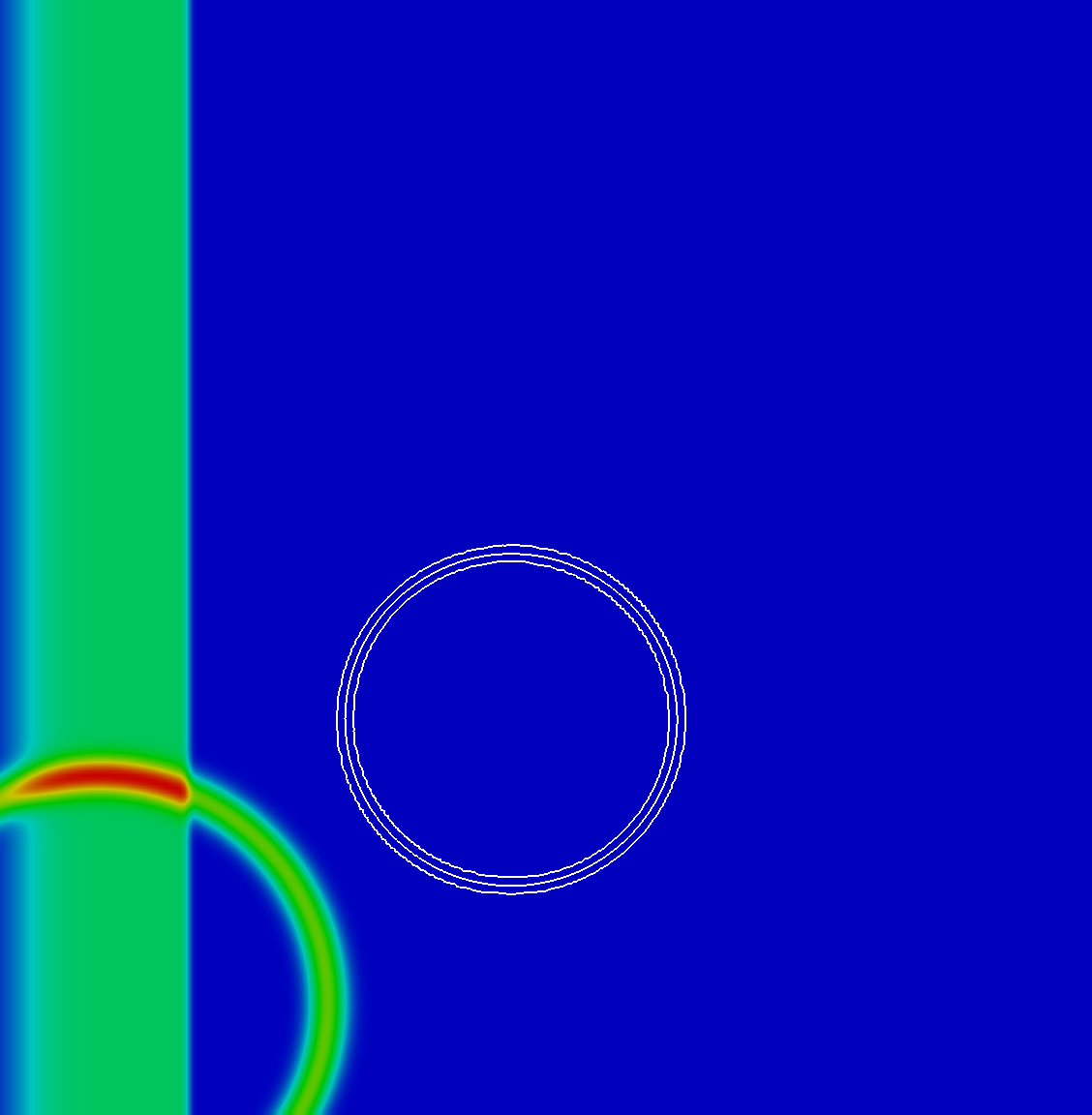}
	\end{minipage}%
	\begin{minipage}{.32\textwidth}
		\centering\includegraphics[width=.825\textwidth,height=.825\textwidth,angle=0]{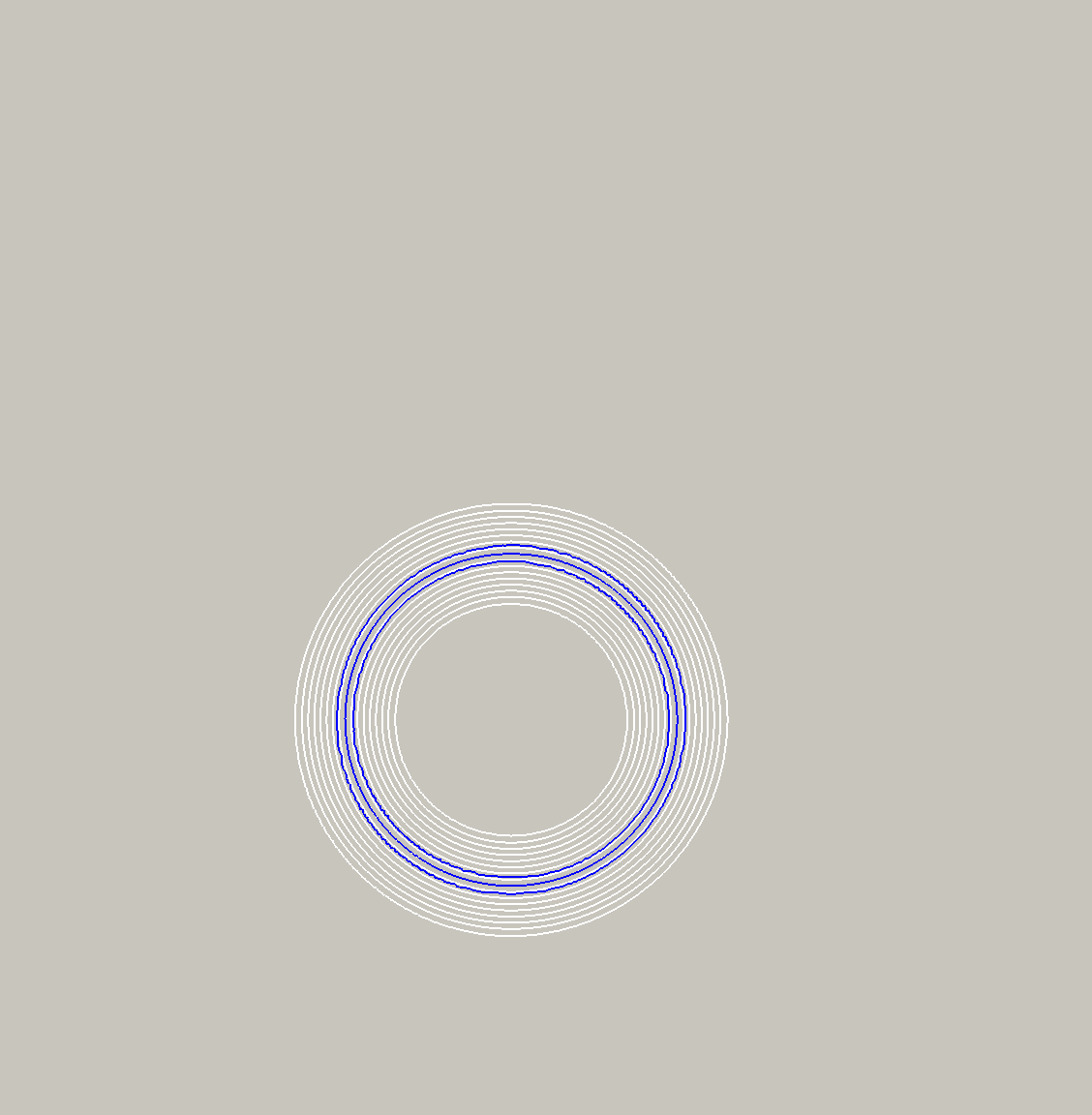}
	\end{minipage}
	\begin{minipage}{.32\textwidth}
		\centering\includegraphics[width=.825\textwidth,height=.825\textwidth,angle=0]{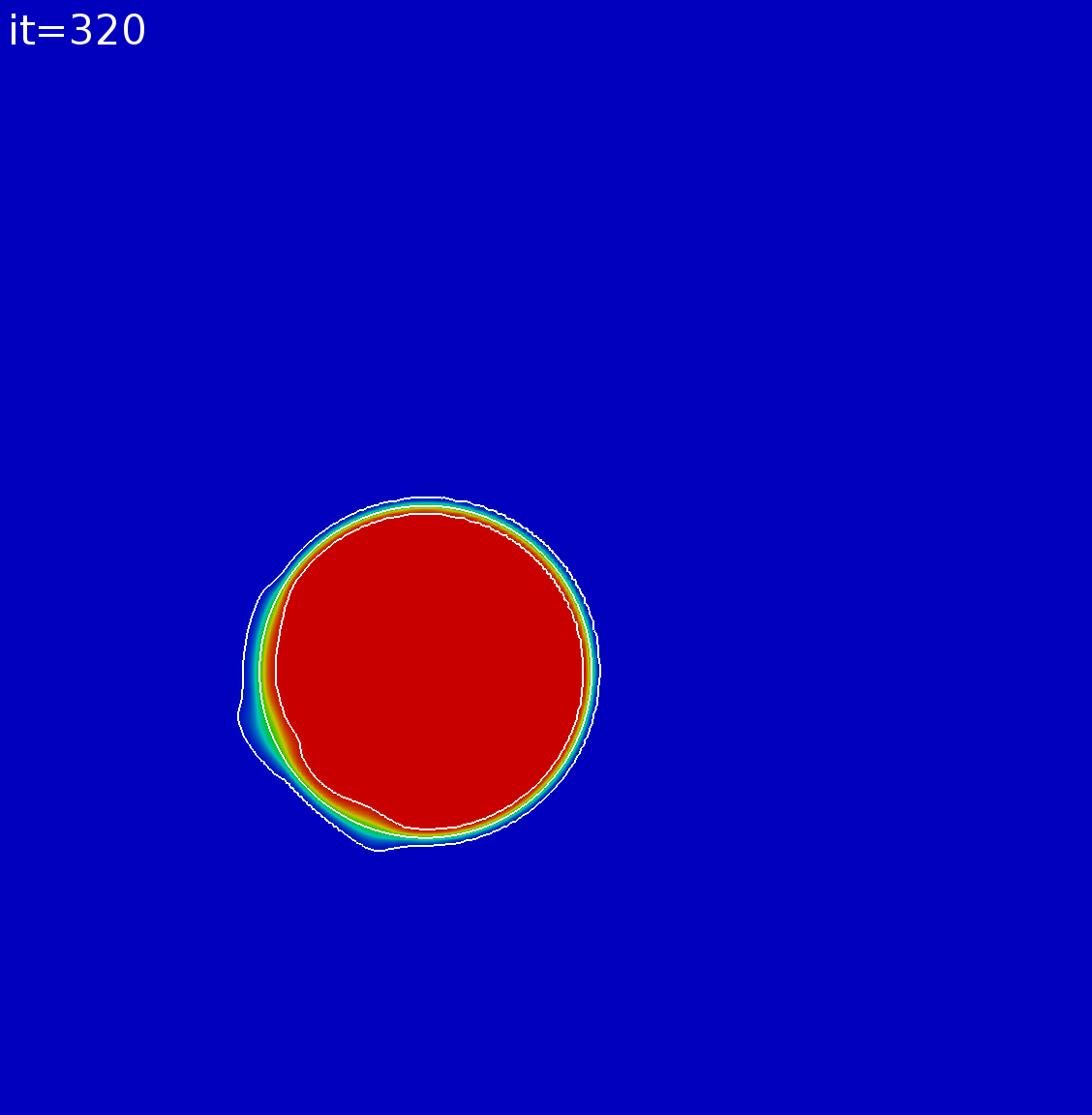}
	\end{minipage}%
	\begin{minipage}{.32\textwidth}
		\centering\includegraphics[width=.825\textwidth,height=.825\textwidth,angle=0]{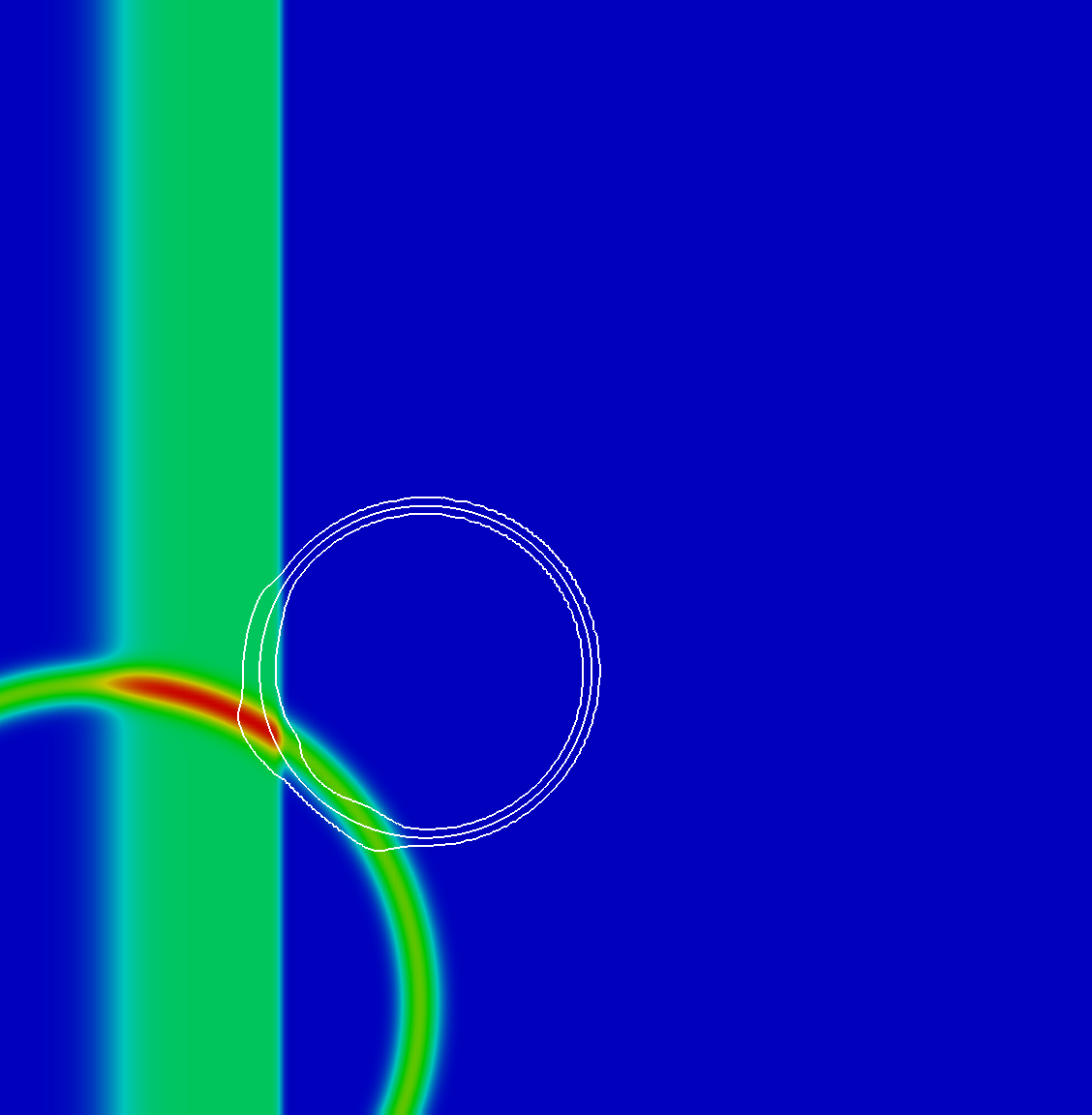}
	\end{minipage}%
	\begin{minipage}{.32\textwidth}
		\centering\includegraphics[width=.825\textwidth,height=.825\textwidth,angle=0]{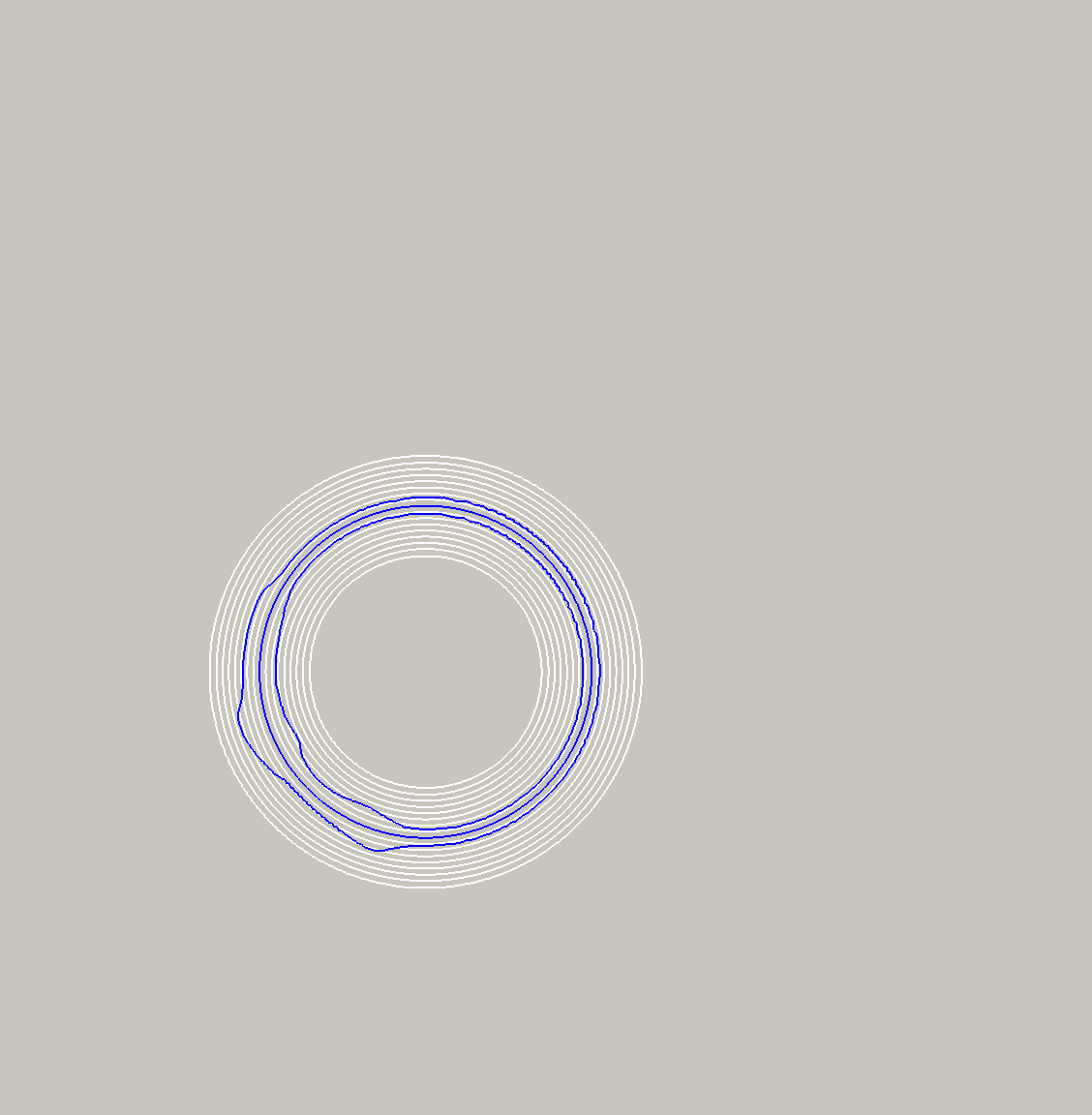}
	\end{minipage}
	\begin{minipage}{.32\textwidth}
		\centering\includegraphics[width=.825\textwidth,height=.825\textwidth,angle=0]{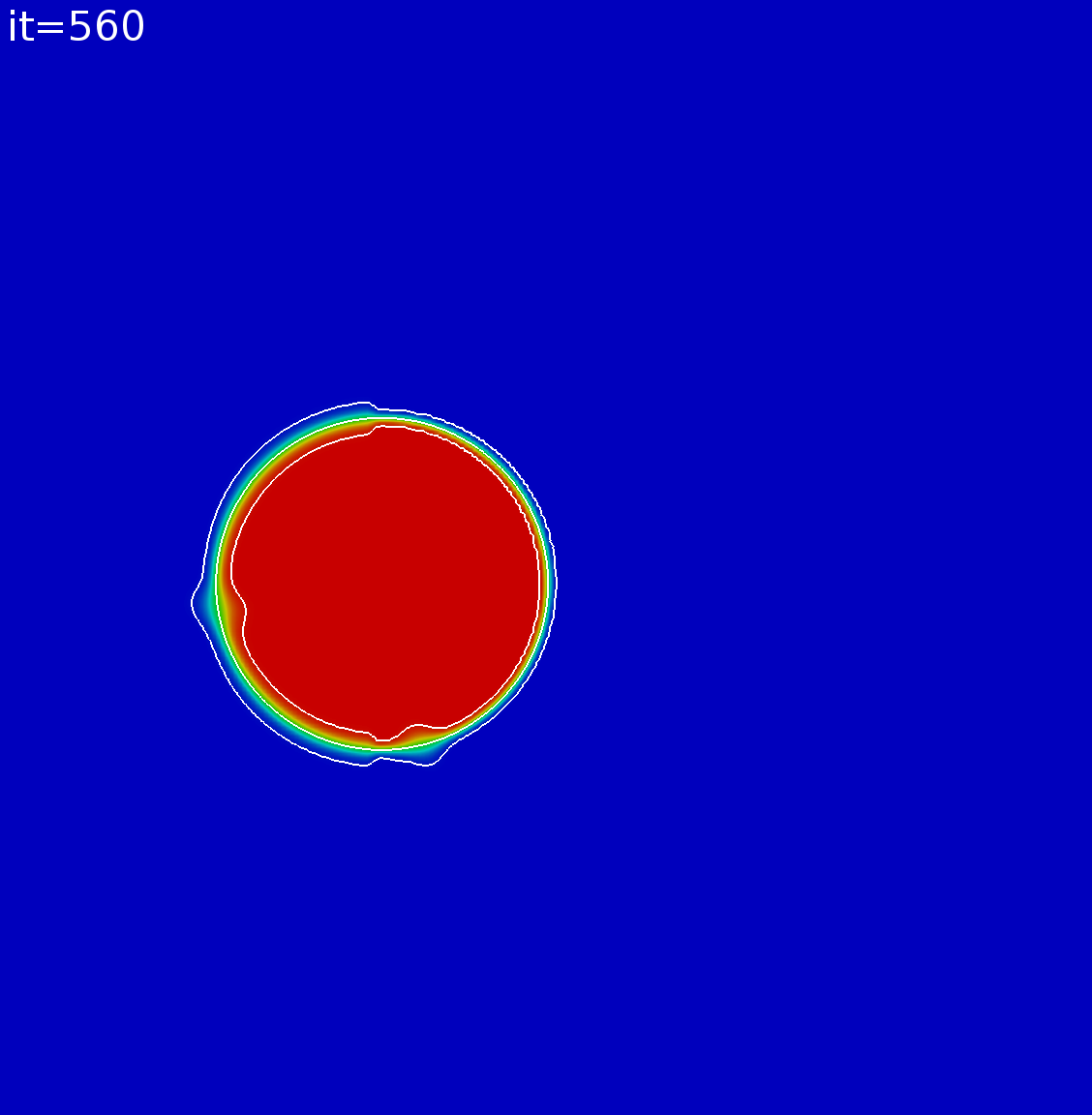}
	\end{minipage}%
	\begin{minipage}{.32\textwidth}
		\centering\includegraphics[width=.825\textwidth,height=.825\textwidth,angle=0]{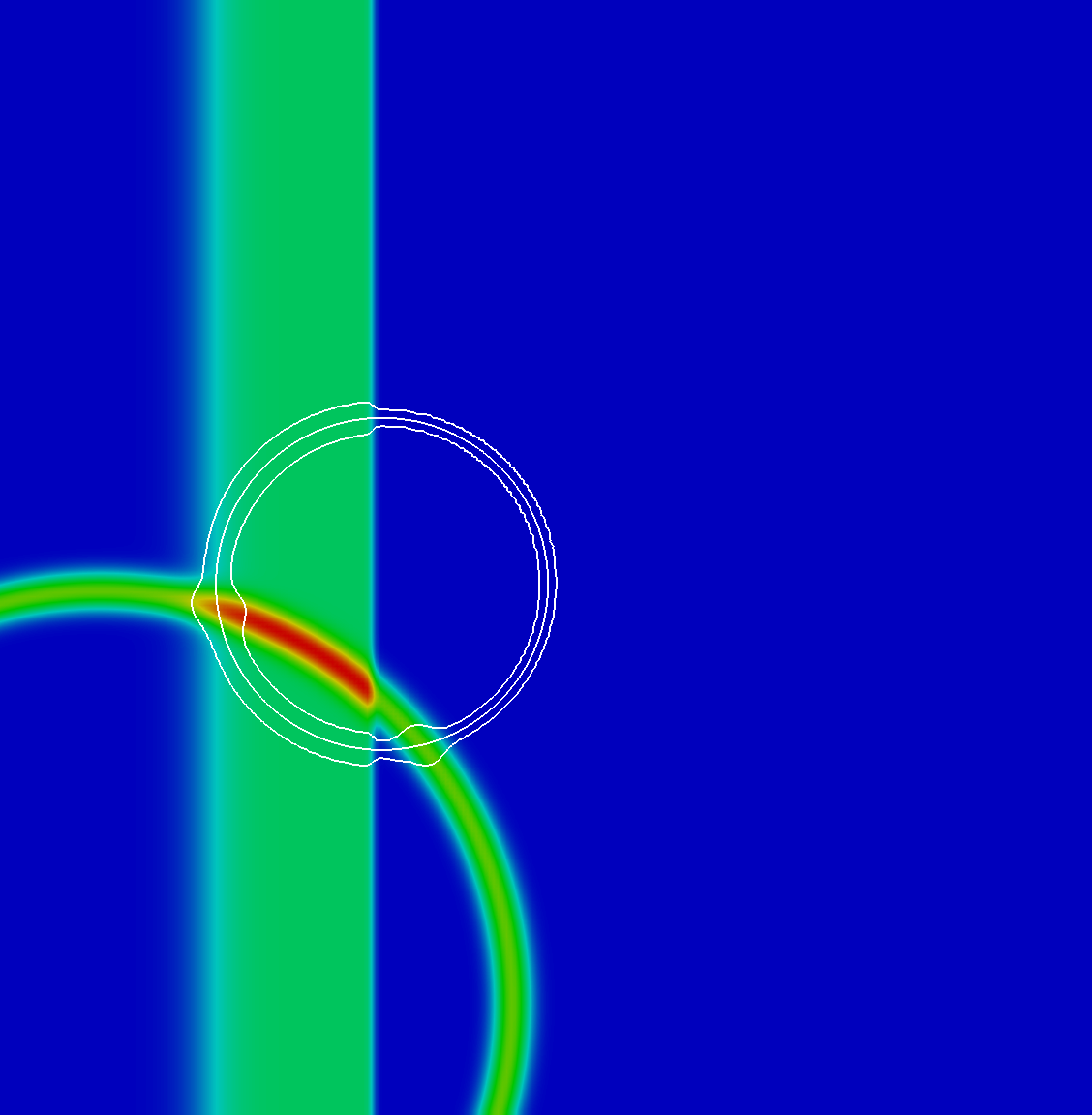}
	\end{minipage}%
	\begin{minipage}{.32\textwidth}
		\centering\includegraphics[width=.825\textwidth,height=.825\textwidth,angle=0]{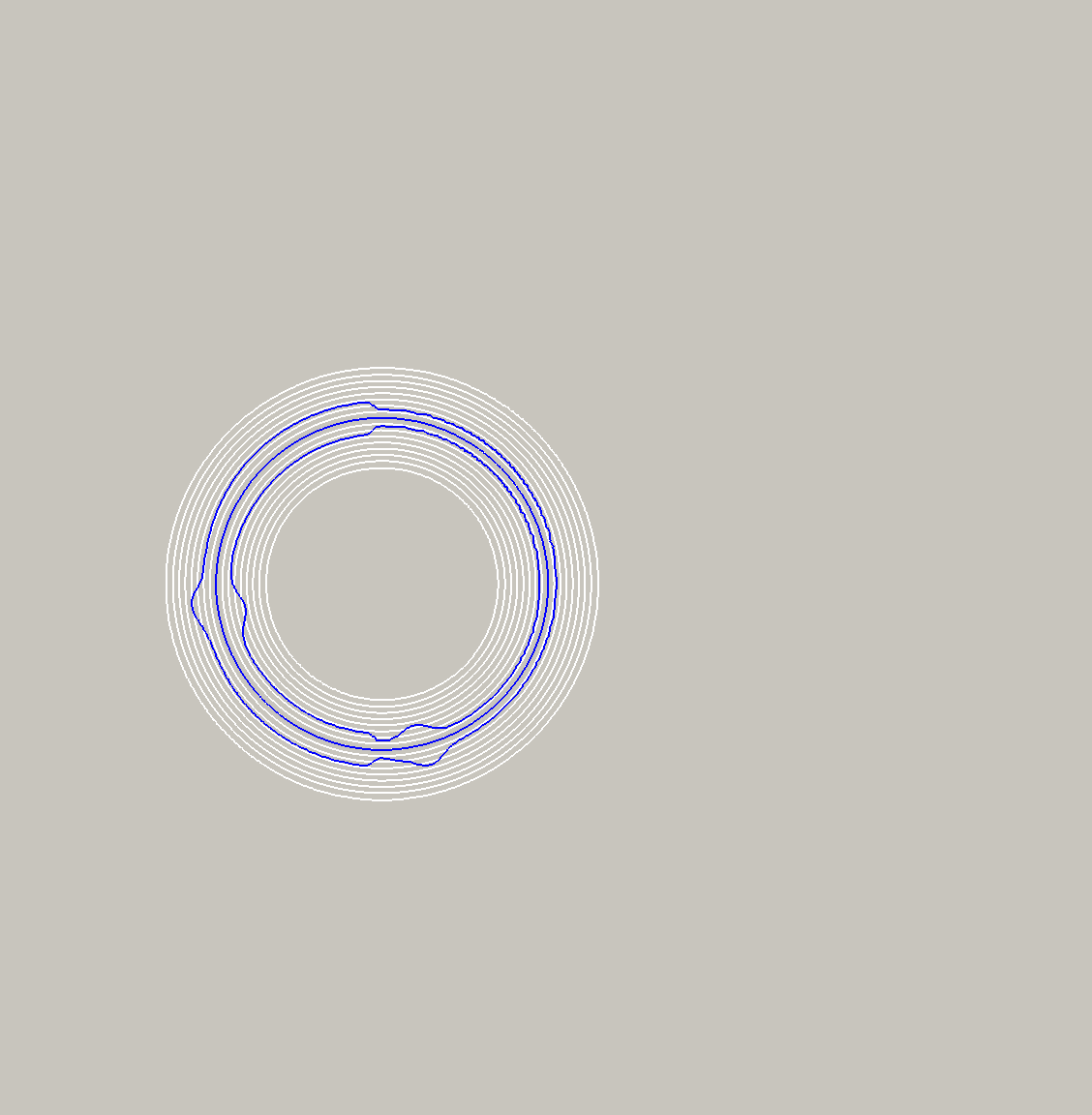}
	\end{minipage}
	\begin{minipage}{.32\textwidth}
		\centering\includegraphics[width=.825\textwidth,height=.825\textwidth,angle=0]{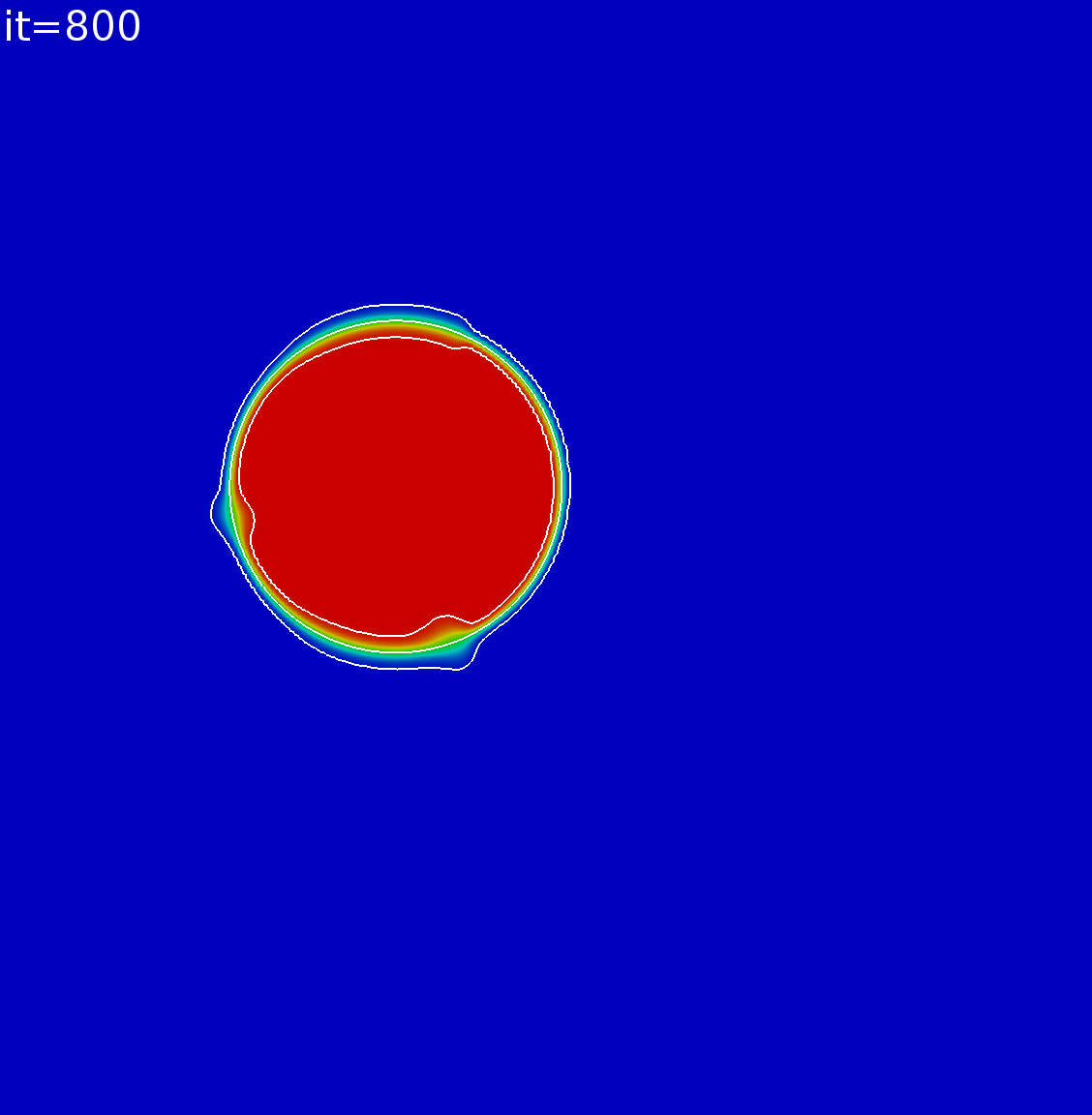}
	\end{minipage}%
	\begin{minipage}{.32\textwidth}
		\centering\includegraphics[width=.825\textwidth,height=.825\textwidth,angle=0]{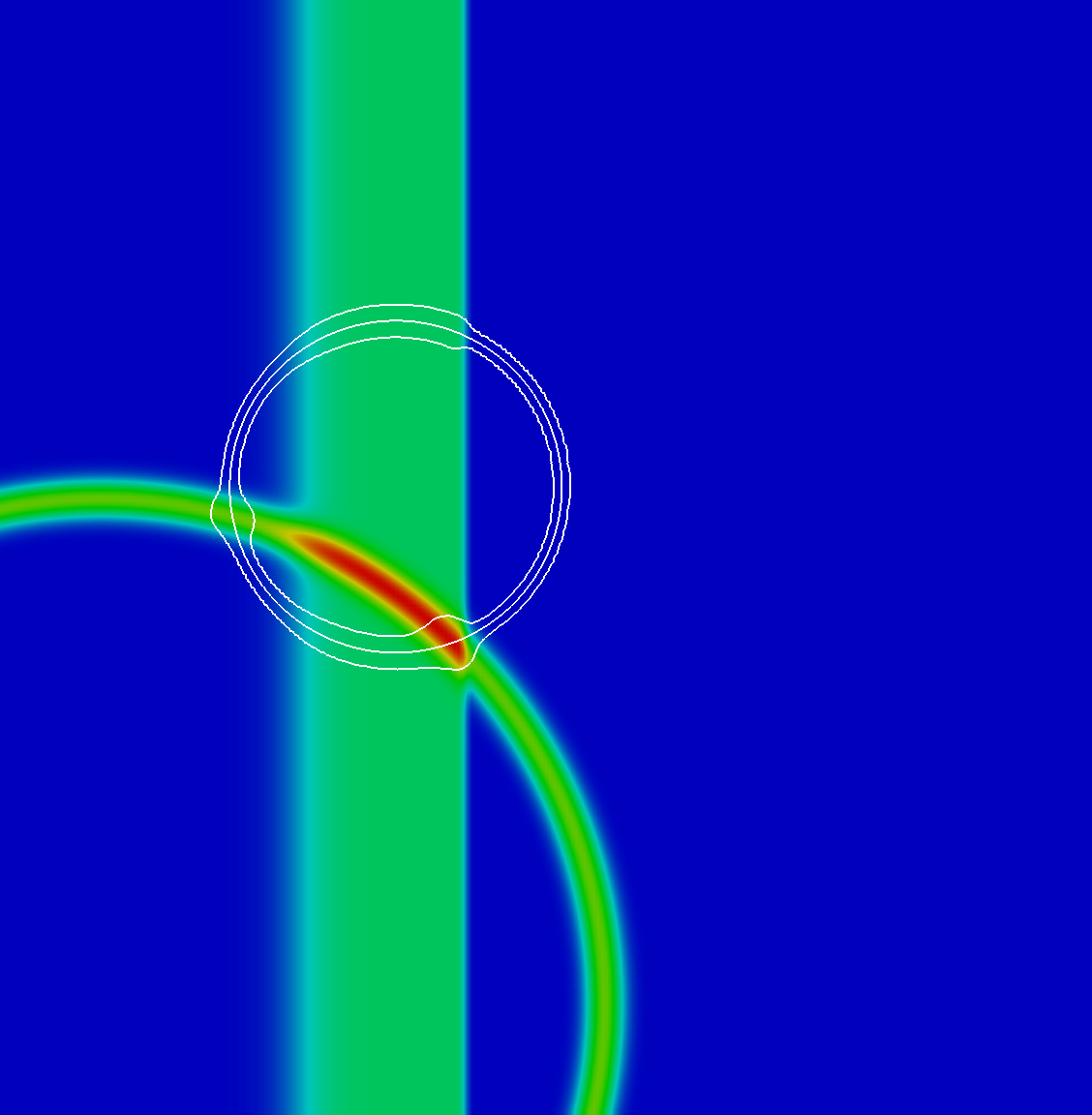}
	\end{minipage}%
	\begin{minipage}{.32\textwidth}
		\centering\includegraphics[width=.825\textwidth,height=.825\textwidth,angle=0]{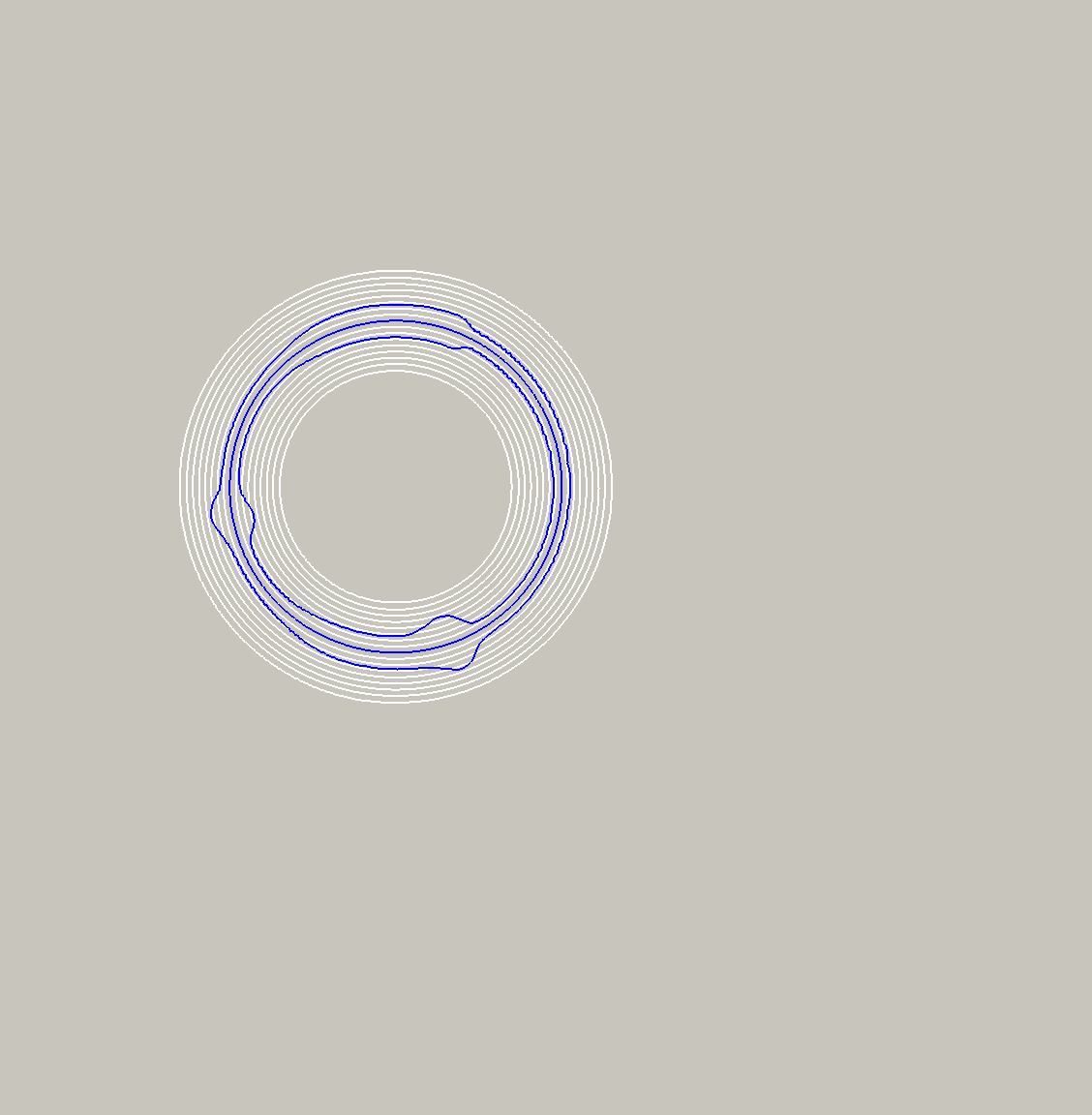}
	\end{minipage}
	\begin{minipage}{.32\textwidth}
		\centering\includegraphics[width=.825\textwidth,height=.825\textwidth,angle=0]{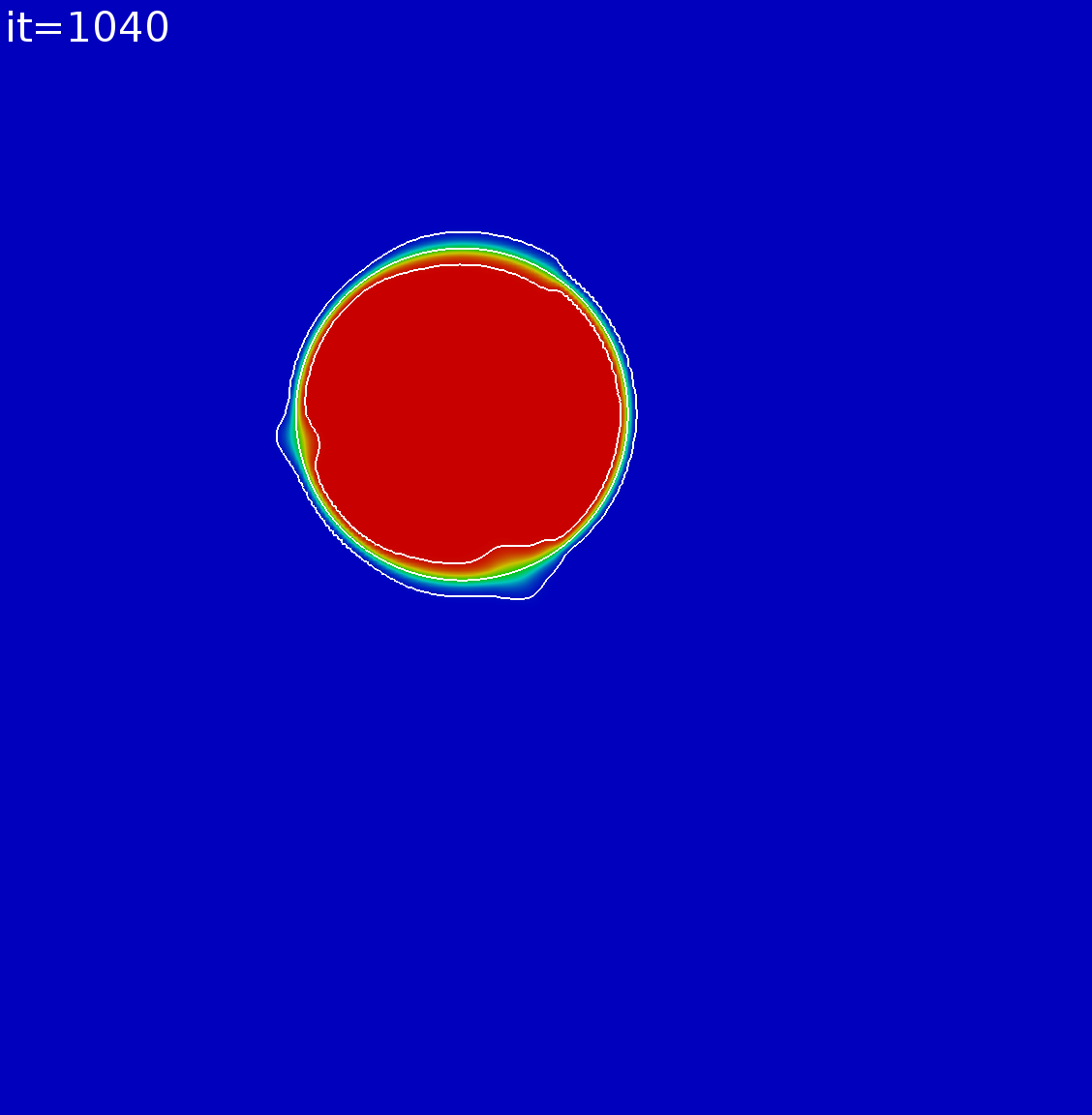}
	\end{minipage}%
	\begin{minipage}{.32\textwidth}
		\centering\includegraphics[width=.825\textwidth,height=.825\textwidth,angle=0]{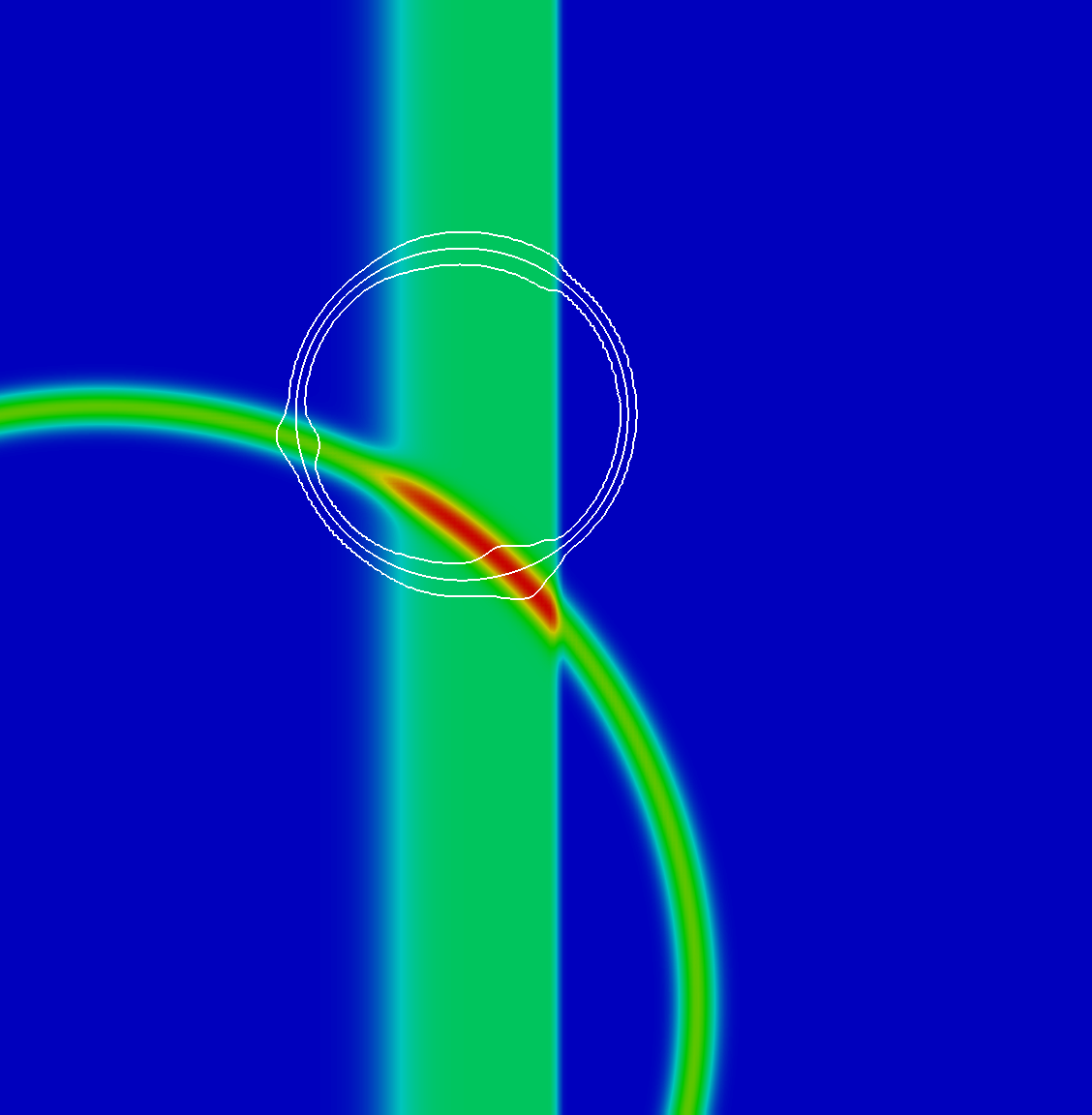}
	\end{minipage}%
	\begin{minipage}{.32\textwidth}
		\centering\includegraphics[width=.825\textwidth,height=.825\textwidth,angle=0]{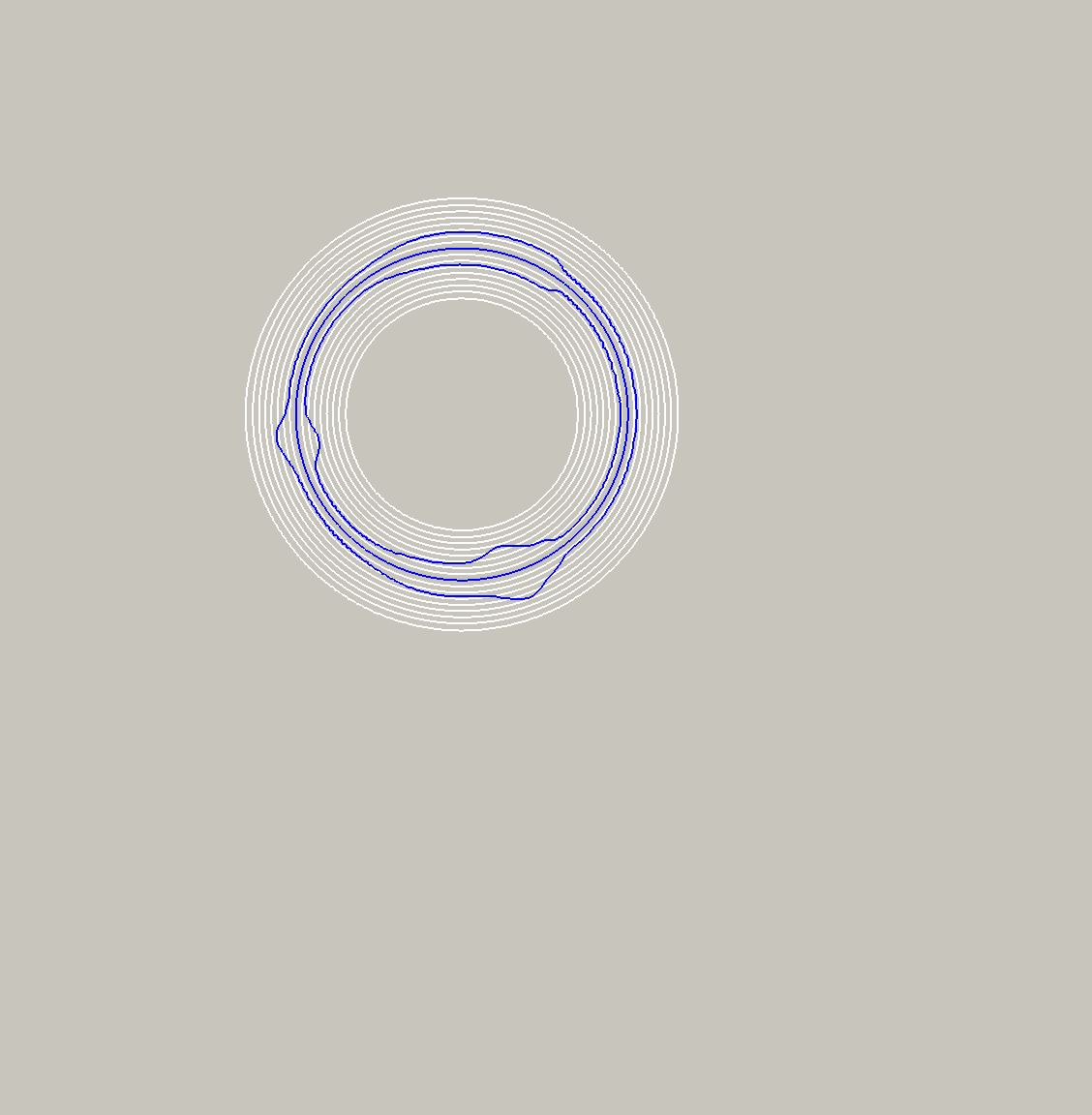}
	\end{minipage}
	\caption{\small{Evolution of the intermittency region affected by
			the variable field $\ephxt$.
			All figures show contours of $\alpha(\psi \!=\! -4\ephb),
			\alpha(\psi \!=\! 0),\alpha(\psi \!=\!4\ephb)$,
			and from left to right fields 
			of $0 \!<\! \alpp \!<\! 1$,
			$\epsilon_{h,b} \!\le\! \ephxt \!<\! 5\epsilon_{h,b}$,
			contours $-32\ephb \!\le\! \psia \!\le\! 32\ephb$. 
                }}				  
		\label{fig10}
\end{figure}
\elnm 
\blnm
\begin{figure}[!ht] \nonumber
	\begin{minipage}{.32\textwidth}
		\centering\includegraphics[width=.825\textwidth,height=.1\textwidth,angle=0]{fig10bar1}
	\end{minipage}%
	\begin{minipage}{.32\textwidth}
		\centering\includegraphics[width=.825\textwidth,height=.1\textwidth,angle=0]{fig10bar2}
	\end{minipage}%
	\begin{minipage}{.32\textwidth}
	\end{minipage}
	\begin{minipage}{.32\textwidth}
		\centering\includegraphics[width=.825\textwidth,height=.825\textwidth,angle=0]{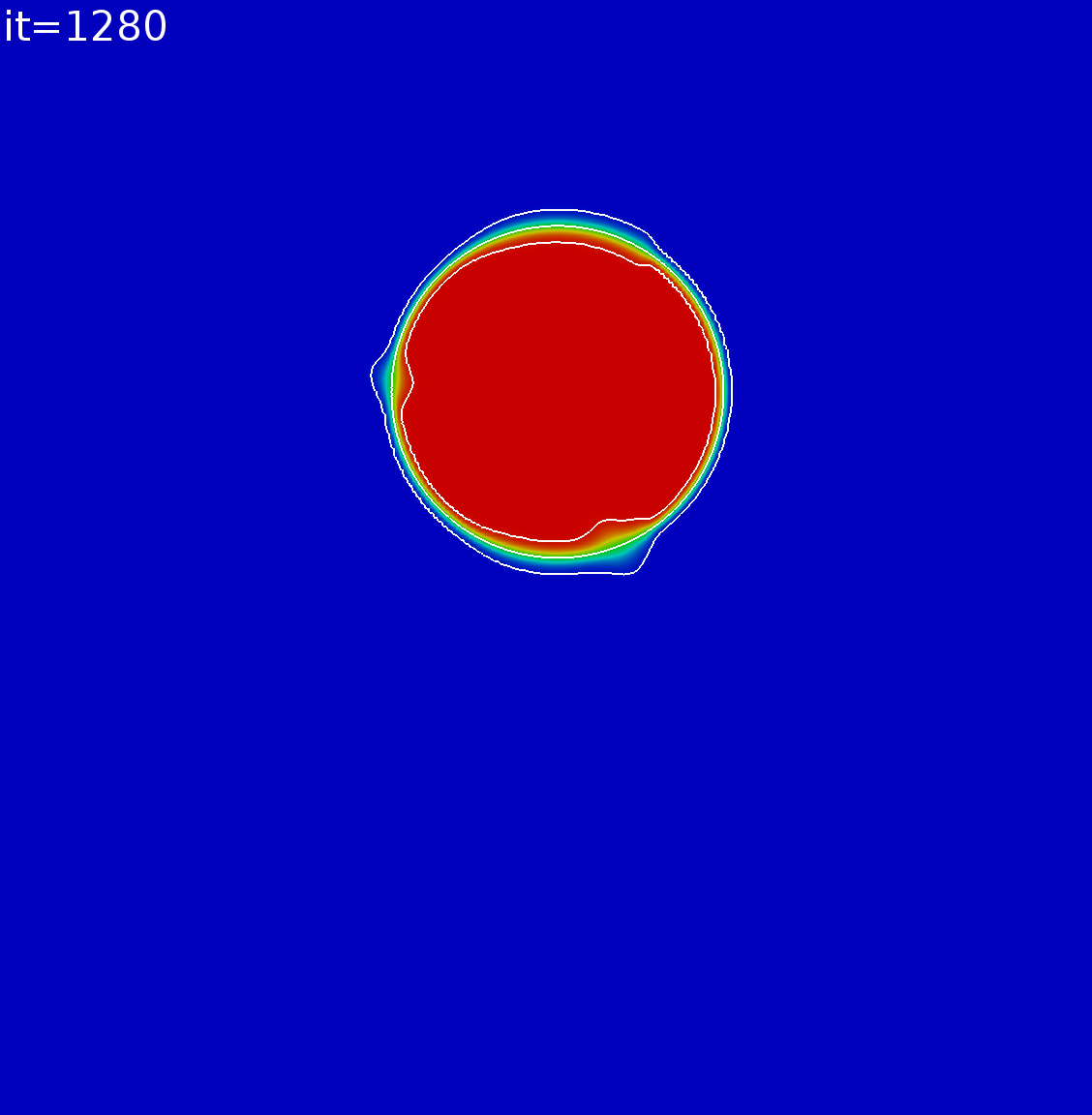}
	\end{minipage}%
	\begin{minipage}{.32\textwidth}
		\centering\includegraphics[width=.825\textwidth,height=.825\textwidth,angle=0]{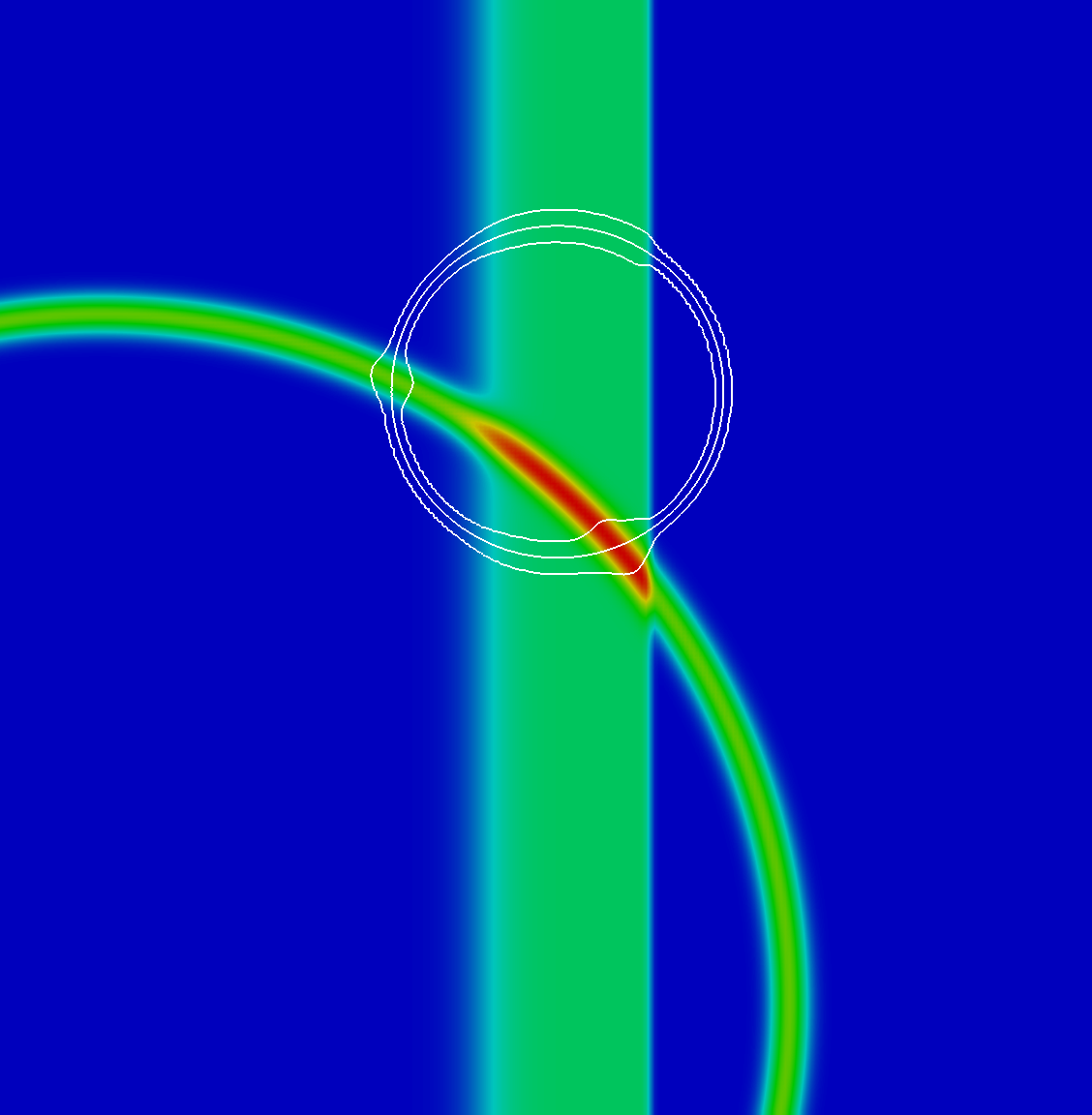}
	\end{minipage}%
	\begin{minipage}{.32\textwidth}
		\centering\includegraphics[width=.825\textwidth,height=.825\textwidth,angle=0]{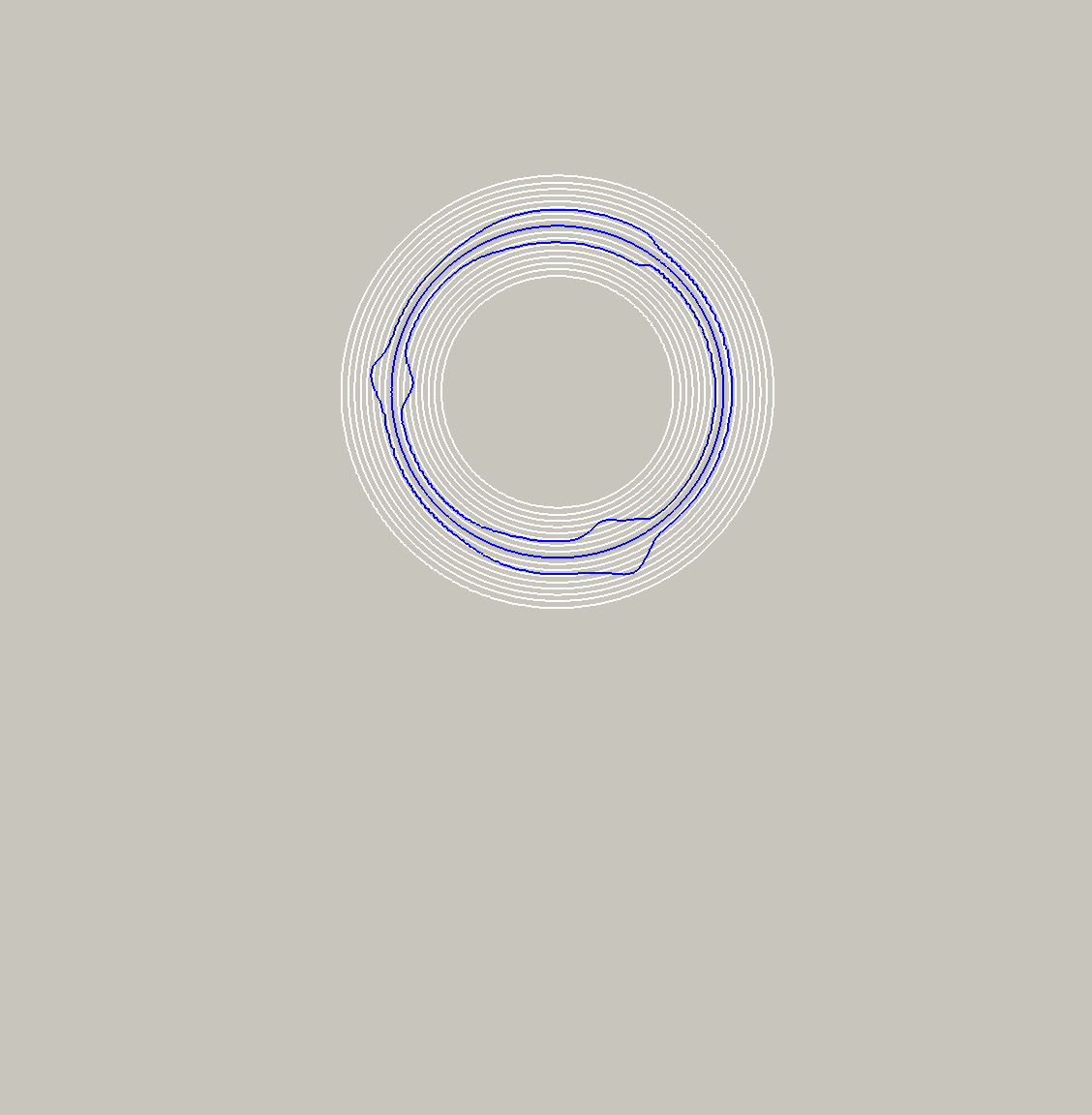}
	\end{minipage}
	\begin{minipage}{.32\textwidth}
		\centering\includegraphics[width=.825\textwidth,height=.825\textwidth,angle=0]{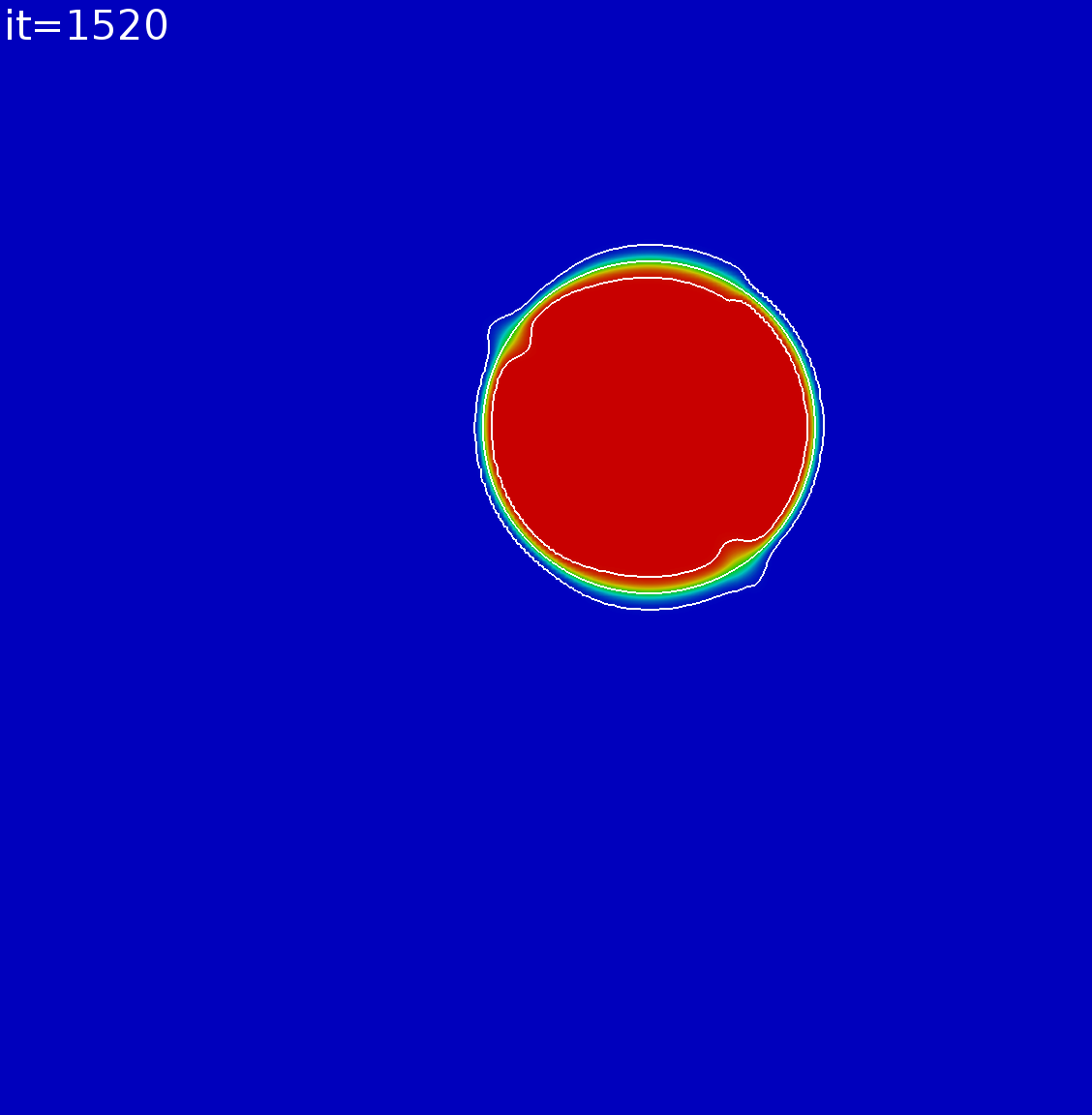}
	\end{minipage}%
	\begin{minipage}{.32\textwidth}
		\centering\includegraphics[width=.825\textwidth,height=.825\textwidth,angle=0]{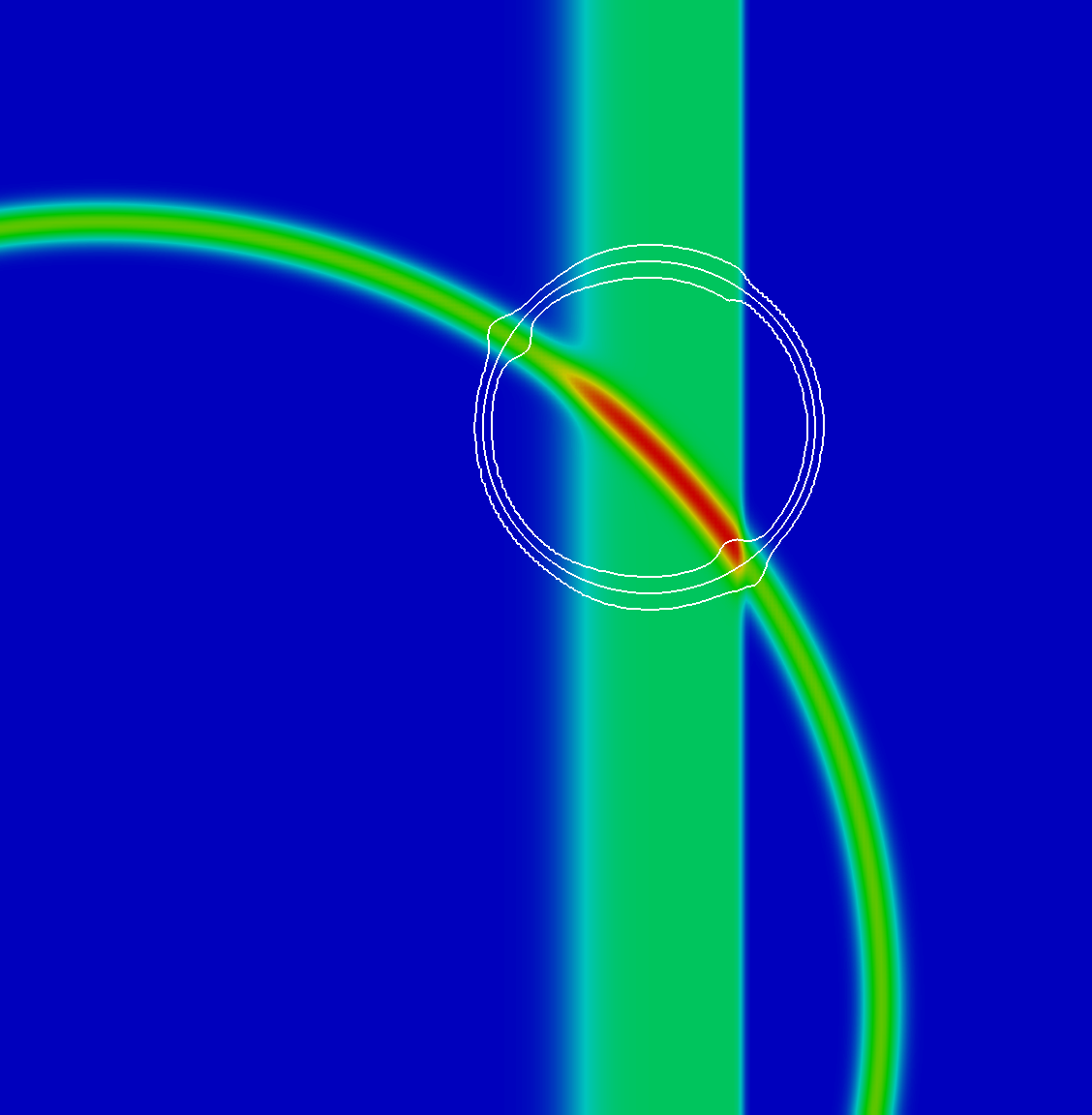}
	\end{minipage}%
	\begin{minipage}{.32\textwidth}
		\centering\includegraphics[width=.825\textwidth,height=.825\textwidth,angle=0]{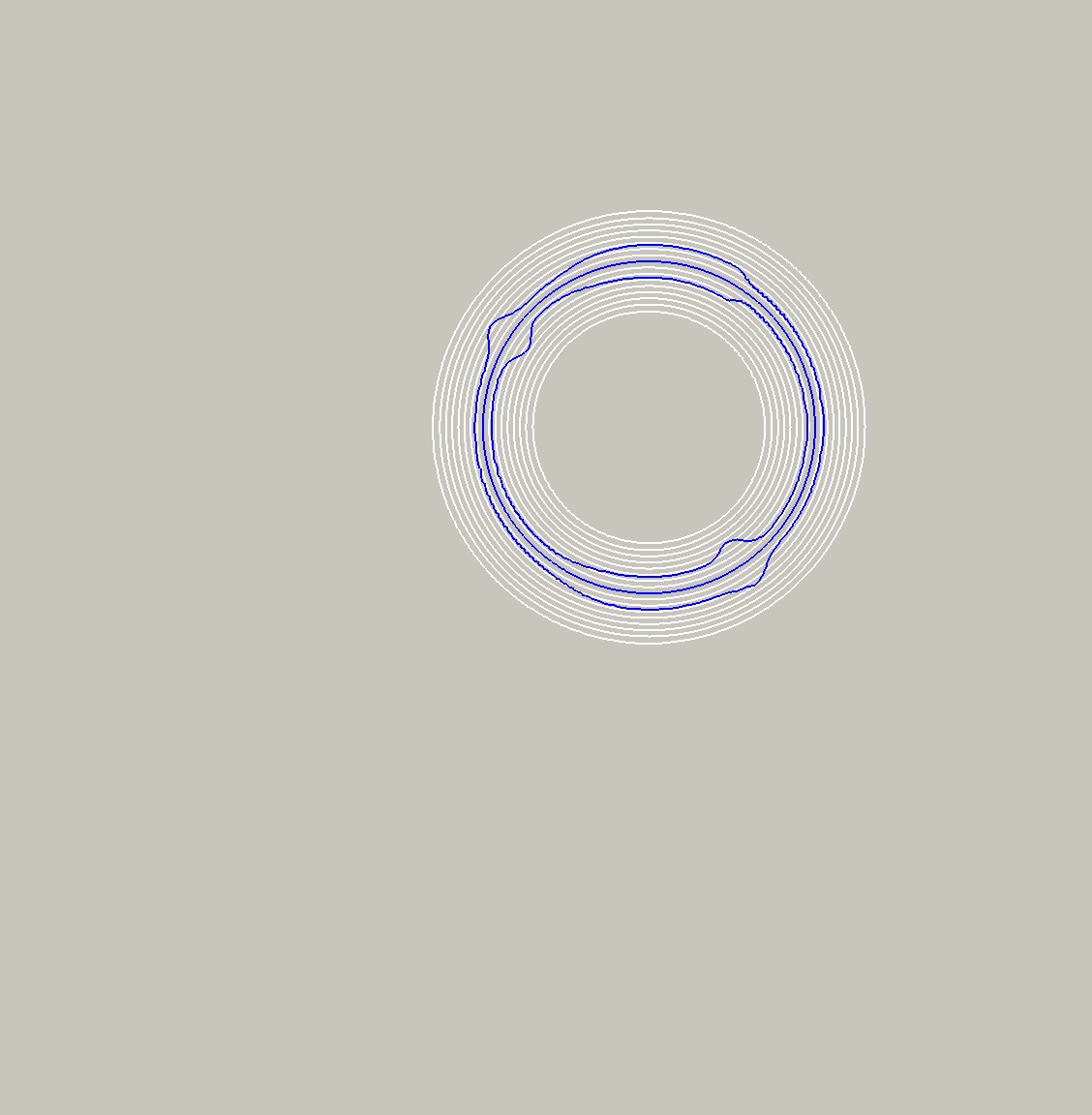}
	\end{minipage}
	\begin{minipage}{.32\textwidth}
		\centering\includegraphics[width=.825\textwidth,height=.825\textwidth,angle=0]{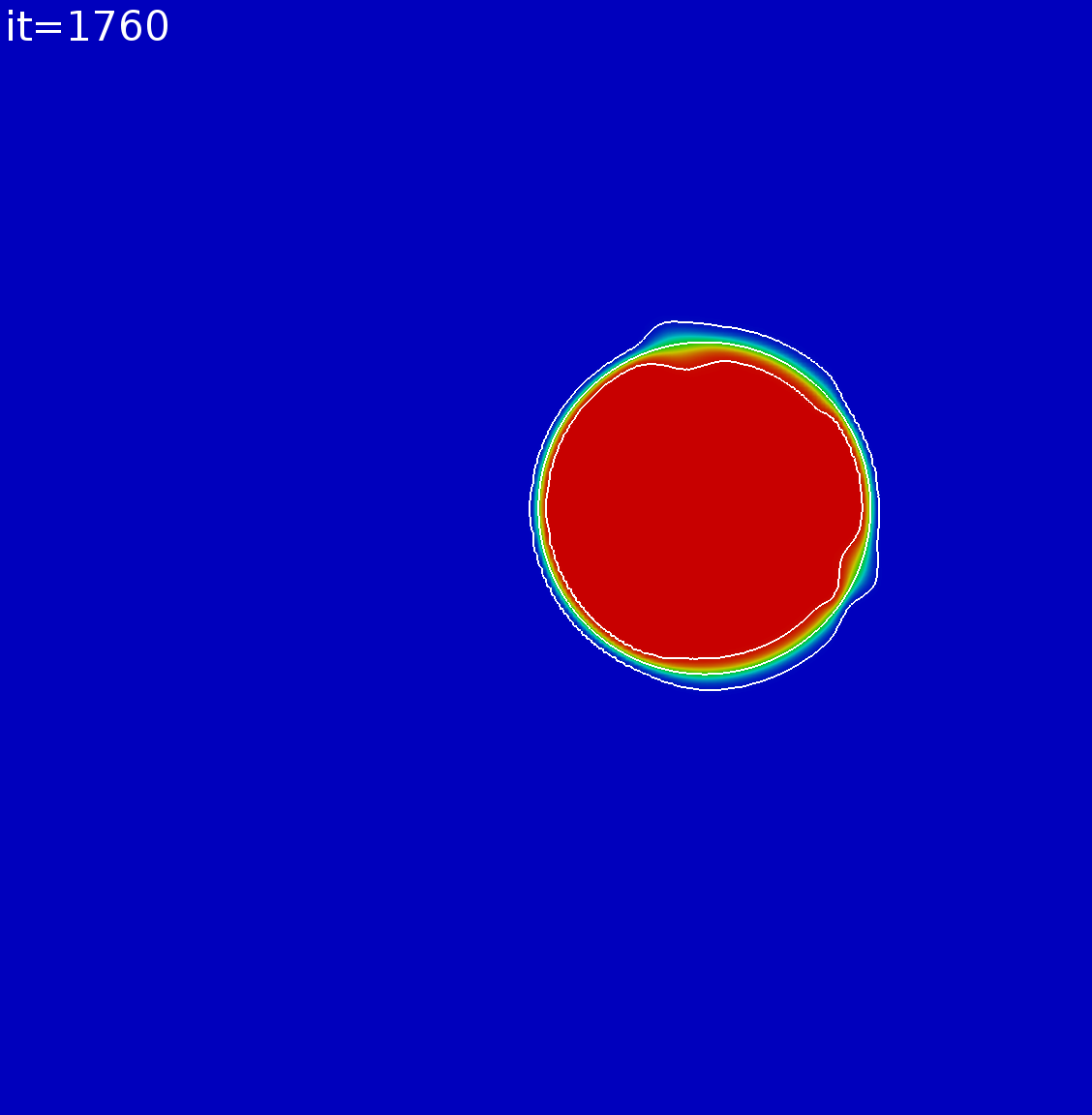}
	\end{minipage}%
	\begin{minipage}{.32\textwidth}
		\centering\includegraphics[width=.825\textwidth,height=.825\textwidth,angle=0]{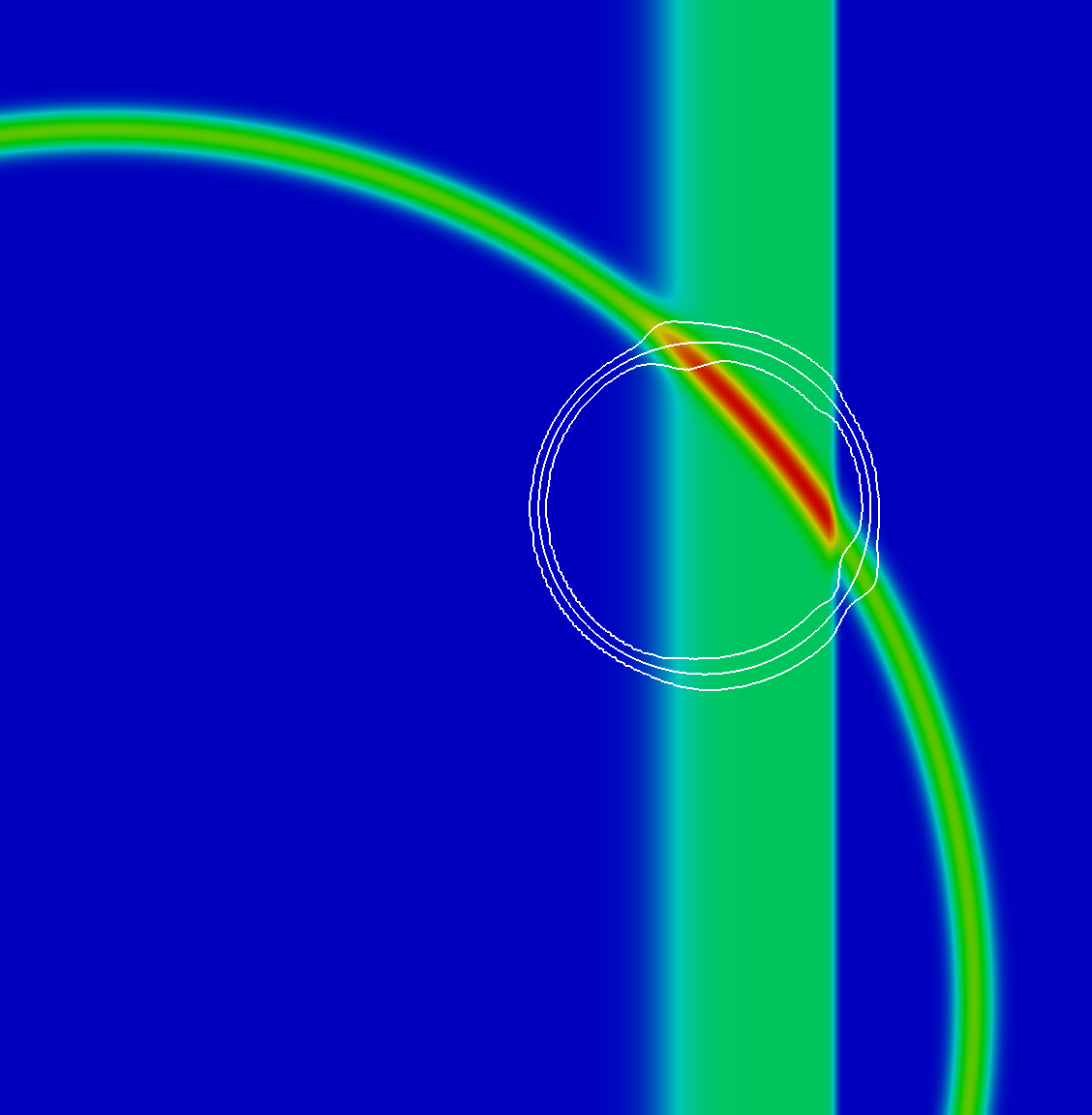}
	\end{minipage}%
	\begin{minipage}{.32\textwidth}
		\centering\includegraphics[width=.825\textwidth,height=.825\textwidth,angle=0]{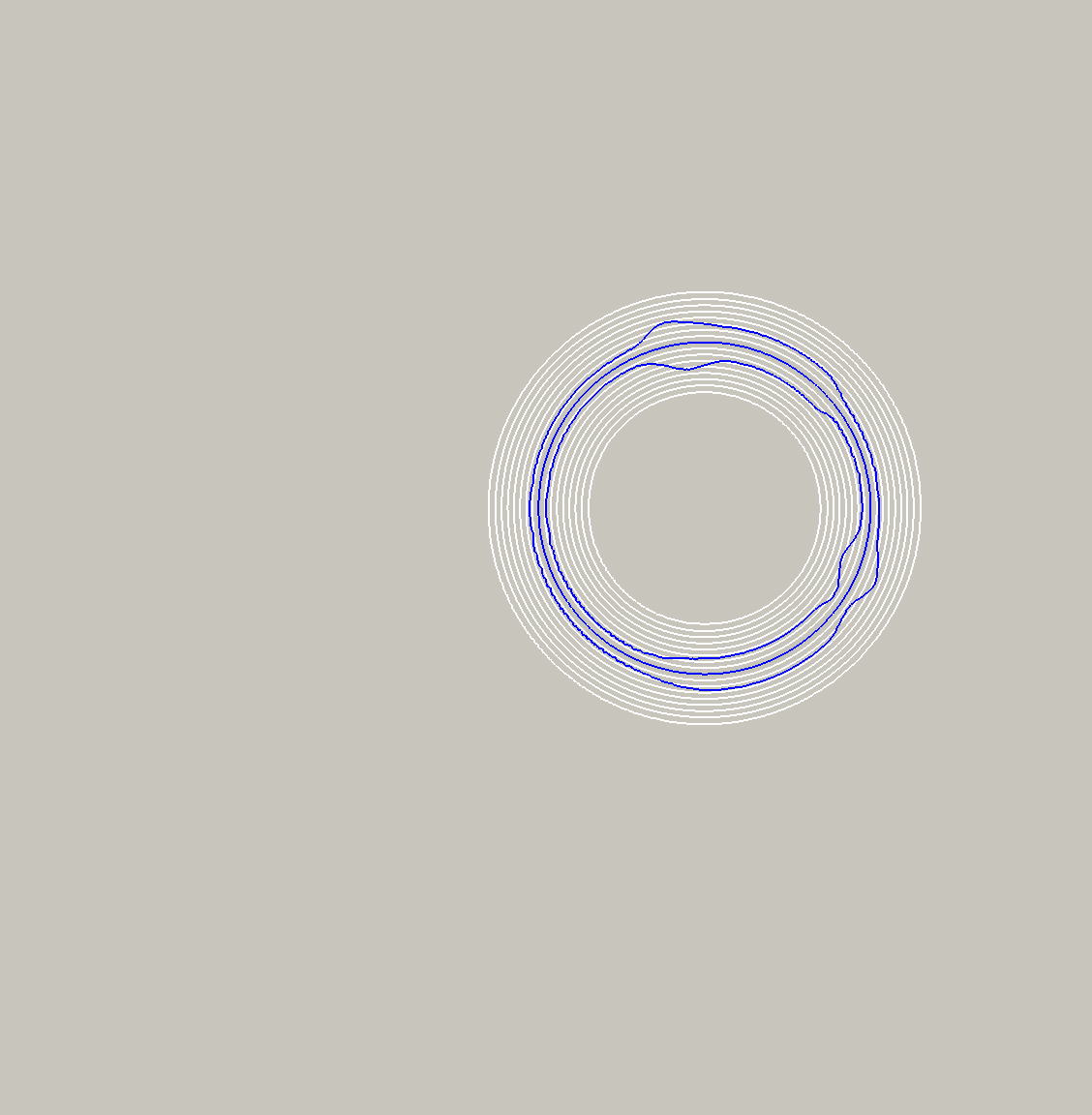}
	\end{minipage}
	\begin{minipage}{.32\textwidth}
		\centering\includegraphics[width=.825\textwidth,height=.825\textwidth,angle=0]{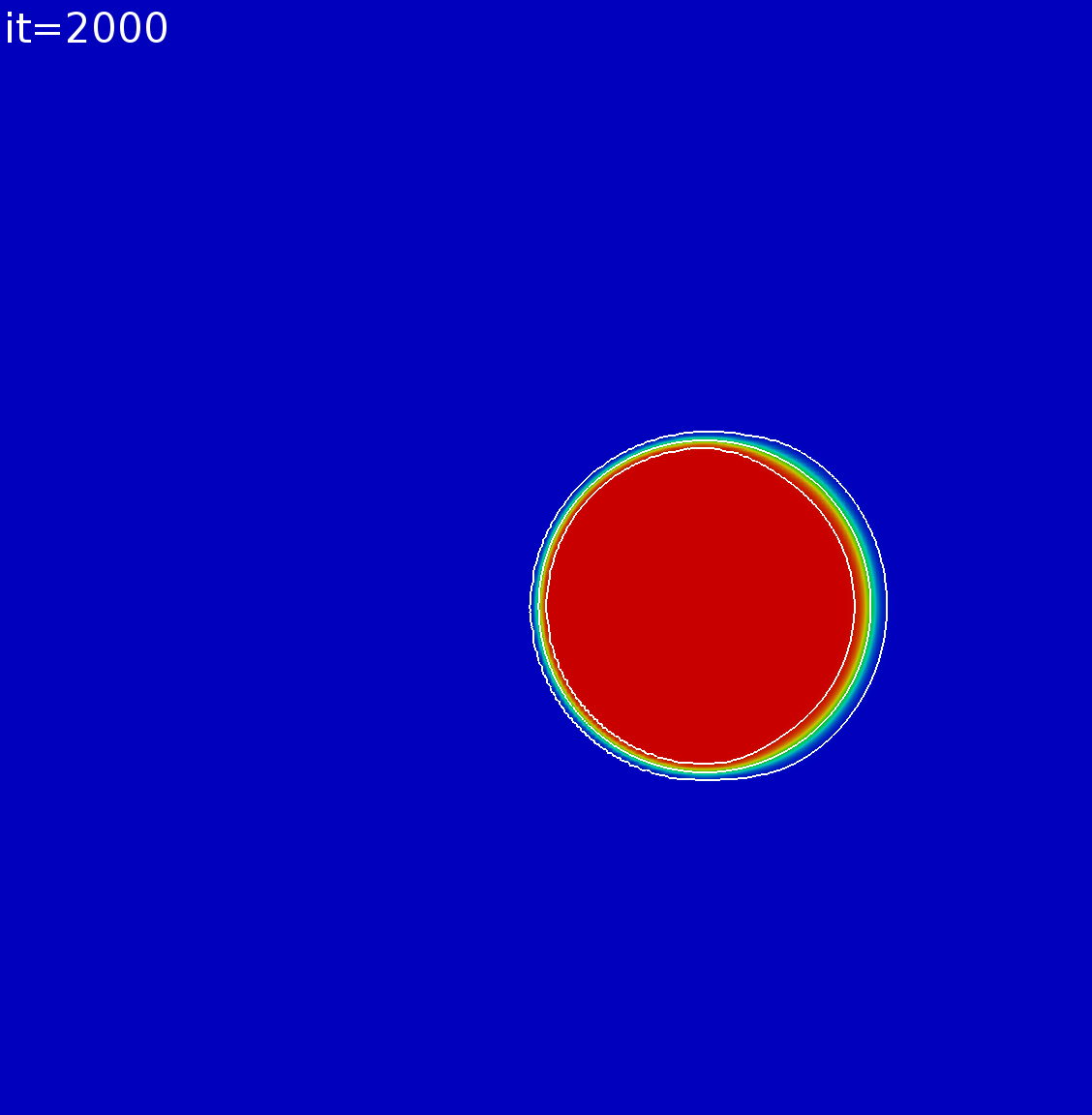}
	\end{minipage}%
	\begin{minipage}{.32\textwidth}
		\centering\includegraphics[width=.825\textwidth,height=.825\textwidth,angle=0]{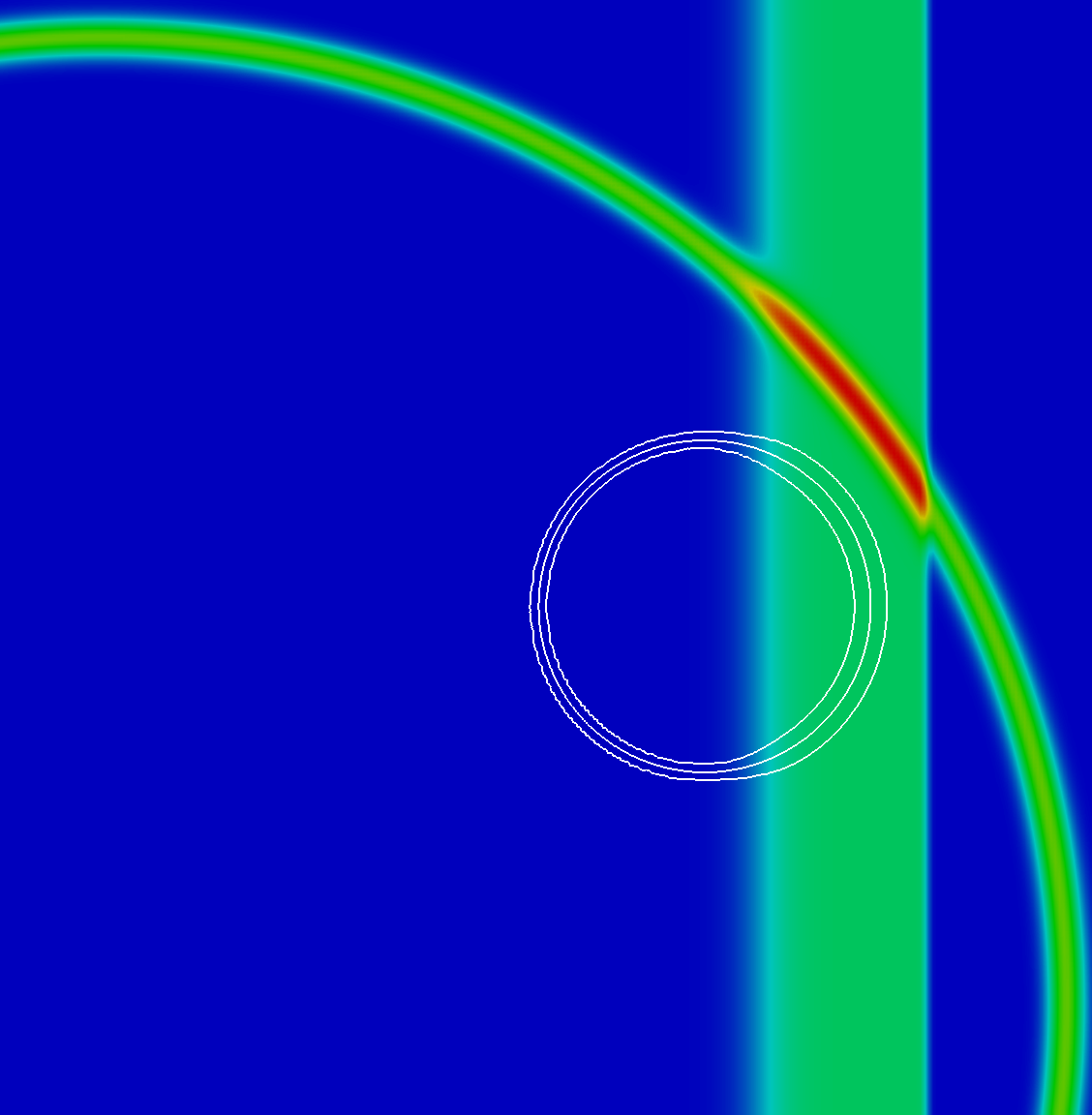}
	\end{minipage}%
	\begin{minipage}{.32\textwidth}
		\centering\includegraphics[width=.825\textwidth,height=.825\textwidth,angle=0]{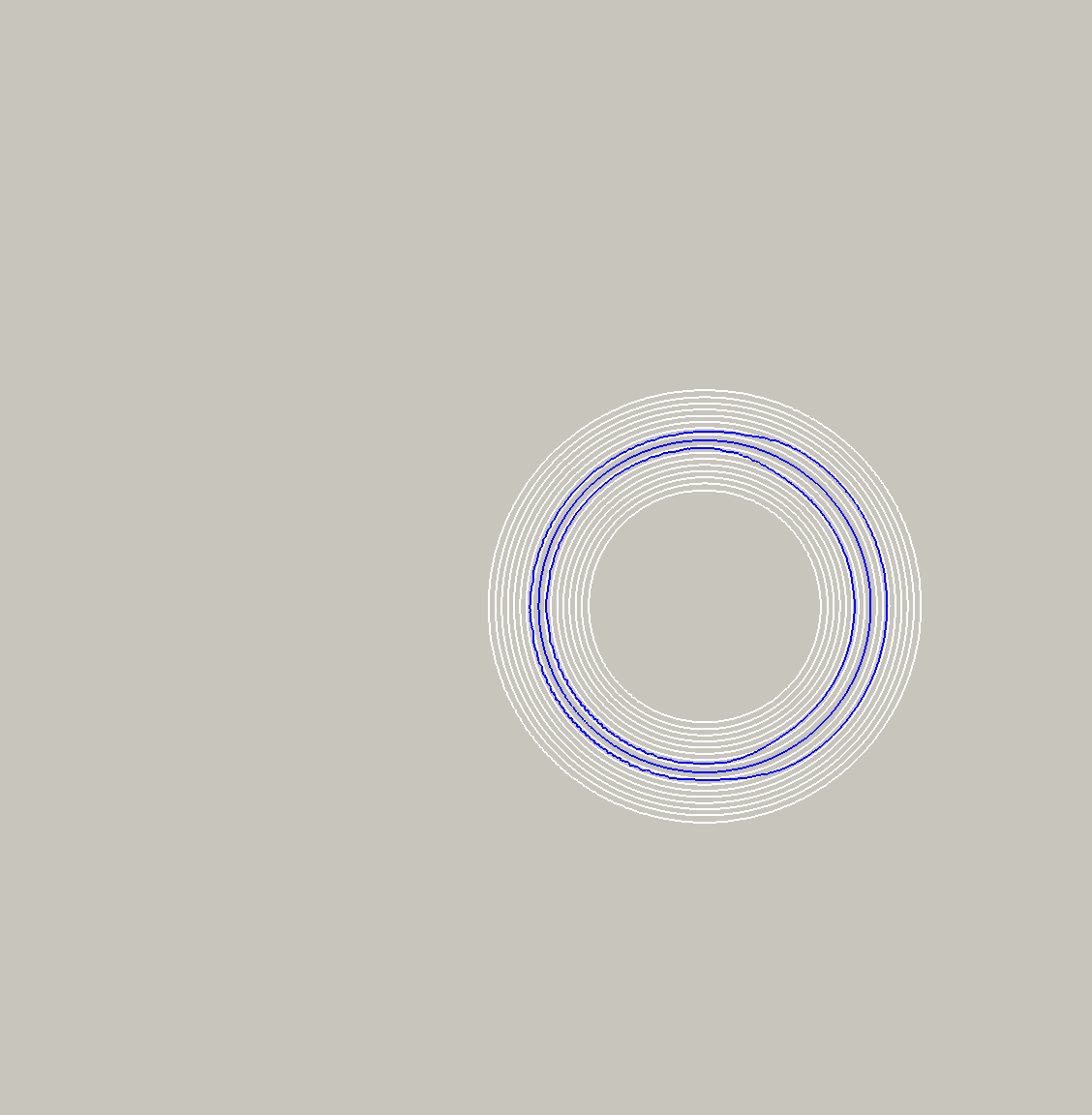}
	\end{minipage}
	\begin{minipage}{.32\textwidth}
		\centering\includegraphics[width=.825\textwidth,height=.825\textwidth,angle=0]{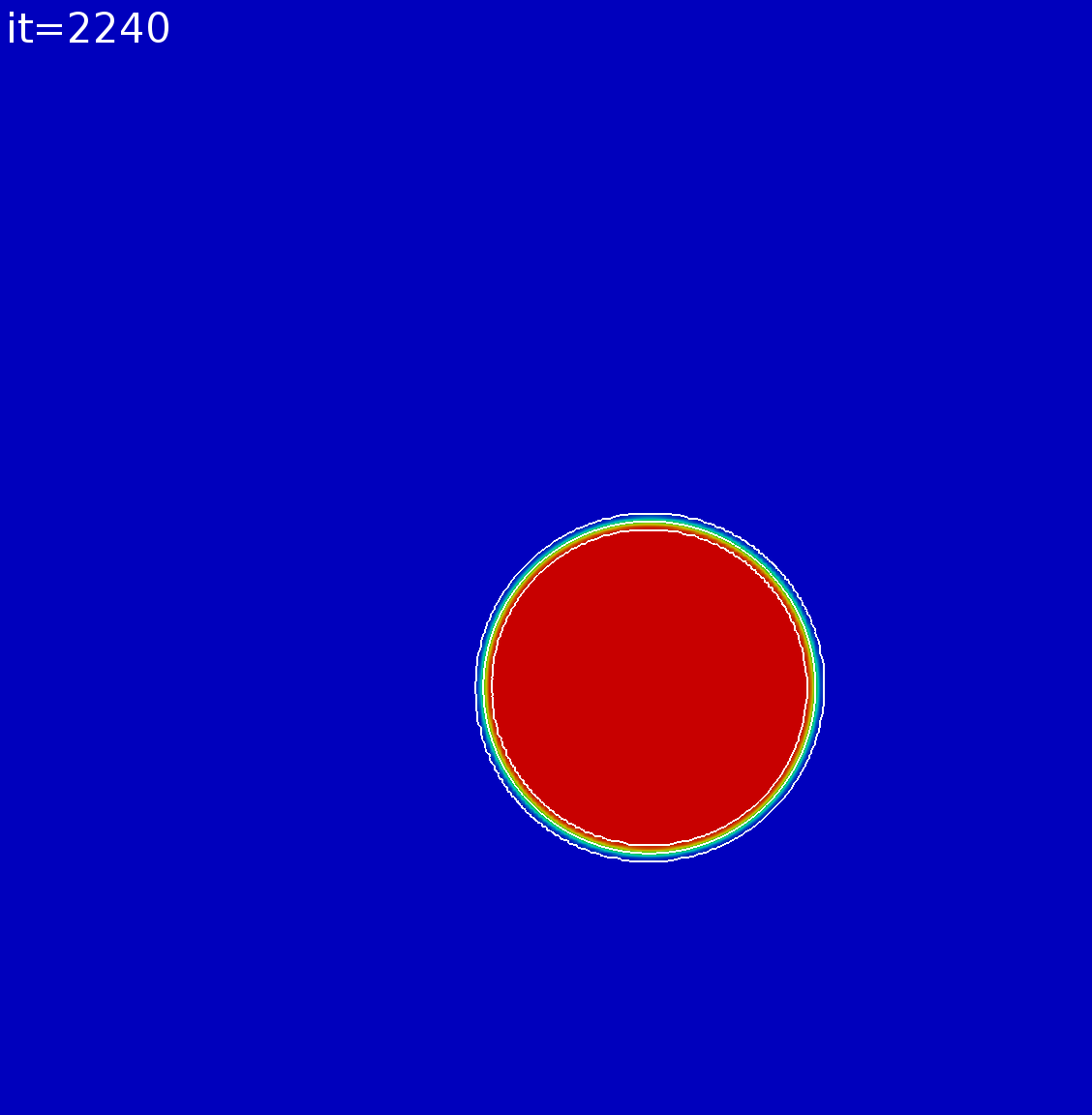}
	\end{minipage}%
	\begin{minipage}{.32\textwidth}
		\centering\includegraphics[width=.825\textwidth,height=.825\textwidth,angle=0]{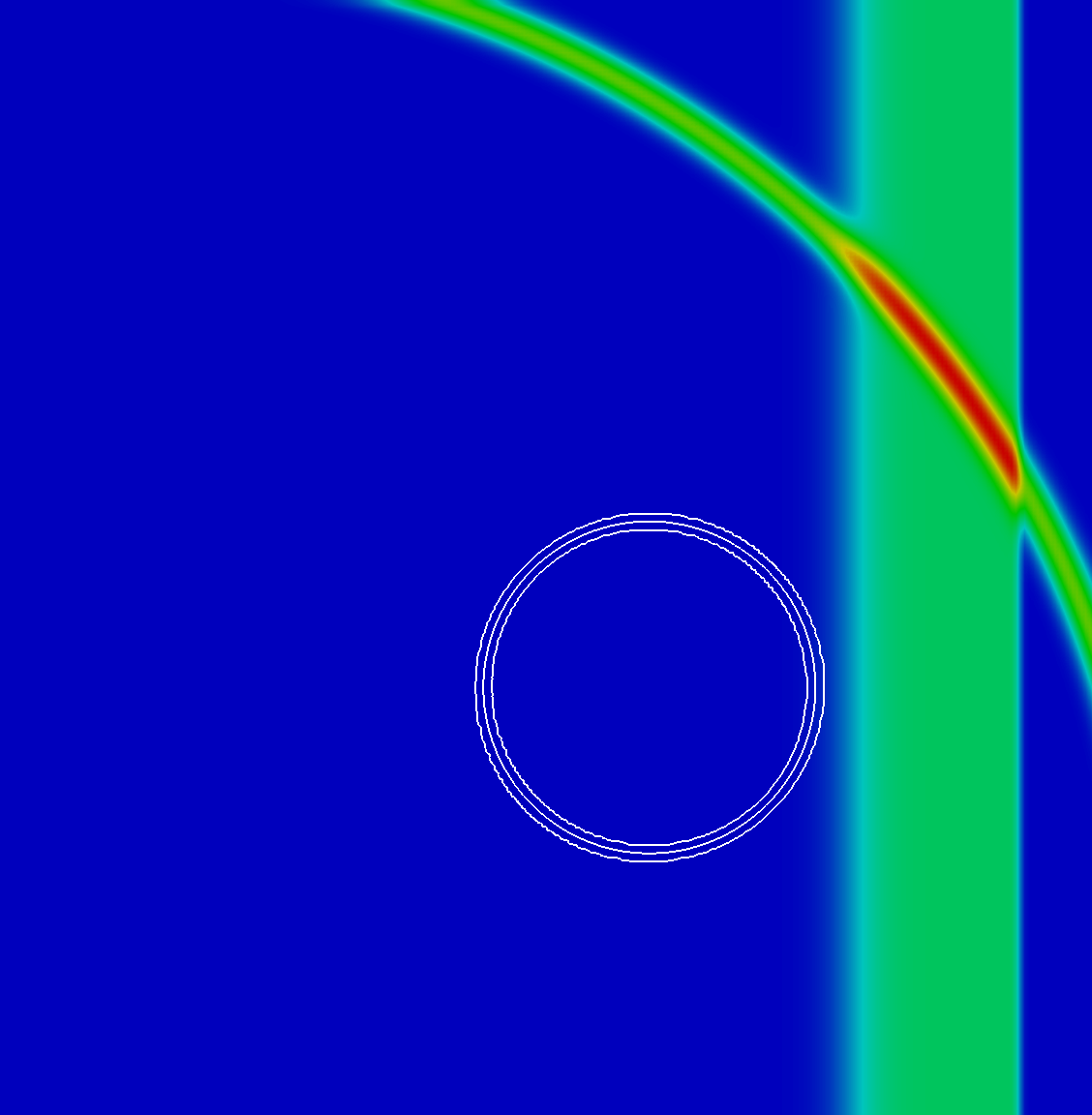}
	\end{minipage}%
	\begin{minipage}{.32\textwidth}
		\centering\includegraphics[width=.825\textwidth,height=.825\textwidth,angle=0]{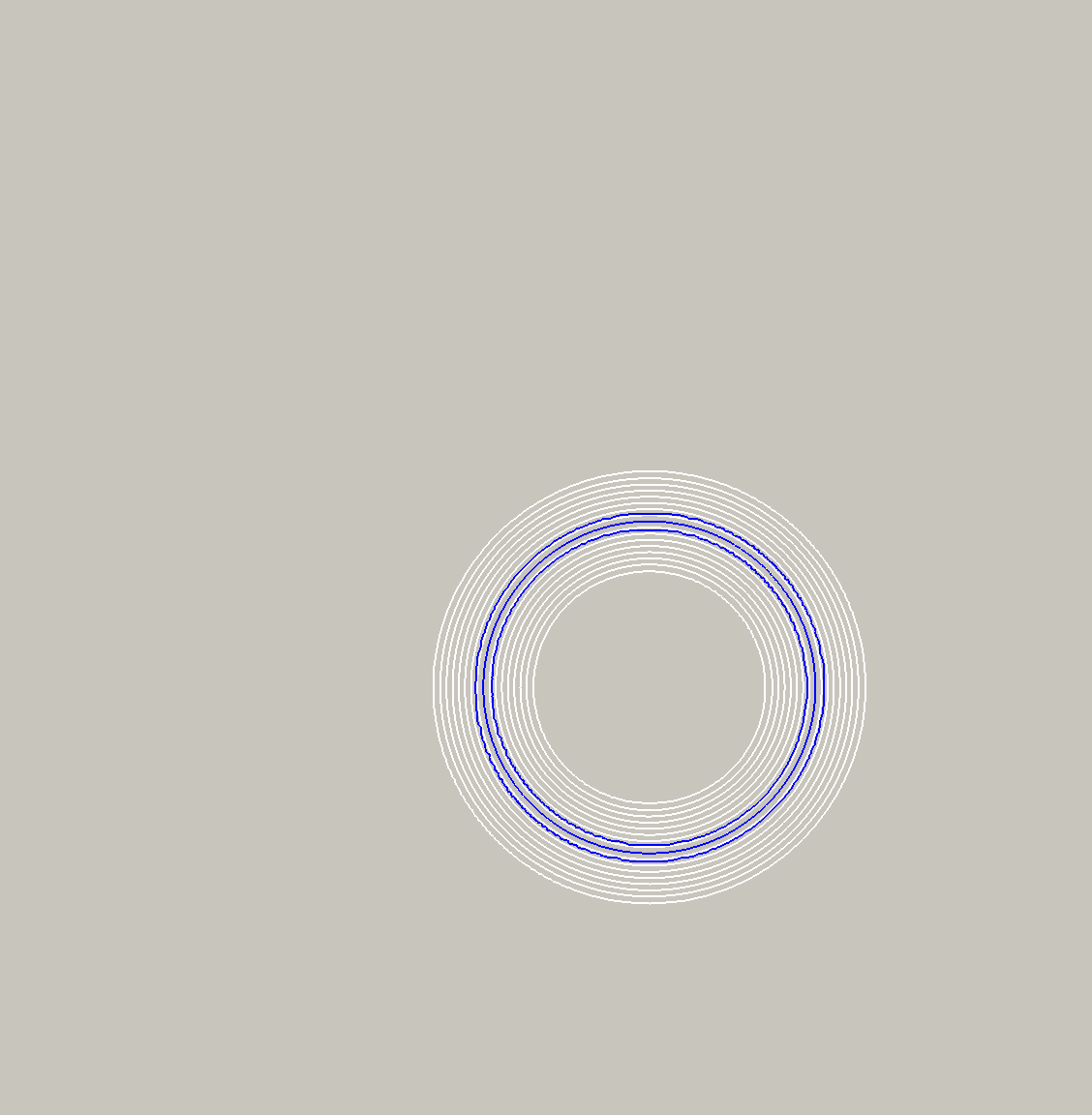}
	\end{minipage}
	\caption{\small{Evolution of the intermittency region affected by
		       	the variable field $\ephxt$.
		       	All figures show contours of $\alpha(\psi \!=\! -4\ephb),
		       	\alpha(\psi \!=\! 0),\alpha(\psi \!=\!4\ephb)$,
		        and from left to right fields 
		        of $0 \!<\! \alpp \!<\! 1$,
		        $\epsilon_{h,b} \!\le\! \ephxt \!<\! 5\epsilon_{h,b}$,
		        contours $-32\ephb \!\le\! \psia \!\le\! 32\ephb $.
	         }}
	\label{fig11}
\end{figure}
\elnm 
In order
to obtain
the full rotation
of the drop 
in the given 
velocity field $\bu$,
$\Delta t =2.5\cdot 10^{-3}\,[s]$
and $N_t=2560$ time
steps are required 
(the Courant
number is $\sim 0.65$).
The verification 
of the numerical methods
and discretization 
of \Eqs{eq17}{eq18}
is described in detail
in the previous works
of the present 
author \citep{twacl15, twacl17}.
Herein, 
in \ref{appD}
the derivation
of the Lagrangian
scheme
used to solve
\Eq{eq17} in
the known
velocity field $\bu$
is recalled.

The intermittency
region 
surrounding
the advected,
circular 
interface
is disturbed
by the variable
characteristic
length scale
field $\ephxt$
defined using
\Eq{eqC6} as
the linear 
superposition
of \Eqs{eqC1}{eqC2}.
The results 
shown
in \Figs{fig10}{fig11}
present
subsequent
time moments $it \!\cdot\! \Delta t$ 
(from top to bottom)
in the 
history
of 
the advected
circular bubble
the regularized
interface of which
is disturbed
by the variable 
$\ephxt$
field.
Each row
in \Figs{fig10}{fig11}
shows the same
three iso-contours
$\alpha \lr \psi\!=\!-4\ephb \rr$,
$\alpha \lr \psi\!=\! 0 \rr$,
$\alpha \lr \psi\!=\! 4\ephb \rr$
set
against
$\alpp$, $\ephxt$, $\psia$
fields,
from left to right  
respectively.
The third column 
in \Figs{fig10}{fig11}
presenting contours
of the signed 
distance function $\psia$
is added to show
the variation
of the cumulative distribution
$\alpp$ 
is predicted
in the region
where $\bng$
is correctly
defined,
see \Sec{ssec41}
for the description
of the numerical
method.

One observes
that the hat-like
profile moving 
in 
the horizontal
direction
has a steep front
and smooth
tail, see \Figs{fig10}{fig11}.
It interferes
with
the bell shaped
axis-symmetrical
characteristic
length scale
variation
resulting
in the
increase of
its local
values,
the red color moving
across 
the computational
domain in
the middle
column 
of \Figs{fig10}{fig11};
the  $\ephxt$
field is 
bounded between
$\ephb \!<\! \ephxt \!<\! 5 \ephb$.
We note
the introduced
numerical 
model  
is sensitive
to 
the rapid
changes 
in the magnitude of $\ephxt$.
For example,
in \Fig{fig10}, 
$it\!=\!320,\,560$
one observes
(along the circumference
of axis-symmetrical variation)
how the width
of the intermittency 
region
is affected
by variable
$\ephxt$.
The proposed
numerical
method  
is 
sensitive
to the local
variations
in the
$\ephxt$
profile
smoothness, too.
In \Fig{fig11}
$it\!=\!1280$,
one notes
(along the 
left-right
borders of
the step disturbance)
how the steepness
of the $\ephxt$
profile 
affects
the
$\alpp$
field.
As it
can be observed
in the last
time moment
presented 
in \Fig{fig11},
in 
the domain
where $\eph \!=\! \ephb$
the intermittency
region returns
back
to its original,
equilibrium
state.
As it
was mentioned
previously
this
is achieved
by
the definition
of  
quadrature
(\ref{eq31})
and
integral
in 
\Eq{eq29}.
%

\section{Conclusions} 
\label{sec5}
%
In
the 
present 
paper,
the  non-equilibrium
model
of the 
intermittency
region between
two weakly
miscible
phases
is introduced.
This new 
multi-scale
model is
planned
for the
framework
of
one-fluid
model
of
the
(turbulent)
two-phase
flow.
At first,
the evolution
equation of
the intermittency
region
is derived 
from the
stochastic
viewpoint,
see \Eq{eq12}.
Next, 
based on 
the mesoscopic 
and macroscopic
interpretations
of the intermittency
region,
conditions
of its equilibrium 
and non-equilibrium
state
are
explicated,
see \Sec{ssec21}.
The  statistical
interpretation
of the solution 
to \Eq{eq12}
is used to argue that
the sharp
interface
tracked
or
captured 
in   
VOF, SLS
models
is localized
inside the
mesoscopic
intermittency
region (gas-liquid
macroscopic
interface)
remaining
in the 
equilibrium
state, 
see \Sec{ssec22}.
It is 
explained
that
the 
level-sets
$H_\gamma \!=\!1/2$, $\psi\!=\!0$
are
two-dimensional
representations
of
the expected
position of
the mesoscopic
interface $\Gamma$
disturbed by
thermal
fluctuations.
This result
unfolds
how 
molecular
effects
are 
taken 
into
account
in  
VOF, SLS
sharp interface
models 
and
answers 
how they 
reconstruct
break up and coalescence
\citep{perumanath19}.

The second
part of 
the present paper
refers
to 
the case
where
the intermittency
region 
could
be in
the non-equilibrium
state as well.
It is demonstrated
that
finding 
the stationary
solution
of \Eq{eq12}
when
the
characteristic
length scale $\ephxt$
characterizing
the local thickness
of the intermittency region
 $\eph\!\ne\!const.$
is equivalent
to minimizing
of 
the
corresponding
free energy
functional,
see \Sec{ssec31}.
This result
qualifies
the stationary
solution to
\Eq{eq12}
as
the local equilibrium 
condition
accounting
for $\ephxt$.
In \Sec{ssec32},
it is used
to derive
the modified
mapping 
between
the
$\alpp\!-\!\psia$ 
functions and
the
semi-analytical
model of the
evolving intermittency
region, see \Eq{eq28}.
In \Sec{sec4}
the new 
semi-numerical
methods
for
the non-equilibrium
solutions of \Eq{eq12} 
are introduced.
In Sections (\ref{ssec42}-\ref{ssec44})
they are  employed
to reconstruct
solutions of
\Eq{eq12} with
variable $\ephxt$
showing 
the complex
behavior
of the intermittency
can be predicted with
the approximate,
semi-analytical model.

It is anticipated
the semi-analytical
approach
introduced
in the present paper
could 
be used
with the existing
numerical 
sharp/diffusive
interface models
to approximate
the effects of
intermittency region
non-equilibrium 
on the flow field.
The only requirement
is reconstruction
of the signed distance
function field
based on the known
expected position
of the gas-liquid
interface $\gamma$.

The modeling
framework
introduced
in the present paper,
is planned 
to be used 
in future
statistical
models of
the macroscopic 
interface agitated
by turbulent fields,
or
mesoscopic interface
affected by variable
thermal energy,
pressure and/or 
concentration
variations.

\section*{Acknowledgments}
This work is supported by the grant
of National Science Center, Poland
(Narodowe Centrum Nauki, Polska)
in the project 
\emph{``Statistical modeling of turbulent two-fluid flows with interfaces''},
ref. no. 2016/21/B/ST8/01010, ID:334165.


\appendix

\section{Exact relations in surface averaging}
\label{appA}

To derive \Eq{eq12}
two exact relations
between the ensemble 
$\av{\cdot}$ 
and surface $\avG{\cdot}\Sigma$
averages
are used.
The first one 
is obtained
directly from
\Eq{eq3} as 
it implies
\blnm
\be
 \av{\bW'\!\cdot\! \nabla H_\Gamma} \!=\! \avG{\bW'\!\cdot\!\bnG} \Sigma.
\label{eqA1} 
\ee
\elnm
The second 
exact relation $\avG{\bnG}\Sigma = \nabla \av{H_\Gamma}$ 
can be derived
starting from
the definition
of the phase 
indicator function \citep{trygg11}
\blnm
\be
H_\Gamma \lr x,y,z,t \rr 
\!=\! \iiint_{V} 
     \delta \lr x-x' \rr  
	 \delta \lr y-y' \rr 
	 \delta \lr z-z' \rr dx'dy'dz' 	 
\label{eqA2}
\ee
\elnm
and its gradient
\blnm
\be
\nabla H_\Gamma \lr x,y,z,t \rr 
\!=\! - \iint_{\Gamma} (-\bnG') 
    \delta \lr x-x' \rr  
	\delta \lr y-y' \rr 
	\delta \lr z-z' \rr  	
	d \Gamma '.
\label{eqA3}
\ee
\elnm
In \Eq{eqA3}
relation  
$\nabla \delta \lr \bx-\bx' \rr \!=\! - \nabla' \delta \lr \bx-\bx' \rr $
is used,
moreover,
it is noticed 
in the divergence
theorem unit vector 
points outwards
surface $\Gamma$ 
unlike the normal
vector 
$\bnG\!=\!\nabla \Psi/|\nabla \Psi|$.
According 
to \citep{pope88},
\Eq{eqA3} 
can be rewritten as
\blnm
\be
\nabla H_\Gamma \lr \bx,t \rr 
\!=\!  \iint_{\Gamma} \bnG \lr  \mu,\lambda,t \rr 
\delta \lr \bx-\bx' \lr \mu,\lambda,t \rr \rr  
A \lr \mu,\lambda, t \rr d \mu d \lambda 
\label{eqA4}
\ee
\elnm
where $\delta \lr \bx-\bx' \lr \mu,\lambda,t \rr \rr\,[1/m^3]$
is the three dimensional
Dirac's delta function.
Therefore,
in the local
orthonormal 
coordinate system
$\mu,\lambda,\Psi$
of infinitesimally 
small surface
element $d\Gamma'\!=\!A \lr \mu,\lambda,t \rr d \mu d \lambda$,
where $\Psi$ is the coordinate
in the normal direction, 
\Eq{eqA4} reads
\blnm
\be
\nabla H_\Gamma \lr \Psi \rr 
\!=\!  \iint_{\Gamma} \bnG 
\delta \lr \mu \rr  
\delta \lr \lambda \rr 
\delta \lr \Psi-\Psi'\lr\mu,\lambda,t \rr \rr  	
A \lr \mu,\lambda,t \rr d \mu d \lambda
\!=\! \deltaGp \bnG
\label{eqA5}
\ee
\elnm
in addition, 
\Eq{eqA5}
let us show 
\blnm
\be
 \pd{H_\Gamma \lr \Psi \rr}{\Psi} \!=\! \deltaGp
\label{eqA6}.
\ee
\elnm
Using the surface average 
definition given
by \Eq{eq4},
the surface average 
of the normal vector $\bnG$
is obtained as
\blnm
\be
\avG{\bnG} \Sigma \!=\! \iint_{\Gamma} 
\av{ \bnG
	\delta \lr \mu     \rr  
	\delta \lr \lambda \rr
	\delta_\Gamma \lr \Psi\!-\!\Psi' \lr \mu,\lambda, t \rr \rr
	A\lr \mu,\lambda, t \rr} d\mu d\lambda
\label{eqA7}
\ee
\elnm
where $\Sigma$
is defined
by \Eq{eq6}.
Finally,
taking the ensemble 
average $\av{\cdot}$
of \Eq{eqA5}
and comparing with \Eq{eqA7}
results in 
the second exact
relation
\blnm
\be
\nabla \av{H_\Gamma} \!=\! \avG{\bnG} \Sigma.
\label{eqA8} 
\ee
\elnm
%

\section{Calculation of the free energy functional derivative}
\label{appB}

In order to compute
the functional derivative 
of \Eq{eq19} with $\ephxt$
we use the following 
definition.
Given a functional
\blnm
\be
 G \ls \alpha \rs = \int_V g \lr \psi, \alpp, \nabla \alpp \rr dV,
\label{eqB1}
\ee
\elnm
its functional
derivative
is obtained as
\blnm
\be
 \pdF{G}{\alpha} \!=\! \pd{g}{\alpha} \!-\! \nabla \!\cdot\! \ls \frac{\partial g}{\partial \nabla \alpha} \rs.
\label{eqB2}
\ee
\elnm
The first
term in \Eq{eqB2}, 
where $G \!=\! F$
and $F$
is given by \Eq{eq19},
results in
\blnm
\be
\pd{f}{\alpha} \!=\! \sigma \ls \frac{2 \alpha \lr 1\!-\!\alpha \rr}{\eph} \lr 1\!-\!2\alpha \rr \!+\! \pdF{k}{\alpha} \rs  .
\label{eqB3}
\ee
\elnm
Since $\pdFl{k}{\alpha}$ 
is given by \Eq{eq20}
and it does not depend
explicitly on
$\ephxt$, 
\Eq{eqB3} reads
\blnm
\be
\pd{f}{\alpha} \!=\! 
2 \sigma \ls \frac{ \alpha \lr 1\!-\!\alpha \rr}{\eph} \lr 1\!-\!2\alpha \rr 
\!+\! \alpha \lr 1-\alpha \rr \nabla \cdot \bng \rs.
\label{eqB4}
\ee
\elnm
The second 
term in \Eq{eqB2},
where $G\!=\!F$
and $F$ is given
by \Eq{eq19},
equals
\blnm
\be
\nabla \!\cdot\! \ls \frac{\partial f}{\partial \nabla \alpha} \rs \!=\!
2\sigma \lr \eph \nabla^2 \alpha + \nabla \eph \!\cdot\! \nabla \alpha \rr.
\label{eqB5}
\ee
\elnm
Therefore, to minimize 
the functional $F \ls \alpha \rs$,
we search for
\be
\pdF{F}{\alpha} \!=\! \eph \nabla^2 \alpha \!+\! \nabla \eph \!\cdot\! \nabla \alpha \!-\! 
\frac{\alpha \lr 1-\alpha \rr}{\eph} \ls  \lr 1\!-\!2\alpha \rr \!+\! \eph \nabla \cdot \bng \rs
\!=\!0.
\label{eqB6}
\ee
%

\section{Variations of the characteristic length scale field}
\label{appC}
%
In the present
work it is assumed 
variations of
the characteristic
length scale 
of the regularized 
interface $\ephxt$
are known and 
given by the
predefined formulas.
The step profile  
\blnm
\be
\ephxtS \!=\!  \ephb  \ls 1 + \frac{H_S}{1 + \exp{ \lr -\frac{ f_S \lr \bx, t \rr }{ W_S \ephb} \rr }}  \rs
\label{eqC1}
\ee
\elnm
or the bell 
shaped profile
\blnm
\be
\ephxtB \!=\!  \ephb  \ls 1 + H_B \exp{ \lr -\frac{ f_B \lr \bx,t \rr }{W_B \ephb} \rr }^2  \rs.
\label{eqC2}
\ee 
\elnm
In the one-dimensional cases 
presented in \Figs{fig4}{fig5},
$H_S \!=\! H_B\!=\!W_S\!=\!1$,
$W_B\!=\!6$
and 
$\ephb \!=\! \Delta x$.
The functions $f_S$, $f_B$ 
in \Eqs{eqC1}{eqC2} 
are both set to 
$f(x,t)\!=\! x-x_\Gamma+\ephb \lr 36-it \rr$
where $x_\Gamma\!=\!0.5$
is position 
of the interface and
$it \!=\!1,\ldots,72$ denotes 
the physical time iteration number.

In the two dimensional case 
studied in \Sec{ssec43},
the variation
of $\ephxt$ presented
in \Figs{fig8}{fig9}
is obtained as 
\blnm
\be
\ephxt = \frac{1}{2} \ls \ephxtS + \ephxtB \rs 
\label{eqC3}
\ee
\elnm
where in \Eq{eqC2}
 \blnm
 \be
  f_B \lr \bx, t \rr = R - R_D + \ephb \lr 72 -it \rr
 \label{eqC4}
 \ee
and $\ephb \!=\! \sqrt{2} \Delta x/4$, 
$R\!=\!\sqrt{\lr \bx-\br \rr^2}$, $\br\!=\!\lr 0.5,0.5 \rr$
determines the center  and $R_D=0.2$
sets initial radius
of axis-symmetrical 
part of  $\ephxt$.
 In \Eq{eqC3} the step, oblique
 variation  is governed by
\be
  f_S \lr \bx, t \rr = -( x \!-\! 0.62 ) \!-\! (y-0.62 ) \!-\! \ephb \cdot it 
\label{eqC5}
\ee
\elnm
moreover 
$H_B\!=\!H_S\!=\!4$, $W_B\!=\!8$,
$W_S\!=\!2$,
 $it\!=\!1,\ldots,16$.

In \Sec{ssec44}
where
the semi-analytical
solution 
with
advection
is studied,
the following 
superposition
of $\ephB$
and $\ephS$ 
is employed
\blnm
\be
\ephxt = \frac{1}{2} \ls \ephb + \ephxtSo -\ephxtSt + \ephxtB \rs 
\label{eqC6}
\ee
\elnm
where $\ephb \!=\! \sqrt{2} \Delta x/4$,
variation 
of $\ephB$
with $H_B\!=\!5$, 
$W_B\!=\!12$,
is 
carried out using
the function $f_B(t)$
given by \Eq{eqC4} where
$R_D\!=\!0.2$ , 
$\br\!=\!\lr 0.1,0.1 \rr$. 
Step profiles
in \Eq{eqC6}
are defined 
using $H_{S1}\!=\!H_{S2}\!=\!3$,
$W_{S1}\!=\!1$
and 
$W_{S2}\!=\!10$,
moreover,
they are driven
in time by
the functions
$f_{S1}(x,t)\!=\! x\!-\!0.15\!-\!0.25 \ephb \!\cdot\! it$
and $f_{S2}(x,t)\!=\!f_{S1}(x,t)\!+\! 0.15$,
respectively.

\section{Derivation of the Lagrangian scheme, constrained interpolation }
\label{appD}
%
The rearrangement
of terms in 
the advection equation
(\ref{eq17}) 
leads to
\blnm
\be
\frac{1}{\alpha \lr 1 \!-\! \alpha \rr}\pd{\alpha}{t} \!=\! -\frac{1}{\eph} |\nabla \psi| \bw \!\cdot\! \bng.
\label{eqD1}
\ee
\elnm
The left hand side is now integrated
between $\alpha^n$ and $\alpha^{n+1}$,
whereas the right hand side between
$t^n$ and $t^{n+1}$ resulting in 
\blnm
\be
\ln{\lr \frac{ \alpha}{1-\alpha} \rr} \Biggr|_{\alpha^n}^{\alpha^{n+1}}
\!=\! - \int_{t^n}^{t^{n+1}} \!\! \frac{1}{\eph}   |\nabla \psi| \bw \!\cdot\! \bng  dt,
\label{eqD2}
\ee
\elnm
where $n,\,n\!+\!1$
denotes old and new time levels,
respectively.
Integration 
given by \Eq{eqD2} 
derives 
the
following scheme 
for advancement
of $\alpp-\psia$ in time $t$,
given by the formula
\blnm
\be
\alpha^{n+1} = \frac{\alpha^n \exp{\ls I\lr t^n \rr \rs}}{1-\alpha^n \lr 1- \exp{\ls I\lr t^n \rr \rs} \rr},
\label{eqD3}
\ee
\elnm
where the RHS integral in \Eq{eqD2}
is denoted as $I(t^n)$.
This integral 
must be approximated
by the appropriate quadrature;
in the present work we adopt the
second-order Adams-Bashfort
method leading  to 
\blnm
\be
I \lr t^n \rr \approx -\ls \frac{3}{2} f\lr t^{n},\psi^{n} \rr
- \frac{1}{2}  f \lr t^{n-1},\psi^{n-1} \rr \rs  \Delta t,
\label{eqD4}
\ee
\elnm
where $f\!=\!|\nabla \psi| \bng \!\cdot\! \bw/\eph$.
The semi-analytical, explicit 
scheme given by \Eqs{eqD3}{eqD4}
is second-order accurate in time 
and no spatial discretization
of $\alpp$ is needed.
It is noted that
in the present
work $\eph^{n+1}\!=\!\eph^{n}$
as the advection equation (\ref{eq17})
is always solved 
with $\ephb\!=\!const.$
in the semi-analytical case.

During
numerical
solution 
of \Eq{eq18}
in time $\tau$,
to obtain $\alpha_{dc}$
shown in \Figs{fig4}{fig5} and \Fig{fig7},
the constrained interpolation
\citep{twacl17}
is used to determine
$\deltaa \!=\! \alpha \lr 1\!-\!\alpha \rr$.
The constrained 
interpolation
in the present
work  is
summarized below
\blnm
\begin{align}
\begin{split}
\psi_f &\approx \frac{1}{2} \lr \psi_P + \psi_F \rr + \mathcal{O} \lr \Delta x^2\rr, \\
\alpha_f &= \alpha \lr \psi_f \rr = \frac{1}{1+\exp{\lr -\psi_f I \lr \psi \rr \rr}},
\label{eqD5}
\end{split}
\end{align}
\elnm
where subscripts $F,f,P$ 
denote  the neighbor control volume $F$ 
and face $f$ of the given 
control volume $P$, 
respectively.
$I\lr \psi \rr$ is
the quadrature defined
by \Eq{eq30} or \Eq{eq31}. 

\section{Error norms}
\label{appE}

To show
convergence
during  
the numerical
solution 
of \Eqs{eq18}{eq27}
where $\alpp$
is disturbed by
$\ephxtS$ or $\ephxtB$ 
(see \Eqs{eqC2}{eqC1}, respectively),
in Figs.~\ref{m1fig6}, \ref{m2fig6}, \ref{m3fig6}
the $L_{1,\tau}$ error 
norm is plotted
after each
physical
time iteration $it$.
This first-order norm
is defined as follows
\blnm
\be
 L_{1,\tau} = \frac{1}{N_c} \sum_{i=1}^{N_{c}} |\alpha_i^{n+1}-\alpha_i^{n}|,
\label{eqE1}
\ee
\elnm
where $N_c$ is the number of
control volumes and $n\!+\!1$ denotes
a new level of the pseudo-time  $\tau$. 

In \Fig{nfig10} 
the norms
$L_{1,S} (\alpha)$
and $L_{1,B} (\alpha)$ 
are used.
They are employed to 
access
the spatial error
of the 
integration
procedure
introduced
in \Sec{ssec41}
during coupled (dc)
and semi-analytical (d)
solutions.
Their 
definition
uses 
the analytical
profiles  
$\alpha_{cnv} \lr \bx, T \rr$,
representing known,
equilibrium
solutions of \Eqs{eq18}{eq27} 
after predefined
physical 
time $T\!=\! it \!\cdot\! \Delta t$
where herein $it=72$, $\Delta t=10^{-3}\,[s]$. 
$\alpha_{cnv} \lr \bx,T  \rr$
is
defined by \Eq{eq13}
with 
$\eph \!=\! 2 \ephb$ or
$\eph \!=\!   \ephb$
in  
the case 
when $\eph\!=\!\ephS$ 
or $\eph\!=\!\ephB$,
respectively.
If discretized
analog
of  $\alpha_{cnv} \lr \bx,T \rr$ is given by $\alpha_{i,cnv} $
the norm $L_{1,D} \lr \alpha \rr$, where $D\!=\!S$ or $B$
is defined at each time iteration $it$, as
\blnm
\be
L_{1,D}^{it} = \frac{1}{N_c} \sum_{i=1}^{N_{c}} |\alpha_i^{it}-\alpha_{i,cnv}|.
\label{eqE2}
\ee
\elnm
%

\bibliography{mybibfile}

\end{document}